\documentclass[11pt]{article}
\usepackage{epsfig}
\usepackage{amsfonts}
\usepackage{amsmath}
\usepackage{bbm,bm}
\usepackage{float}
\allowdisplaybreaks[1]

\usepackage{amsthm}
\usepackage{amssymb}
\usepackage{graphicx}
\usepackage{tikz}
\usepackage{pgfplots}
\usepackage{chngcntr}

\usepackage[
  colorlinks=true,
  linkcolor=black,
  citecolor=black,
  filecolor=black,
  urlcolor=black,
  breaklinks=true]{hyperref}

\usepackage[utf8]{inputenc}

\usepackage[
    backend=biber, 
    style=numeric-comp, 
    giveninits=true,
    sorting=none, 
]{biblatex}
\AtEveryBibitem{\clearfield{title}}

\usepackage{ulem}
\def\lsi{\raise0.3ex\hbox{$<$\kern-0.75em\raise-1.1ex\hbox{$\sim$}}}
\newcommand{\lsim}{\mathop{\lsi}}
\newcommand{\rmi}[1]{{\mbox{\scriptsize #1}}}
\renewcommand{\vec}[1]{{\bf #1}}
\newcommand{\eqal}[2]{\begin{equation}\begin{aligned}  #2 \end{aligned}\end{equation}}

\makeatletter \@addtoreset{equation}{section} \makeatother

\makeatletter
\renewcommand\section{\@startsection{section}{1}{\z@}%
  {-5.5ex \@plus -1ex \@minus -.2ex}
  {2.3ex \@plus.2ex}%
  {\normalfont\large\bfseries}}
\renewcommand\subsection{\@startsection{subsection}{2}{\z@}%
  {-3.25ex\@plus -1ex \@minus -.2ex}%
  {1.5ex \@plus .2ex}%
  {\normalfont\normalsize\bfseries}}
\renewcommand\thesection{\@arabic\c@section}
\renewcommand\thesubsection{\thesection.\@arabic\c@subsection}
\renewcommand{\@seccntformat}[1]{%
  \csname the#1\endcsname.\hspace{1.0em}}
\makeatother

\graphicspath{
  {./figures/}
}

\begin{document}

\flushbottom

\begin{titlepage}

\begin{flushright}
HIP-2025-8/TH
\end{flushright}
\begin{centering}

\vfill

{\Large{\bf
An Effective Sphaleron Awakens
}}

\vspace{0.5cm}

\renewcommand{\thefootnote}{\fnsymbol{footnote}}
Xu-Xiang Li$^{\rm a}$%
\footnotemark[3]%
, 
Michael J.~Ramsey-Musolf$^{\,\rm{}b,c,d,e}$%
\footnotemark[1]%
, \\  
Tuomas V.~I.~Tenkanen$^{\rm b,c,f,g}$%
\footnotemark[2],
and 
Yanda Wu$^{\rm b,c}$%
\footnotemark[4]

\vspace{0.5cm}

$^\rmi{a}$%
{\em 
Department of Physics and Astronomy,
University of Utah, \\
Salt Lake City, Utah 84112, USA
\\}

\vspace*{0.11cm}

$^\rmi{b}$%
{\em 
Tsung-Dao Lee Institute,\\
Shanghai Jiao Tong University, Shanghai 201210, China
\\}

\vspace*{0.11cm}

$^\rmi{c}$%
{\em
Shanghai Key Laboratory for Particle Physics and Cosmology,
Key Laboratory for Particle Astrophysics \& Cosmology (MOE),
Shanghai Jiao Tong University,
Shanghai 200240, China
\\}

\vspace*{0.11cm}

$^\rmi{d}$%
{\em
Amherst Center for Fundamental Interactions, Department of Physics,\\
University of Massachusetts, Amherst,
MA~01003, USA
\\}

\vspace*{0.11cm}

$^\rmi{e}$%
{\em
Kellogg Radiation Laboratory, California Institute of Technology,\\ 
Pasadena, CA~91125, USA
\\}

\vspace*{0.11cm}

$^\rmi{f}$%
{\em
Nordita,
KTH Royal Institute of Technology and Stockholm University,\\
Hannes Alfv\'ens v\"ag 12,
SE-106 91 Stockholm,
Sweden\\}

$^\rmi{g}$%
{\em
Department of Physics and Helsinki Institute of Physics,
PL 64, FI-00014 University of Helsinki,
Finland\\}

\vspace*{0.6cm}

\mbox{\bf Abstract}

\end{centering}

\vspace{0.3cm}

\noindent

Using thermal effective field theory, we present a self-consistent perturbative formulation of the Higgs phase sphaleron rate after a radiatively-induced first-order phase transition. This gauge-invariant formulation is based on dimensionally reduced effective field theory (3D EFT) at high temperatures and paves a way for including higher order corrections within the 3D EFT perturbation theory without double counting. Concretely, we compute the Higgs phase sphaleron rate in a semi-classical approximation within the two-loop resummed 3D EFT. We find compact results for the sphaleron rate and the baryon washout factor as well as criteria for the baryon number preservation by providing a clear connection to the results obtained using a similar 3D EFT description for the bubble nucleation. We demonstrate these calculations for the real triplet-extended Standard Model, and conclude that when all two-loop thermal effects for the matching are accounted, no sufficiently strong one-step electroweak phase transitions exist within the parameter space regime that can be mapped onto the 3D EFT we have considered. 

\vfill
\end{titlepage}

\tableofcontents
\renewcommand{\thefootnote}{\fnsymbol{footnote}}
\footnotetext[1]{mjrm@sjtu.edu.cn, mjrm@physics.umass.edu}
\footnotetext[2]{tuomas.tenkanen@helsinki.fi}
\footnotetext[3]{xuxiang.li@utah.edu}
\footnotetext[4]{yanda.wu7@sjtu.edu.cn}
\clearpage

\renewcommand{\thefootnote}{\arabic{footnote}}
\setcounter{footnote}{0}


\section{Introduction}
\label{sec:intro}

To this date, the origin of the observed matter-antimatter asymmetry of the universe still remains a mystery. 
The Standard Model (SM) of particle physics is unable to explain the small but non-vanishing ratio of net baryon over anti-baryon to photon number density%
\footnote{
Or alternatively, the baryon-to-entropy ratio $Y_B \equiv n_B/s = (8.70\pm 0.04)\times 10^{-11}$ \cite{Planck:2018vyg,Fields:2019pfx}. 
}
\begin{align}
\label{eq:BAU}
\eta \equiv \frac{n_B - n_{\bar{B}}}{n_\gamma} = (6.10 \pm 0.40)\times 10^{-10} \,,
\end{align}
observed from the Cosmic Microwave Background by the Planck satellite \cite{Planck:2018vyg} and the Big Bang Nucleosynthesis measurements \cite{Fields:2019pfx}.
Among the most widely studied mechanisms to explain the observed value for $\eta$ 
is the electroweak baryogenesis (EWBG) \cite{Kuzmin:1985mm} in which the baryon asymmetry is generated dynamically during a first-order electroweak phase transition (EWPT).
The EWBG is often considered attractive, as it links the baryon asymmetry to the Higgs mechanism and the origin of elementary particle masses, and is experimentally testable in particle colliders \cite{Ramsey-Musolf:2019lsf}.
For reviews on the EWBG, see e.g.~\cite{Rubakov:1996vz,Bernreuther:2002uj,Morrissey:2012db,Bodeker:2020ghk}, and for its alternative competitors, such as leptogenesis e.g.~\cite{Biondini:2017rpb} and references therein.

A successful mechanism to explain the observed value for $\eta$ requires three general ingredients identified by Sakharov \cite{Sakharov:1967dj}: (i) a baryon number violating process, (ii) charge (C) and charge-parity (CP) violating processes and (iii) departure from thermal equilibrium.
The condition (i) is obvious for generating a net baryon number from zero; the condition (ii) guarantees difference between the production rates of 
fermions and anti-fermions;
the condition (iii) is necessary because the thermal average of the baryon number always vanishes in thermal equilibrium, enforced by the CPT symmetry.

The condition (i) can be realized in the Standard Model (SM) due to the chiral nature of weak interactions, via the electroweak \textit{sphaleron} \cite{Manton:1983nd,Klinkhamer:1984di}.
In the semi-classical picture the electroweak sphaleron is an unstable topological field configuration of the SU(2) gauge-Higgs theory, which has half-integer Chern-Simons number separating topologically different vacua of the SU(2) Yang-Mills theory. 
A dynamic transition between vacua across a sphaleron, i.e. sphaleron transition, can induce an integer variation in the Chern-Simons number, leading to a variation of baryon ($B$) and lepton ($L$) numbers (while preserving $B-L$) through the Adler-Bell-Jackiw anomaly \cite{Adler:1969gk,Bell:1969ts}.
The rate of these transitions is referred to as the sphaleron rate, or more generally in non-abelian gauge theories, as the Chern-Simons diffusion rate of topological charge density \cite{Laine:2022ytc}. 

Conditions (ii) and (iii), on the otherhand, fall short: the SM does not have a first-order phase transition (in which the plasma surrounding expanding bubble walls would be driven out of equilibrium) but rather a smooth crossover \cite{Kajantie:1996mn,Csikor:1998eu}. 
In addition, the CP-violating processes in the SM (observed in neutral kaon oscillations, and recently, in baryon decays \cite{LHCb:2024exp}), associated with the complex phases of the Cabibbo-Kobayashi-Maskawa matrix, are too heavily suppressed at high temperatures \cite{Shaposhnikov:1987pf, Farrar:1993hn, Farrar:1993sp, Gavela:1994dt, Brauner:2011vb, Brauner:2012gu}.

Therefore, realizing the EWBG mechanism requires ingredients of physics beyond the Standard Model (BSM) that can make the conditions (ii) and (iii) viable. 
In principle, additional scalar fields in the Higgs sector could accomplish both: catalyze a first-order electroweak phase transition and introduce enhanced C- and CP-violating processes,%
\footnote{
We emphasize that the electron electric-dipole moment (EDM) precision measurements \cite{ACME:2018yjb} severely constrain the possibility of the generation of C- and CP-violations from BSM sectors \cite{Keus:2017ioh}.
Mechanisms for spontaneous violation occurring via dark sectors only at high temperatures, hence avoiding the EDM constraints, have been explored in \cite{Keus:2019szx,Huber:2022ndk}.  
}
see e.g.~\cite{Laine:1998wi,Alanne:2016wtx,Cline:2017jvp}.
In the present article, we do not discuss new sources for C- and CP-violation, but focus on the conditions (i) and (iii), which provide \textit{necessary}, yet not sufficient, conditions for the baryogenesis.

To accommodate a first-order electroweak phase transition, new BSM physics is required to be light enough and cannot interact with the SM Higgs boson too feebly \cite{Ramsey-Musolf:2019lsf}. 
These features make such models testable at the High-Luminosity run in the LHC \cite{Apollinari:2017lan} and in next-generation collider experiments, such as ILC \cite{ILC:2013jhg}, CLIC \cite{CLIC:2018fvx}, FCC \cite{FCC:2018byv} and CEPC \cite{CEPCStudyGroup:2018ghi}. 
This relatively short-term testability of the EWBG makes it an appealing candidate for the baryogenesis, and is under active studies, see e.g.~\cite{Liu:2011jh,Shu:2013uua,Inoue:2015pza,Chiang:2016vgf,deVries:2017ncy,Ramsey-Musolf:2017tgh,Bell:2019mbn,Cline:2020jre,Zhou:2020irf,Enomoto:2021dkl,Kainulainen:2021oqs,Cline:2021dkf,Modak:2021vre,Enomoto:2022rrl,Liu:2023sey,Aoki:2023xnn} and \cite{Ahriche:2014jna,Fuyuto:2014yia,Fuyuto:2015jha,Gan:2017mcv,Kharzeev:2019rsy,Wu:2023mjb,Hong:2023zrf,Hu:2023gbp,Qin:2024idc,Matchev:2025ivr,Matchev:2025irm,Tanaka:2025cpw} 
for sphaleron-focusing studies.
Furthermore, the existence of a first-order electroweak phase transition can also be probed by next-generation gravitational wave observatories, such as LISA \cite{Audley:2017drz,LISACosmologyWorkingGroup:2022jok} and other similar experiments \cite{Kawamura:2011zz,Guo:2018npi,TianQin:2015yph}, also c.f.~\cite{Kudoh:2005as,Crowder:2005nr,Yagi:2011wg,Gong:2014mca,Musha:2017usi,Hu:2017mde,Hu:2017yoc,Liang:2021bde}. 

 A modern approach to thermal first-order phase transitions and the bubble nucleation in perturbation theory in terms of thermal effective field theory (3D EFT) \cite{Kajantie:1995dw,Braaten:1995cm}
 have been formulated in \cite{Gould:2021ccf,Hirvonen:2021zej,Ekstedt:2022zro,Gould:2022ran,Lofgren:2023sep,Gould:2023ovu}.
 In this approach, a perturbative expansion is consistently organized based on strict power counting which accounts for relevant 
 mass hierarchies. For perturbative analysis of a first-order electroweak phase transition, the key element for a consistent perturbative expansion is that the leading order effective potential includes a barrier in-between its minima, generated by the gauge fields \cite{Moore:2000jw}.
In this article, we make a perturbative computation of the sphaleron rate in the Higgs phase compatible with this idea. 

The key aspects of our present analysis and findings include:
\begin{itemize}
    \item We investigate a minimal 3D EFT for the electroweak phase transition consisting of SU(2) gauge and Higgs fields, which serves as a universal 3D EFT for a broad class of Higgs-portal BSM theories where the new scalars are sufficiently heavy to be integrated out at high temperatures. It captures large two-loop thermal corrections (akin to \cite{Burnier:2005hp}) and allows one to implement a non-perturbative criterion for first-order phase transitions \cite{Kajantie:1995kf}. Moreover, key features of bubble nucleation can be computed with comparable rigor \cite{Gould:2022ran}.

    \item Using the semi-classical approximation for the Higgs phase sphaleron rate, we demonstrate that the leading order effective action of the sphaleron can be 
    obtained
    by retaining only the Yang-Mills term and the kinetic term of the Higgs field.
    The scalar potential enters solely the boundary condition,
    for which it is crucial to use the leading order effective potential where the potential barrier is generated by the gauge field \cite{Ekstedt:2022zro}. 
    We argue that since the energy scale of the sphaleron is higher than that of the critical bubble
    \cite{Gould:2021ccf,Hirvonen:2021zej,Gould:2022ran}, perturbative expansion for the sphaleron rate converges relatively faster compared to the bubble nucleation rate.

    \item After rescaling the sphaleron effective action, we find compact and gauge invariant results for the sphaleron rate as well as the washout factor, ultimately solving all concerns raised in \cite{Patel:2011th}. 
    In addition, we argue that our setup provides conceptual clarity for a consistent perturbative expansion when compared to the approach in \cite{Burnier:2005hp}. 

    \item Combining our results for the sphaleron rate and the washout factor with consistent perturbative calculation of the bubble nucleation rate, we provide a  unified, consistent description (Fig.~\ref{fig:yplot}) for the baryon number preservation criteria (BNPC) \cite{Kajantie:1995kf,Moore:1998swa,Patel:2011th} that incorporates EWPT thermodynamics, the bubble nucleation rate, and the sphaleron rate. In doing so, we provide a framework that allows one to track the sphaleron dynamics from the onset of nucleation to the completion of the transition. Self-consisency is ensured through adoption of the power counting in couplings given in Refs. 
    \cite{Ekstedt:2022zro,Hirvonen:2021zej,Lofgren:2021ogg}
    and renormalizaton group improvement as outlined in Ref.~\cite{Gould:2021oba}. With this framework in hand, we obtain a range of requirements -- \lq\lq strong\rq\rq\, to \lq\lq weak\rq\rq\, -- for preservation of any baryon asymmetry produced at the onset of nucleation.
    
    This analysis is encapsulated in Fig.~\ref{fig:yplot} and related discussion, which considers two relevant dimensionless parameters $x$ and $y$ constructed from the Higgs thermal self-interaction coupling, $\lambda_3$; the effective gauge coupling, $g^2_3$; and thermal mass-squared parameter, $\mu_3^2$. In this context,
    a strong version of this condition is given by%
    \footnote{
    We stress, that this upper bound on $x(T_c)$ is obtained in the semi-classical approximation, and is subject to higher order corrections.
    }
    \begin{align}
    x(T_c) \lesssim 0.025 \quad \text{(BNPC)} \qquad \text{with} \qquad 
    x(T) \equiv \frac{\lambda_3(T)}{g^2_3(T)} \,,
    \end{align}
    which guarantees to prevent the wash-out entirely. In addition, we obtain a weaker version of the BNPC as $0.025 \lesssim x(T_c) \lesssim 0.036$ for potential survival from exponential wash-out (for details, see Eq.~\eqref{eq:weak-BNPC}). 
    These conditions are universal for all theories that can be mapped onto the 3D EFT considered. Hence, our result generalizes some discussions in \cite{Moore:1998swa} by combining the result of the sphaleron rate with modern treatment for the bubble nucleation (at leading order in 3D EFT perturbation theory). 

    \item 
    We apply our analysis to one-step phase transitions in the real-triplet-extended Standard Model, focusing on a parameter regime where the triplet is heavy at high temperatures and can be integrated out. 
    While one-loop calculation suggests the existence of strong first-order transitions sufficient to evade sphaleron washout, we find that sizable two-loop thermal corrections significantly alter the picture.
    In fact, all identified first-order transitions become very weak at two-loop order. This underscores the critical role of two-loop thermal effects in Higgs-portal models and aligns with \cite{Niemi:2024vzw}.
    
\end{itemize}

In short, we are making sure that a first-order phase transition exists in accord with previous, universal non-perturbative lattice simulation results we can utilize -- as well as provide comparison to non-perturbative determination of the sphaleron rate, when possible -- and 
then determine the Higgs phase sphaleron rate perturbatively.
Concretely, in our computations in this article at hand, we formulate a consistent perturbative expansion that resolves issues of the gauge dependence and the slow convergence, that have plagued the previous literature (see Sec.~\ref{sec:discussion} for further discussion). 

The remainder of this article is organized as follows. 
In Sec.~\ref{sec:review} we provide a qualitative overview of the role of the sphaleron in the EWBG and a brief review of the developments in understanding the electroweak sphaleron rate.
In Sec.~\ref{sec:formulation} we compute the sphaleron rate in the Higgs phase, using SU(2) + Higgs 3D EFT, and compare our perturbative computation with lattice results from the literature for temperatures below the SM crossover. 
In Sec.~\ref{sec:BNPC} we find a perturbative, gauge-invariant formulation for the washout factor and baryon number preservation.
In Sec.~\ref{sec:bsm} we present an application to the real-triplet extended Standard Model that admits first-order phase transitions that can be described using the 3D EFT of previous sections. In Sec.~\ref{sec:conlusions} we summarize, provide further discussions and outlook future directions, while several technical details of our computations are relegated to appendices.
%

\section{Electroweak Baryogenesis and Sphaleron}
\label{sec:review}

We start with a qualitative tour to the EWBG and the sphaleron rate, before presenting our computations and quantitative discussions in sections that follow. 

\subsection{Qualitative Overwiew and Effective Descriptions}

\begin{figure}[ht!]
  \centering
\includegraphics[width=0.7\textwidth]{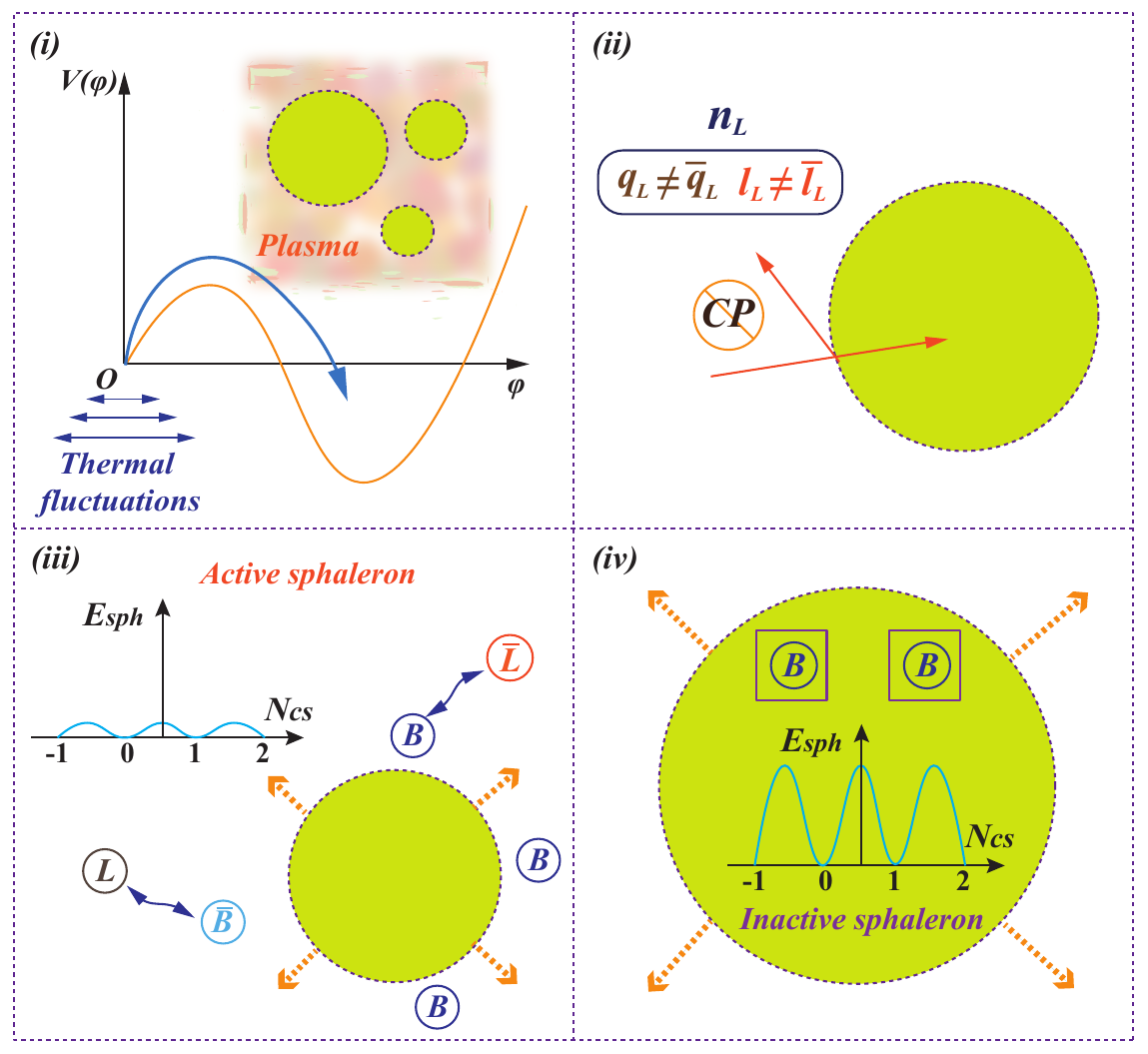}%
  \caption{
  A schematic illustration of four stages of the electroweak baryogenesis:
  (i) In a first-order phase transition the thermal fluctuations drive the system from a false to a true vacuum and transitions occurs via the bubble nucleation;
  (ii) The generation of a non-zero left-handed fermion number density
  from the C- and CP-violating scattering between particles and the expanding bubble wall;
  (iii) In the confinement phase outside the bubbles, the non-zero left-handed fermion number density would bias the rapid sphaleron process to generate more baryons than anti-baryons; such baryon excess is then devoured by the expanding bubble.
  (iv) As the sphaleron energy increases inside the bubble, the baryon number violating process is suppressed. Nevertheless, the produced net baryon number may be preserved, partly washed out, or completely erased, depending on the value of the washout exponent factor (See Eq.~(\ref{eq:washout_equation_sec2})).
}
\label{fig:EWBG_big_picture}
\end{figure}

Should underlying BSM physics help to catalyze a first-order electroweak phase transition, generation of baryon asymmetry could be realized in the following processes \cite{Morrissey:2012db} (illustrated in Fig.~\ref{fig:EWBG_big_picture}),%
\footnote{
See also \cite {Turok:1990in} for an alternative mechanism.
}
for which we highlight the effective descriptions for the bubble nucleation and the sphaleron rates in terms of semi-classical pictures.

\begin{itemize} 
    \item[(i)] In a thermal phase transition in the expanding universe,
    at temperatures ($T$) below the critical temperature, thermal fluctuations drive the system over the potential energy barrier 
    separating the high temperature ``symmetric'' phase and the low temperature ``broken'' electroweak Higgs phase.%
    \footnote{
    We note that these phases are analytically connected to each other \cite{Fradkin:1978dv}, as it is not possible to spontaneously break local, gauge symmetry \cite{Elitzur:1975im}. 
    Instead of using the term ``symmetric'' phase,  we stick to a more suitable terminology, e.g. the confinement phase. See also \cite{Damgaard:1985nb,Dosch:1995uz}.
    }
    A phase transition occurs through nucleation of bubbles, which eventually expand over macroscopic length scales and are subject to relativistic hydrodynamics. In microscopic scales, expanding bubbles disrupt the surrounding hot particle plasma out of equilibrium due to the released latent heat.
    The plasma can also be driven out of equilibrium by interactions of plasma particles with the bubble walls, turbulence and shock waves.
    
    In an effective description for the bubble nucleation \cite{Gould:2021ccf}, short-distance thermal fluctuations screen long-distance, highly occupied bosonic modes, hence modifying their interactions in a thermal medium, i.e. the hot plasma. Due to the high occupation these modes are effectively classical, yet subject to quantum corrections: this brings in a semi-classical approximation, in which the scalar field configuration undergoing the transition, or the critical bubble, is described by a classical \textit{bounce} solution $\varphi_\text{b}$. 
    
    In thermal equilibrium, the rate of the bubble nucleation is then described by a Boltzmann factor $e^{-S_{\text{eff}}(\varphi_\text{b})}$ with effective action evaluated on the bounce, and quantum corrections are described through the fluctuation determinant. 
    On the otherhand, the effective description for a dynamic, out-of-equilibrium contribution to the nucleation rate (related to damping and dissipation when the system crosses the energy barrier) is significantly more complicated.
    Once the bubble nucleation rate is rapid enough compared to the expansion rate of the universe, the entire system undergoes a phase transition.

    \item[(ii)]
    The C- and CP-violating scattering processes between the plasma particles and the expanding bubble wall lead to different diffusion rates among particles and their antiparticles and therefore a non-zero left-handed fermion number density in front of the bubble wall.
    An opposite asymmetry in the number density of right-handed particles $n_R$ is also generated by the CP violation, but it is irrelevant to the baryogenesis since the sphaleron process happens only between $\mathrm{SU}(2)_{L}$ vacua.
    The time scale for the process of $n_L$ creation is typically much shorter than the electroweak sphaleron transitions, so the sphaleron bias process in step (iii) is decoupled from (ii).
    For studies of CP-violating transport properties of the plasma see e.g.~\cite{Lee:2004we,Bodeker:2004ws,Cirigliano:2006wh,Fromme:2006wx,Fromme:2006cm,Cirigliano:2009yt,Cirigliano:2011di,Konstandin:2013caa,Kainulainen:2021oqs,Postma:2022dbr,Jukkala:2022jzy,Kainulainen:2024qpm,Li:2024mts} and references therein. 

    \item[(iii)]
   Outside the bubble, the rapid sphaleron process can induce a violation of 
   baryon plus lepton number, $B+L$, while keeping $B-L$ unchanged. The non-zero number of left-handed fermions created in (ii) would bias sphaleron towards creating more baryons than anti-baryons in front of the bubble. 
   Such baryon excess is then devoured by the expanding bubble.
    
    For each fermion generation, 
    the divergence of the left-handed baryonic and leptonic currents is proportional to the Yang-Mills field strength $F$ (here and below we omit the group indices for simplicity) and its dual \cite{Adler:1969gk,Bell:1969ts,tHooft:1976rip}:
\begin{equation}
\label{eq:anomaly}
	\partial_\mu j^{\mu}_B=\partial_\mu j^{\mu}_L \propto F \cdot \widetilde{F} \,.
\end{equation}
The temporal variation of the Chern-Simons number is then proportional to \cite{Moore:2000mx}
\begin{align}
    \Delta N_{\text{CS}}(t)=N_{\text{CS}}(t)-N_{\text{CS}}(t=0) \propto \int_{\Omega} F \cdot \widetilde{F} \, ,
\end{align}
where the integral is performed over both space and time.
The integer variation of the Chern-Simons number will pass over the sphaleron field configuration which has a characteristic half-integer Chern-Simons number (c.f.~\cite{Klinkhamer:1984di,Wu:2023mjb}). This implies that 
the Chern-Simons diffusion rate, $\Gamma_{\text{diff}} = \frac{\langle \Delta N_\text{CS}^2\rangle_T}{Vt}$ (in the large volume and time limit) or twice the sphaleron rate, $\Gamma_{\text{diff}}=2 \Gamma_\text{sph}$ (c.f.~\cite{Moore:1998ge, Burnier:2005hp})
controls the rate of change for the baryon number density \cite{Khlebnikov:1988sr,Quiros:1999jp,Patel:2011th} 
\begin{align}
\frac{\partial}{\partial t }n_B \propto -\frac{\Gamma_\text{sph}}{T^3}{} \, n_B \,,   
\end{align}
which we discuss in more detail in Sec.~\ref{sec:BNPC}.
We note that in the confinement phase,
the sphaleron rate $\Gamma_\text{sph}$ is unsuppressed, and a sophisticated combination of analytic effective descriptions and numerical simulations have found \cite{Bodeker:1999gx,Moore:2000mx}
\begin{align}
\Gamma_\text{sph}\propto \alpha_w^5T^4 \,,
\end{align}
where $\alpha_w=g^2/4\pi$, in terms of the SU(2) gauge coupling $g$. 
We stress that the thermal Yang-Mills theory is strongly coupled and confining,
and its dynamics -- dictated by the so-called ``magnetic'' mass scale -- are inherently non-perturbative \cite{Linde:1980ts,Shaposhnikov:1993jh,Braaten:1994na,Dosch:1996ex}.

    \item[(iv)] 
    
    Inside the bubble, due to the activation of the Higgs mechanism, length scale of gauge boson fluctuations becomes smaller compared to the confinement phase (i.e. the Higgs mechanism introduces a new, larger mass scale compared to the magnetic mass), and this situation is reflected in the sphaleron  dynamics. 
     The energy of the sphaleron increases, and this suppresses the Chern-Simons number diffusion.
    
    Three scenarios may occur: 

    \begin{itemize}
    
   \item[(I)] Immediately after the bubble nucleation, the baryon number violation rate (see Eq.~(\ref{eq:relation_B_var_and_sph})) inside the bubble falls below the Hubble rate, and no washout takes place (strong BNPC).
   
   \item[(II)] The baryon number violation rate remains above the Hubble rate for a finite interval after the nucleation temperature, $T_n$, erasing part of the generated baryon asymmetry (weak BNPC): 
    \begin{align} \label{eq:washout_equation_sec2}
    n_B(T_*) = e^{-\mathcal{W}} n_B(T_n) \,,
    \end{align}
    where $T_*<T_n$ is the decoupling temperature at which the baryon number violation rate and Hubble rate coincide, and $\mathcal{W}$ is the washout exponent (see Eq.~\eqref{eq:washout_exponent_definition}). In this case, C- and CP-violating interactions must overproduce baryon asymmetry, to offset the washout. 
    
    \item[(III)] The washout exponent $\mathcal{W}$ is so large that essentially all of the generated baryon asymmetry is erased.

    \end{itemize}

    With respect to the thermal medium (with short-distance thermal fluctuations), the sphaleron length scale scale is still large, and
    this implies, that the sphaleron in the Higgs phase admits a semi-classical approximation akin to the critical bubble. Yet we stress, that the energy scale of the sphaleron in the Higgs phase is much larger than that of the critical bubble, and consequently these two effective descriptions turn out to have important differences.
    
    In a semi-classical picture, the sphaleron is a classical, spatially varying scalar $\varphi_{\text{sph}}$ and gauge field $\vec A_{\text{sph}}$ configuration that can be found by extremizing the effective action. The sphaleron rate in thermal equilibrium can be computed from the Boltzmann factor
    \begin{align}
       \Gamma_\text{sph} \propto e^{-S_{\text{eff}}(\varphi_{\text{sph}},\vec A_{\text{sph}} )} \,, 
    \end{align}
    and quantum effects result in corrections through a fluctuation determinant. Again, out-of-equilibrium contributions to the rate are more complicated. 
    
\end{itemize}

\subsection{Effective Descriptions for Sphaleron Rate: A Brief Review}

To set the stage for our own computations, we start with a brief review of the chronology of computations of the sphaleron rate focusing on the 3D EFT and the effective 
hard thermal loop (HTL) approaches.

\subsubsection{Early Computations of Sphaleron Rate}
The existence of the semi-classical sphaleron configuration within the Standard Model was first realized in \cite{Manton:1983nd, Klinkhamer:1984di} and later extended by \cite{Klinkhamer:1993hb, Akiba:1988ay}.
The physics of the sphaleron in the EWBG was first studied in \cite{Kuzmin:1985mm}. 
It was shown that the sphaleron transitions between different vacua can occur at high temperatures due to thermal fluctuations which leads to far less suppressed rate compared to mere zero temperature tunneling \cite{tHooft:1976rip,tHooft:1976snw}. 
In such thermal escape at temperature $T$, the sphaleron rate is classically described by a Boltzmann factor $\exp(-E_{\text{sph}}/T)$ where $E_{\text{sph}}$ is the energy of a static sphaleron solution.

Higher order quantum corrections to this statistical (time-independent) exponential part are described by a prefactor -- the fluctuation determinant -- which were first discussed in \cite{Arnold:1987mh,Arnold:1987zg} and further approximated and fully computed in \cite{Carson:1989rf,Carson:1990jm,Baacke:1993jr,Baacke:1993aj,Baacke:1994ix}. 
Notably, in these early-day computations high-temperature expansions were applied at leading order.  

\subsubsection{Effective Field Theory for Static Modes}
To describe higher order thermal corrections to the EWPT thermodynamics one has
account for the Debye screening that modifies interactions in a thermal medium. This can be achieved by resumming a class of low energy or large distance contributions that are enhanced at high temperatures \cite{Arnold:1992rz,Farakos:1994kx}. 
The most efficient way to incorporate effects of such thermal screening is by means of the effective field theory (EFT) approach \cite{Kajantie:1995dw,Braaten:1995cm}, which utilizes the technique of high-temperature dimensional reduction \cite{Ginsparg:1980ef,Appelquist:1981vg} (see also \cite{Nadkarni:1982kb,Landsman:1989be}).

The essence of dimensional reduction is the scale hierarchy at high temperatures: in the imaginary time formalism, which is used to compute static, equilibrium thermodynamics, 
all fields are composed as series of three-dimensional (3D), purely spatial Matsubara modes \cite{Matsubara:1955ws}. 
Characteristic mass scales for the non-static modes, i.e. the modes with non-zero Matsubara frequency, are dictated by high temperature and thereby heavier compared to modes with zero Matsubara frequency. 

These light 3D zero Matsubara modes are responsible for large distance, \textit{infrared} properties of the phase transition, including 
static properties of the critical bubble and the sphaleron (both of which in addition have real time dynamics). 
The ultraviolet (UV) scale of the heavy, non-zero Matsubara modes effectively screens the IR modes, and this can be captured by integrating out the UV scale \cite{Hirvonen:2022jba}. 
This leads to a (resummed) 3D EFT. For previous perturbative studies of the sphaleron rate within 3D EFTs, see \cite{Moore:1995jv, Burnier:2005hp}.

Resulting 3D EFT can be studied at few first orders in perturbation theory \cite{Gould:2023ovu,Ekstedt:2024etx}, but eventually perturbation theory breaks down for the modes deep at IR \cite{Linde:1980ts}, since the electroweak theory is confining and strongly coupled at high temperatures \cite{Gross:1980br,Shaposhnikov:1993jh}. 
To comprehensively study the IR physics, one has to resort to non-perturbative techniques, for example the lattice Monte Carlo simulation of the 3D EFT \cite{Kajantie:1993ag,Farakos:1994kj,Farakos:1994xh,Kajantie:1995kf}. 
Ultimately, it was the use of 3D EFT lattice simulations that lead to a realization that there is no first-order electroweak phase transition but a smooth crossover in the Standard Model \cite{Kajantie:1996mn,Rummukainen:1998as} (see also \cite{Csikor:1998eu}). 

\subsubsection{Real Time Dynamics and Non-Perturbative Simulations}
\label{sec:dynamic-part}
In the high temperature confinement phase, the sphaleron rate is purely non-perturbative and one needs to resort to numerical methods. 
At early days, plenty attempts were made to determine dynamical sphaleron transitions by solving time evolution of fields using equations of motion, within the 3D EFT at classical level 
\cite{Grigoriev:1988bd,Grigoriev:1989ub,Grigoriev:1989je,Ambjorn:1990pu,Ambjorn:1990wn,Ambjorn:1995xm,Tang:1996qx,Ambjorn:1997jz,Moore:1997cr}. 
Indeed, at high temperatures bosonic excitations at IR (low momenta) are highly occupied, and can be described as classical fields, opposed to high momenta UV modes, which exhibit particle-like behaviour. 
These attempts were, however, troubled by the fact that in such analyses the gauge field dynamics are UV divergent due to Landau damping and results are cut-off dependent \cite{Bodeker:1995pp,Arnold:1996dy}, and as such identifying the physical continuum limit of the classical theory requires care \cite{Laine:2022ytc}.

Some amelioration to such problems were provided by adding hard thermal loop effects \cite{Braaten:1989mz,Frenkel:1989br,Braaten:1991gm}%
\footnote{
See also \cite{Ekstedt:2023anj,Ekstedt:2023oqb} and references therein, for recent developments on the higher order HTL corrections.
}
to otherwise classical theory, resulting in expensive simulations \cite{Moore:1997sn,Bodeker:1999gx}. 
Shortly after, however, it was realized that these HTL and damping effects can be captured in an effective approach based on Langevin dynamics for the real-time evolution \cite{Bodeker:1998hm}: bosonic modes with energy much less than the temperature $E \ll T$ are highly occupied, with Bose-Einstein distribution $n_b(E) \approx T/E \gg 1$, and their dynamics are, to first approximation, classical. 
Thermal loop corrections to the classical picture result from high energy modes with $E \sim T$, but when these are integrated out, one is left with effective parameters, damping and noise, described by the Langevin dynamics.%
\footnote{
For a compact, yet illuminating summary, see also Sec. 2 of \cite{Eriksson:2024ovi}.
}

In essence, the HTL effective theory \cite{Bodeker:1998hm} is the real-time analogue of the dimensionally reduced 3D EFT for static quantities.
This formulation resulted in a lot of progress \cite{Arnold:1998cy,Arnold:1999ux,Arnold:1999uy,Bodeker:1999ey,Bodeker:2002gy} and eventually non-perturbative simulations \cite{Moore:1998ge,Moore:1998zk,Moore:1998swa,Moore:2000mx} based on a combination of multicanonical and real time lattice techniques (similar non-perturbative simulations for the bubble nucleation rate were developed slightly after in \cite{Moore:2000jw,Moore:2001vf}).

In these lattice simulations, the sphaleron rate, or Chern-Simons number diffusion rate, is determined by defining a surface in-between topologically distinct vacua (a ``separatrix'' that sphaleron transitions will cross when changing the Chern-Simons number), and by computing a probability flux across this surface, multiplied by a dynamical prefactor describing what fraction of crossings lead to settling to a new vacuum (with HTL effects included in the dynamical prefactor). 
For details, see \cite{Moore:1998swa} and more recent lattice simulations \cite{DOnofrio:2012phz,DOnofrio:2014rug,Annala:2023jvr}.%
\footnote{
Other past numerical studies include 
\cite{Moore:1999fs,Moore:2010jd,Altenkort:2020axj,Laine:2022ytc}.
}

\subsection{Towards a Self-Consistent Perturbative Formulation}

In \cite{Moore:1998swa} it was concluded that the effect of the ``dynamical prefactor'' to the Higgs phase sphaleron rate, which depends on HTL effects, is not large. This suggests that one can compute the statistical, time-independent part of the sphaleron rate in the 3D EFT perturbation theory, and as the first approximation entirely neglect the most complicated, dynamical prefactor. Such computation was performed in \cite{Burnier:2005hp}, using a two-loop effective potential within the 3D EFT, and later compared against results of lattice simulations in \cite{DOnofrio:2014rug,Annala:2023jvr}, showing remarkable agreement at temperatures below the SM crossover (c.f.~Fig.~\ref{fig:sphaleron_rate_SM_compare}). We emphasize here that both of these approaches rely on the exactly same 3D EFT construction by two-loop dimensional reduction \cite{Kajantie:1995dw}. 

This indicates that purely non-perturbative effects on the statistical part of the sphaleron rate are rather minor, and consequently prospects for purely perturbative calculations based on the use of 3D EFT are promising.
However, despite its apparent success, the approach of \cite{Burnier:2005hp} includes approximations that could be improved. First, the setup of \cite{Burnier:2005hp} is not based on a clear perturbative expansion, and as such is prone to double count loop corrections to the Boltzmann factor that are included via the effective potential as well as to the fluctuation determinant (statistical prefactor).%
\footnote{
This is analogous to a danger of double counting similar contributions in the context of vacuum tunneling \cite{Strumia:1998nf}, as well as thermal bubble nucleation \cite{Croon:2020cgk,Gould:2021ccf}. 
}
To avoid this double counting, Ref.~\cite{Burnier:2005hp} sets the positive mode contributions of the determinant (factor $\kappa$ therein) to unity by hand and argues that the use of the two-loop effective potential in the Boltzmann factor already (approximately) includes such contributions. 
We argue, that such manipulation by hand lacks a rigorous justification, and  could be entirely avoided in a properly formulated perturbative expansion.
Furthermore, use of the two-loop effective potential (which is computed around a spatially constant scalar background) cannot account for contributions of spatially varying scalar and gauge fields around the sphaleron background. 
Second, calculations (for the effective potential) in \cite{Burnier:2005hp} are performed simply in Landau gauge, and the lack of well-defined perturbative expansion results in a gauge-dependent outcome. 

Hence, in this work we implement a well-defined perturbative expansion for the statistical part of the sphaleron rate that maintains gauge-invariance.
This was attempted previously in \cite{Patel:2011th}, for both the effective potential and equilibrium properties such as the critical temperature derived from it and the sphaleron energy (for the Boltzmann factor).
Yet, a comprehensive framework for a consistent perturbative expansion incorporating the full set of higher-order thermal corrections remained to be addressed.
See also similar discussions in \cite{Wainwright:2011qy, Garny:2012cg,Morrissey:2012db,Chiang:2017zbz}. 
Nevertheless, multiple works \cite{Patel:2012pi,Profumo:2014opa,Chao:2014ina,Blinov:2015sna,Kotwal:2016tex,Chiang:2017nmu,Chiang:2018gsn,Ramsey-Musolf:2019lsf} that followed have focused on using the methodology of \cite{Patel:2011th}, and other works \cite{Kozaczuk:2015owa,Chao:2017vrq,Chen:2017qcz,Kang:2017mkl,Alves:2018jsw,Bell:2019mbn,Zhou:2019uzq,Kozaczuk:2019pet,Huber:2022ndk,Zhang:2023jvh,Zhang:2023mnu} have simply truncated their perturbative expansions to not include thermal resummations altogether.

Consistent perturbative expansions for the equilibrium properties of phase transitions, that reconcile the gauge-invariance and thermal resummations were found in \cite{Croon:2020cgk,Schicho:2022wty,Ekstedt:2022zro,Hirvonen:2022jba,Gould:2023ovu} using the 3D EFT perturbation theory (see also \cite{Gould:2021dzl} that does not include gauge fields, \cite{Ekstedt:2020abj} that does not utilize 
the 3D EFT, and discussions in \cite{Lofgren:2023sep}). 
These approaches have shown great agreement with non-perturbative lattice simulations in the cases where comparisons have been possible \cite{Gould:2021dzl,Ekstedt:2022zro,Gould:2022ran,Gould:2023ovu,Niemi:2020hto,Ekstedt:2024etx}. 
Similar consistent expansions for computing the thermal bubble nucleation rate -- which is analogous to the calculation of the sphaleron rate -- have been developed using the 3D EFT in \cite{Moore:2000jw,Moore:2001vf,Gould:2021ccf, Hirvonen:2021zej} building upon \cite{Kramers:1940zz,Langer:1969bc,Langer:1974cpa, Linde:1980tt,Linde:1981zj}, with discussions on higher order corrections \cite{Baacke:1999sc,Ekstedt:2021kyx,Ekstedt:2022tqk,Ekstedt:2022ceo,Ekstedt:2023sqc} (including dynamical part in \cite{Ekstedt:2022tqk,Hirvonen:2024rfg}) supported by fresh lattice simulations \cite{Gould:2022ran,Gould:2024chm,Hallfors:2025key}.

Thereby, resolutions to concerns raised in \cite{Patel:2011th} have been found in 3D EFT perturbation theory, apart from a similar, consistent perturbative expansion for the sphaleron rate and for quantities derived from it, and that is what we will deliver in following sections.

\section{Sphaleron Rate under Static 3D EFT Formulation}
\label{sec:formulation}

In this section, we get into technical details of our formulation. 
Our goal is to compute the Higgs phase sphaleron rate, i.e. the sphaleron rate inside the bubble, which is essential in determining the baryon-number-wash-out factor required for the baryon erasure bound.

\subsection{Higgs Phase Sphaleron Rate}

In analogy to the bubble nucleation rate in \cite{Ekstedt:2022tqk}, we decompose the sphaleron rate as
\begin{equation}
\label{eq:Gamma-sph-factorisation}
\Gamma_{\text{sph}} = A_{\text{dyn}} \times A_{\text{static}} \,,
\end{equation}
which is factorized into a non-equilibrium, dynamic part $A_{\text{dyn}}$ and an equilibrium, time-independent statistical part $A_{\text{static}}$.

The calculation of the dynamical part $A_{\text{dyn}}$ requires a real time -- out of equilibrium -- formalism \cite{Bodeker:1998hm,Bodeker:1999ey}, where hard thermal loops are integrated out in order to obtain an effective description based on Langevin dynamics.%
\footnote{
$A_{\text{dyn}}$ is often misinterpreted as $\omega_-$, i.e. a negative, unstable mode of the fluctuation determinant \cite{Akiba:1988ay,Carson:1990jm}, see discussion in \cite{Arnold:1987mh,Ekstedt:2022tqk}.
}
The formalism captures non-equilibrium dynamics, such as Landau damping effects, which cannot be described solely within the 3D EFT for the equilibrium quantities.
Incorporating such effects goes beyond the scope of our computation in this article.
In the following calculations, we just estimate the dynamical part as $A_{\text{dyn}} \sim T$ by the dimensional analysis.

The statistical part, on the other hand, can be computed solely within the 3D EFT.
In the semi-classical picture, the statistical part is calculated perturbatively in saddle point approximation as \cite{Carson:1989rf,Carson:1990jm,Baacke:1993jr,Baacke:1993aj,Baacke:1994ix}
\begin{equation}
\label{eq:static}
    A_{\text{static}} \simeq \text{[det]}_{\text{sph}} e^{-S_{\text{3D}}(\phi_{3,\text{sph}},\vec{A}_{3,\text{sph}})} \,,
\end{equation}
where the leading-order contribution is described by the exponential Boltzmann factor, the prefactor -- a one-loop fluctuation determinant $\text{[det]}_{\text{sph}}$ -- denotes the next-to-leading order corrections within the 3D EFT perturbation theory and higher loop corrections are not written down explicitly.
The 3D EFT action $S_{\text{3D}}$ in the Boltzmann factor is evaluated at the classical sphaleron solution of the equations of motion of the Higgs and gauge fields $\phi_{3,\text{sph}}$, $\vec{A}_{3,\text{sph}}$ within the 3D EFT.
We emphasize that here higher order corrections (corresponding to long-distance fluctuations compared to short-distance fluctuations at length scales $\sim (\pi T)^{-1}$ related to the thermal scale) in the determinant (and beyond) refer to corrections computed in the 3D EFT perturbation theory \cite{Farakos:1994kx}, while the 3D EFT construction itself includes corrections up to next-to-leading order (two loops) from short distance fluctuations, captured in the effective parameters of the (resummed) 3D EFT \cite{Braaten:1995cm}. This delineation will be detailed in the next section.

The fluctuation determinant $\text{[det]}_{\text{sph}}$, however, is an involved calculation.
At the present stage, our aim is to compute the leading behaviour of the sphaleron rate within the 3D EFT, i.e. the Boltzmann factor in terms of the sphaleron energy.
Hence, we do not compute the fluctuation determinant (or even higher-loop corrections beyond it) here but leave them to future work as well.%
\footnote{
Previously, fermionic contributions to the fluctuation determinant have been computed in \cite{Moore:1995jv} using higher-dimensional operators within the 3D EFT. Recently, similar studies \cite{Chala:2024xll,Bernardo:2025vkz,Chala:2025aiz,Chala:2025oul} have surfaced that study such marginal operator effects within the 3D EFT, albeit not computing their effects to the sphaleron rate. 
}

To organize our perturbative calculations, we employ a formal power counting parameter $g$, such that 
\begin{align}
\label{eq:power-counting}
-\ln \frac{\Gamma_{\text{sph}}}{T^4} \simeq 
\underbrace{S_{\text{3D}}}_{\sim g^{-1}} 
- \underbrace{\ln \frac{ \text{[det]}_{\text{sph}}}{T^3}}_{\sim \pi^{-1}} 
- \ln \frac{A_{\text{dyn}}}{T} \,,
\end{align}
where the contribution of the fluctuation determinant is suppressed compared to the leading order by $g/\pi$, which corresponds to the magnitude of an expansion parameter within the 3D EFT. 
We demonstrate below in detail how to arrive to the power counting in Eq.~\eqref{eq:power-counting}.
In the following we will compute the sphaleron 3D EFT action exactly meanwhile merely estimate the fluctuation determinant and simply omit the dynamical prefactor. 
%

\subsection{Dimensional Reduction and 3D EFT}

Starting from a BSM theory with an extended scalar sector, defined by a Lagrangian 
\begin{align}
\mathcal{L}_{\text{BSM}}(A^a_\mu,B_\mu, C^\alpha_\mu, \Psi , \phi, S),
\end{align}
where $A^a_\mu$, $B_\mu$ and $C^\alpha_\mu$  are the SU(2), U(1) and SU(3) gauge fields, respectively, $\Psi$ collectively denotes the all SM fermions, $\phi$ is the Higgs doublet and $S$ describes a new scalar field, one can construct a dimensionally reduced effective field theory at high temperature \cite{Ekstedt:2022bff}, in terms of a corresponding bosonic 3D Lagrangian $\mathcal{L}_{\text{3D}}(A^a_{i,3}, \phi_3, S_3)$ that respects the same symmetries as its parent theory. Note that we neglect the U(1) sector within the 3D EFT (i.e. set 3D U(1) gauge-coupling $g_3' = 0$ for simplicity), while the SU(3) sector essentially decouples (and only affects the dimensional reduction mapping, see Appendix~\ref{sec:temporal-gluon-effect}).

We denote parameters of the 3D EFT with subscript ``3''.
By further assuming that the scalar $S_3$ is much heavier than $\phi_3$ (in the 3D EFT) around the temperature scale of the electroweak phase transition, by integrating out the new scalar one arrives to an EFT solely for the SU(2) gauge and Higgs fields, described by three parameters: the effective gauge coupling $g^2_3$, the Higgs thermal mass $\mu^2_3$ and the Higgs thermal self-coupling $\lambda_3$ that are functions of temperature, known SM parameters, and unknown new parameters related to the field $S$.    
As a concrete example, at leading order the EFT parameters have simple expressions
\begin{align}
\mu^2_3 &= \mu^2 + \frac{T^2}{16} \Big(3g^2 + {g'}^2 + 4 g^2_Y + 8 \lambda + C a_2 \Big) \,,
\end{align}
which is the familiar scalar one-loop thermal mass, in terms of the SU(2) and U(1) gauge couplings $g$ and ${g'}$, the top quark Yukawa coupling $g_Y$ and the Higgs self-coupling $\lambda$. A possible BSM contribution in Higgs portal models is proportional to the portal coupling $a_2$ between the new scalar $S$ and the Higgs field, given by contribution 
\begin{equation}
V(\phi,S) = \frac{a_2}{2}\phi^\dag\phi S^\dag S \, ,
\end{equation}
to the scalar potential.
The numerical coefficient $C$ depends on the representation of $S$.   
At leading order couplings have only trivial temperature-scaling $g^2_3 = g^2 T$ and $\lambda_3 = T \lambda$, with possible BSM physics contributing at next-to-leading order (for details, see \cite{Kajantie:1995dw} and \cite{Ekstedt:2022bff}). In order to make first-order phase transitions possible, contributions from BSM physics to $\lambda_3$ are crucial, and we discuss them in detail in Sec.~\ref{sec:bsm}.

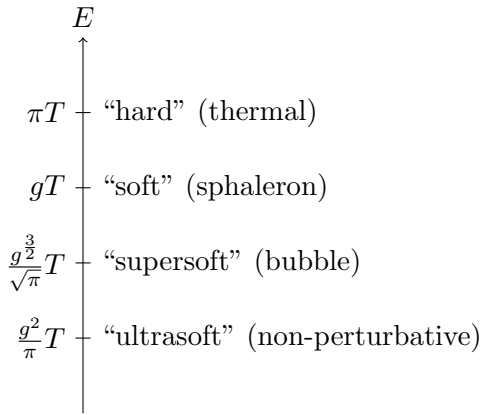
\begin{figure}
\centering
\begin{tikzpicture}
\draw[->] (0,0) -- (0,5) node[above] {$E$};
\foreach \y/\yleft/\yright in {4/$\pi T$/``hard'' (thermal),3/$g T$/``soft'' (sphaleron),2/$\frac{g^{\frac{3}{2}}}{\sqrt{\pi}}T$/``supersoft'' (bubble),1/$\frac{g^2}{\pi}T$/``ultrasoft'' (non-perturbative)}{
\draw (0,\y) -- (-0.1,\y) node[left] {\yleft}; 
\draw (0,\y) -- (0.1,\y) node[right] {\yright};
}
\end{tikzpicture}
\caption{
Hierarchy of energy scales $E$ at high temperature $T$, in terms of formal weak expansion parameter $g$.
}
\label{fig:energy-scales}
\end{figure}

To understand the construction of the dimensionally reduced effective theory, let us start with relevant energy scales of the problem, i.e. hierarchy of energy scales of using the imaginary-time formalism (or Matsubara formalism) of thermal field theory \cite{Laine:2016hma}.
In terms of a formal power counting parameter%
\footnote{Conventionally $g$ coincides with SU(2) gauge coupling in electroweak theories, but strictly speaking, in BSM theories it is the largest dimensionless coupling or ratio of dimensionful couplings, used to organize the perturbative expansion.
}
$g$ these scales are (see Fig.~\ref{fig:energy-scales})

\begin{itemize}
    \item[(i)] ``Hard'' scale $\pi T$: thermal scale of the non-static (non-zero) Matsubara modes. 
    These ``ultraviolet'' (UV) modes are integrated out in dimensional reduction, including all fermionic modes. 
    \item[(ii)] ``Soft'' scale $g T$: scale of the purely spatial 
    zero Matsubara modes, described by the 3D EFT.  
    This is the characteristic energy scale of Debye screened temporal gauge field components (and the BSM scalar $S_3$ in our setup), as well as the spatial gauge fields and sphaleron in the Higgs phase.
    \item[(iii)] ``Supersoft'' scale $\frac{g^{\frac{3}{2}}}{\sqrt{\pi}} T$: within the 3D EFT, characteristic scale for first-order phase transitions and the bubble nucleation is deeper in the IR 
    compared to soft scale, see \cite{Gould:2023ovu, Gould:2021ccf}. 
    \item[(iv)] ``Ultrasoft'' scale $\frac{g^2}{\pi} T$: the most deep IR scale, which is strongly coupled due to spatial gauge fields developing a non-perturbative ``magnetic mass'' due to confining nature of SU(2) \cite{Linde:1980ts}. 
    This is the characteristic energy scale of gauge fields and the sphaleron in the confinement phase \cite{Moore:2000mx}.      
\end{itemize}

Construction of the 3D EFT by matching the actions between hard and soft scales is done by means of the standard dimensional reduction.%
\footnote{
We comment a following subtlety in the standard literature, related to temporal gauge field components: these soft fields are often integrated out in a ``2nd step'' of dimensional reduction \cite{Kajantie:1995dw}, to construct a final EFT below the soft scale. In our computation, however, the sphaleron lives at the soft scale, so it might not seem immediately clear whether temporal components such as $A^a_0$ triplet field can be integrated out. However, in the Higgs phase they are much heavier than other mass scales, including the sphaleron, and can hence be safely integrated out \cite{Moore:2000mx}. For a related discussion, see \cite{Farakos:1994kx,Kajantie:1997ky,Kierkla:2023von}.
}
Formally in powers of $g$, the matching results in an accuracy \cite{Kajantie:1995dw,Ekstedt:2022bff,Hirvonen:2022jba}
\begin{align}
\label{eq:high-T-expansion}
S_{\text{3D}} \sim \int \mathrm{d}^{3} x \Big( \underbrace{\mathcal{O}(g^2)}_{\text{tree-level + hard 1-loop}} + \underbrace{\mathcal{O}(g^3)}_{\text{soft 1-loop}} + \underbrace{\mathcal{O}(g^4)}_{\text{hard/soft 2-loop}} + \,\mathcal{O}(g^5)\, \Big) T^3 \,,
\end{align}
for the 3D EFT action. 
The leading order (LO) $\mathcal{O}(g^2)$ terms are those of tree-level and one-loop thermal masses, the next-to-leading order (NLO) $\mathcal{O}(g^3)$ terms originate from the one-loop soft scale contributions and the next-to-next-to-leading order (NNLO) $\mathcal{O}(g^4)$ terms result from two-loop thermal masses and one-(two)-loop corrections to couplings from hard (soft) scale.  

 As explained in \cite{Arnold:1992rz,Farakos:1994kx, Kajantie:1995dw,Gould:2021oba}, the renormalization group improvement effect enters first at $\mathcal{O}(g^4)$: running of the LO terms is cancelled by explicit logarithms of the renormalization scale that appear at NNLO in high-$T$ expansion, i.e. running of the couplings inside one-loop thermal mass are cancelled by the logarithms in the two-loop thermal mass.
We truncate the matching relations at $\mathcal{O}(g^4)$ when deriving the effective action, and concretely demonstrate in Sec.~\ref{sec:SM-crossover} that it is crucial to perform the computation at this order. Furthermore, the matching relations at $\mathcal{O}(g^4)$ are gauge-invariant \cite{Croon:2020cgk,Hirvonen:2021zej,Schicho:2022wty}.

Higher order corrections at $\mathcal{O}(g^5)$ originate from three-loop soft contributions of temporal gauge field components. 
However, hard scale contributions to higher dimensional or marginal operators appear at $\mathcal{O}(g^6)$, and in the SM these are dominant compared to neglected soft contributions, which formally appear at lower orders \cite{Kajantie:1995dw}.

Concretely, the 3D EFT action at the soft scale of the SU(2)-Higgs theory in the Euclidean space reads as
\begin{equation}
\label{eq:S3d}
S_{\text{3D}} = \int \mathrm{d}^{3} x \biggl[ \frac{1}{4}F_{ij,3}^{a}F^{a}_{ij,3}+(D_{i}\phi)_{3}^{\dagger}(D_{i}\phi)_{3}
+\mu^{2}_{3}\phi^{\dagger}_{3}\phi_{3}+\lambda_{3} (\phi_{3}^{\dagger}\phi_{3})^{2} 
\biggr] \,,
\end{equation}
where $a$ and $i$ denotes the adjoint isospin index and the spatial Lorentz index respectively, and we have truncated the order at $\mathcal{O}(g^4 T^3)$.%
\footnote{
We note, that beyond $\mathcal{O}(g^4)$ accuracy that we work on, Eq.~\eqref{eq:S3d} will include higher dimensional operators, including marginal operators. 
These are schematically build from kinetic pieces $F^2$, $(D\phi)^2$ as well as pure scalar terms.
The effects of such operators within the 3D EFT framework have been discussed in \cite{Croon:2020cgk,Camargo-Molina:2021zgz,Ekstedt:2021kyx,Ekstedt:2022ceo,Chala:2024xll,Chala:2025aiz,Chala:2025oul} and in \cite{Moore:1995jv} in the context of the sphaleron rate, but for simplicity we do not include marginal operators, but assume that their effects are negligible. 
For example, a sextic operator of form $(\phi_{3}^{\dagger}\phi_{3})^{3}$ appears at $\mathcal{O}(g^6)$ in terms of contributions from the hard modes. 
When soft scale BSM physics is integrated out, such an operator can have significantly larger Wilson coefficient, since the soft scale contributions are enhanced. This is also the case for the effect of temporal gauge field component, which contributes formally at $\mathcal{O}(g^3)$ to $(\phi_{3}^{\dagger}\phi_{3})^{3}$, but this effect is still subleading compared to $\mathcal{O}(g^6_Y)$ effect from the top quark loop \cite{Kajantie:1995dw}. 
}
The subscript ``3'' is used to highlight quantities living in the 3D space: in the 3D EFT, the SU(2) doublet Higgs $\phi_3$ and the gauge field $A^a_{i,3}$ have mass dimension $T^{\frac12}$ and couplings $\lambda_{3}$ and $g^2_3$ have mass dimension $T$. 
The field strength tensor is defined as 
\begin{align}
F_{ij,3}^{a} &= \partial_{i} A_{j,3}^{a} - \partial_{j} A_{i,3}^{a} + g_3 f^{abc}A_{i,3}^{b}A_{j,3}^{c} \,,
\end{align}
where $f^{abc}$ is the SU(2) structure constant and the scalar covariant derivative reads as
\begin{align}
(D_{i}\phi)_3 &= \partial_{i} \phi_3 - i g_3 A_{i,3}^{a} T^{a} \phi_3 \,,
\end{align}
with SU(2) generators $T^a = \sigma_a/2$ in terms of Pauli matrices $\sigma_a$. 
For simplicity, we 
neglect (minor) effects of the U(1) subgroup within the 3D EFT \cite{Kajantie:1996qd}. 
This choice of the specific 3D EFT allows us to formulate a perturbative framework for the Higgs phase sphaleron rate after the SM crossover (for which 3D EFT effects of the U(1) subgroup can be ignored up to good accuracy \cite{Annala:2023jvr}).
It is also the same EFT we can apply when new BSM fields are heavy enough to be integrated out from the EFT at high temperatures.  
A specific application is provided in Sec.~\ref{sec:bsm}. 

Thermodynamic properties of the 3D EFT in Eq.~\eqref{eq:S3d} are governed by two dimensionless ratios
\begin{align}
\label{eq:xy-definition}
y \equiv \frac{\mu^2_3}{g^4_3}, \qquad x \equiv \frac{\lambda_3}{g^2_3} \,.
\end{align}
The critical temperature $T_c$ occurs when relation $y(T_c) = y_c(x)$ holds, where the curve $y_c(x)$ has been found non-perturbatively in \cite{Kajantie:1995kf,Gould:2022ran} (and perturbatively in \cite{Ekstedt:2022zro,Ekstedt:2024etx}). 
The character (as well as the strength) of the phase transition is controlled by $x_c \equiv x(T_c)$, and for $x_c \lesssim 0.098$ the character is of first order, 
second-order exactly at $x_c \approx 0.098$ (with $y_c \approx -0.0175$). 
For $x_c \gtrsim 0.098$ there is no phase transition, but merely a smooth crossover between confinement phase and the Higgs phase
\cite{Kajantie:1995kf,Gould:2022ran}.%
\footnote{
In smooth crossover, the free-energy, or the pressure, does not have a discontinuity, and hence strictly speaking there is only a single continuously connected phase. 
}
In the minimal Standard Model, $x_c \approx 0.29$ corresponding to a crossover \cite{Kajantie:1996mn}. For BSM physics to alter this story (within validity of the 3D EFT of Eq.~\eqref{eq:S3d}) it has to reduce $x_c$, and we discuss a concrete example of this in Sec.~\ref{sec:bsm}.

We stress that it is the (gauge-invariant) ratio of the thermal Higgs self-interaction coupling to the effective gauge coupling that sets the strength of the phase transition (which is proportional to $x^{-2}$). This can be intuitively understood in a clear perturbative picture of \cite{Ekstedt:2022zro}: a strong phase transition is a result of a mass hierarchy between a light transitioning-field (the Higgs scalar) and a heavy transition-inducing field (the gauge field), see also \cite{Kajantie:1997hn,Kajantie:1997tt,Moore:2000jw}. This relation between a first-order phase transition and mass hierarchy is general to all (perturbative) first-order phase transitions \cite{Gould:2023ovu}.%
\footnote{
As another example, see \cite{Gould:2023jbz}: in the scalar-Yukawa model, a Dirac fermion plays the role of a transition-inducing heavy field, and consequently the transition strength is controlled by the ratio of scalar self-coupling to Yukawa coupling, smaller ratio corresponding to stronger transitions.  
}

Indeed, already in seminal Ref.~\cite{Moore:1998swa} the baryon number preservation condition is given in terms of $x$ by $x_c < 0.037$, which is gauge-invariant, includes all non-perturbative physics of the sphaleron rate including the dynamical part, and can account two-loop thermal resummations via dimensional reduction. Unfortunately, major of the past studies have failed to recognize these features and the physics behind it.

However, the condition $x_c < 0.037$ for sufficiently strong transition to prevent the sphaleron washout can be improved. Not by improving the determination of the sphaleron rate (which, we stress, was determined on the lattice in  \cite{Moore:1998swa}), but by providing a more accurate connection to the bubble nucleation, compared to what was done in \cite{Moore:1998swa}.

In analogy to $y_c(x)$ curve, Ref.~\cite{Gould:2022ran} has determined a condition $y(T_p) = y_p(x)$ (both in perturbation theory and non-perturbatively) for percolation temperature $T_p$, defined by a condition that a phase transition completes by percolation, i.e. that eventually the expanding bubbles of the Higgs phase fill the universe.%
\footnote{
In this work at hand, we refer the same curve, obtained in leading order perturbation theory within the 3D EFT, as $y(T_n) = y_n(x)$ at the onset of nucleation at temperature $T_n$, since in practice $T_n \approx T_p$ in this class of models.
}
This condition is obtained by connecting the bubble nucleation rate with cosmological expansion in terms of the Hubble parameter (for details, see~\cite{Gould:2022ran}). 

In this work at hand, one of our goals is to find a similar curve $y_f(x)$ that describes the sphaleron freeze-out inside the bubbles, that we can obtain by computing the sphaleron rate at leading order in perturbation theory within the 3D EFT. 
Yet, we stress that since we work in the 3D EFT that accounts for large two-loop thermal corrections from the hard scale, in that regard our following discussion is greatly improved upon in contrast to e.g.~\cite{Arnold:1987mh,Arnold:1987zg,Carson:1989rf,Carson:1990jm,Baacke:1993aj,Baacke:1993jr,Baacke:1994ix} that work at leading order in the high-temperature expansion. 

In this work at hand, for the first time we combine information from both the nucleation and sphaleron freeze-out in terms of the curves $y_n(x)$ and $y_f(x)$ (c.f. Fig.~\ref{fig:yplot}), and hence we can generalize the discussion of \cite{Moore:1998swa}. We note that our leading order results in perturbation theory can then be upgraded in the future, by determining $y_f(x)$ non-perturbatively \cite{Annala:2025xxx} to match the level of rigor in \cite{Moore:1998swa}.      

To achieve this goal, we start by discussing power counting for a suitable perturbative expansion.

\subsection{Power Counting within the EFT Perturbation Theory}
\label{sec:power_counting}

In pure 3D Yang-Mills theory, the only appearing mass scale is the non-perturbative, magnetic mass of the gauge boson (i.e. the ultrasoft ``glueball'' mass) and this sets the length scale for the sphaleron, i.e. $(g^2 T)^{-1}$. 
In the gauge-Higgs theory, in the confinement phase (``symmetric'' phase) there is an another mass scale coming from the Higgs thermal mass, and the effect of the Higgs field has been studied in \cite{Moore:2000mx}. 
We note that the semi-classical approximation for the sphaleron cannot be used in the confinement phase, as the Higgs field does not acquire its non-zero vacuum expectation value, so a semi-classical sphaleron solution does not exist.  

In the Higgs phase, on the other hand, the gauge boson acquires a major contribution to its mass through the Higgs mechanism. 
This contribution, lying at the soft scale (Fig.\,\ref{fig:energy-scales}), is larger compared to the magnetic mass. 
Intuitively, the length scale of the sphaleron is controlled by the gauge boson mass, and therefore parametrically smaller in the Higgs phase compared to the confinement phase. 
Since the soft scale can be treated perturbatively, we should be able to compute a relatively reliable sphaleron rate in the Higgs phase.

Concretely, let us go forth and count powers of terms in the soft scale action in eq.~\eqref{eq:S3d}, again using $g$ for counting. 
All the terms in the action should be of equal size, for them to contribute at the leading order. 
Omitting Lorentz indices, we formally count $F_3^2 \sim (\partial A_3)^2 \sim m^2_{W,3} A_3^2$ by balancing the Yang-Mills term with the 3D W-boson mass term from the scalar kinetic term.
Using $\phi_3 \sim T^{\frac12}$, the characteristic momentum scale $k$ for the sphaleron schematically obeys $k \sim \partial \sim m_{W,3} \sim g_3 \phi_3 \sim g T$.
Thus $k$ is at the soft scale, and the sphaleron size $R_{\text{sph}} \sim (g T)^{-1}$, set by the gauge field mass scale.
Balancing the other terms in the Yang-Mills term results in $g_3 A^2_3 \partial A_3 \sim g^2_3 A^4_3$ and, therefore, $A_3 \sim \phi_3 \sim T^{\frac12}$. 

By this formal exercise, we find that the kinetic terms in Eq.~\eqref{eq:S3d} scale as $\sim g^2 T^3$. 
The scaling of the scalar potential, on the otherhand, is slightly more complex.
In a wide temperature range, generally the scalar thermal mass and the thermal self-coupling have scaling
\begin{equation}
\label{eq:2nd-order-scaling}
\mu_3^2 \sim (g T)^2, \quad \lambda_3 \sim g^2 T \,, 
\end{equation}
inherited from the full parent theory with $\mu^2 \sim g^2 T^2$ which follows from the high-temperature expansion and $\lambda \sim g^2$ which simply assigns a same size for scalar and gauge loops at zero temperature (this latter assumption, however, is not necessary, as we shall discuss shortly). Scaling of Eq.~\eqref{eq:2nd-order-scaling} implies that the scalar potential scales as $g^2 T^3$, and this leads to our power counting in Eq.~\eqref{eq:power-counting} by noting that the characteristic length scale for the soft scale is $g T$, i.e. $\mathrm{d}^3x \sim (g T)^{-3}$ in the action. The scaling of the fluctuation determinant is discussed in further detail in appendix~\ref{sec:sphaleron-determinant}. 
In particular, we apply the scalings in Eq.~\eqref{eq:2nd-order-scaling} to compute the sphaleron rate below the SM crossover.

However, it has been long known, that the general scalings of Eq.~\eqref{eq:2nd-order-scaling} do \textit{not} apply around and below the critical temperature of a first-order phase transition \cite{Arnold:1992rz,Kajantie:1997hn,Kajantie:1997tt,Gynther:2005av,Moore:2000jw, Laine:2015kra, Eriksson:2024ovi}. Instead, an appropriate scaling is 
\begin{equation}
\label{eq:1st-order-scaling}
\mu_3^2 \sim (g^{\frac32} T)^2, \quad \lambda_3 \sim g^3 T \,, 
\end{equation}
which follow from assumptions that around the critical temperature partial cancellations in the thermal mass make it smaller compared to Eq.~\eqref{eq:2nd-order-scaling}, and furthermore to accomodate a first-order phase transition, scalar loops have to be relatively suppressed compared to gauge field loops, and hence $\lambda \sim g^3$ in the parent theory.%
\footnote{
We note, that any convention for the parent theory scaling relating $\lambda$ and $g$ is not of thermal nature (unlike the scaling for the thermal mass), but merely a convention how to set up a perturbative expansion that can accommodate first-order phase transitions, in perturbation theory \cite{Arnold:1992rz}. 
}

In perturbation theory, scaling of Eq.~\eqref{eq:1st-order-scaling} has striking implications for computations of both equilibrium thermodynamics and bubble nucleation: the leading order potential is not described by the mere tree-level scalar potential, but in addition includes the one-loop contribution from the gauge bosons \cite{Kajantie:1997hn,Moore:2000jw}. The appearance of the gauge boson contributions at leading order potential (that we discuss in more technical detail in Sec.~\ref{sec:fopt}) has been made rigorous in \cite{Ekstedt:2022zro,Gould:2022ran,Hirvonen:2022jba} in terms of a new EFT specific to the Higgs phase, at the scale $g^{\frac32 } T$ (c.f.~ Fig.~\ref{fig:energy-scales}), where the gauge fields have been integrated out. This scale has been referred as the ``nucleation scale'' in the context of the bubble nucleation \cite{Gould:2021ccf,Hirvonen:2021zej,Lofgren:2021ogg} and as the ``supersoft scale'' in \cite{Gould:2023ovu} which demonstrates that this scale below the soft scale ($g T$) is widely applicable to first-order phase transitions in more general context. In particular, scaling of Eq.~\eqref{eq:1st-order-scaling} is necessary for a perturbative description of nucleation to be possible, and furthermore ensures a consistent perturbative expansion, including gauge-invariance.

Returning back to the sphaleron, we emphasize that the mass scale of the sphaleron (in the Higgs phase) is at the soft scale, and this is independent of the scalar potential, i.e. whether the scaling of scalar parameters follows Eq.~\eqref{eq:2nd-order-scaling} or \eqref{eq:1st-order-scaling}. This implies, that while a perturbative computation of the sphaleron rate in the semi-classical approximation is similar to the computation of the bubble nucleation rate, it is not based on a construction of any further EFT below the soft scale. In the following sections, we describe in detail how the scalar potential is accounted when computing the sphaleron rate.

Finally, we point out that power countings of Eqs.~\eqref{eq:2nd-order-scaling} and \eqref{eq:1st-order-scaling} are only a necessary tool to organize perturbation theory. On the lattice, the 3D EFT of Eq.~\eqref{eq:S3d} is directly discretized and thermodynamics as well as bubble- and sphaleron-dynamics are calculated by non-perturbative Monte-Carlo evaluation of the underlying path integral.

\subsection{Sphaleron Action and Scaling Properties} 
\label{sec:sphaleron_ansatz_action_equation_of_motion}

Now that we have established our perturbative setup based on the 3D EFT at the soft scale, we compute the sphaleron rate.
To solve the sphaleron equations of motion and calculate the sphaleron energy, it is advantageous to scale the fields and parameters in the 3D action (Eq.~\eqref{eq:S3d}) into dimensionless form. 

In the literature, Ref.~\cite{Carson:1990jm} scales the fields with the (temperature-dependent) Higgs vacuum expectation value $v$ (directly in the full theory, with high-temperature expansion). After such scaling, the resulting sphaleron energy is proportional to $v$. As such, $v$ corresponds to the minimum of the scalar potential with a second-order phase transition. 
Instead of using $v$, one might be tempted to use the 3D mass parameter, $\bar{\mu} \equiv \sqrt{-\mu_3^2}$ 
to scale the fields (see also discussion related to gauge-invariance in \cite{Patel:2011th}). 
However, the scaling with $\bar{\mu}$ can only be used in a situation in which $\mu_3^2 <0$.
Hence it is not suitable for describing a first-order phase transition, 
since in the Higgs phase at temperatures below the critical temperature $\mu_3^2$ can vary from positive to negative values (and hence also vanish at some temperature). 

We shall use the dimensional quantity $g_3^2$ for scaling, which is positive definite at all temperatures (as well as gauge-invariant \cite{Croon:2020cgk}). 
Therefore, we use following scaling of fields and coordinates 
\begin{align}
A_3=g_3\hat{A}_3,\quad \phi_3=g_3 \hat{\phi}_3,\quad x_i=\frac{\xi_i}{g_3^2},
\end{align}
and consequently the action in Eq.~(\ref{eq:S3d}) is scaled into
\begin{align} \label{eq:S3D_dimensionless_version}
   S_\text{3D}\rightarrow \hat{S}_{\text{3D}} = \int  \mathrm{d}^{3}\xi \left[\frac{1}{4}\hat{F}_{ij,3}^{a}\hat{F}^{a}_{ij,3}+(\hat{D}_{i}\hat{\phi})_{3}^{\dagger}(\hat{D}_{i}\hat{\phi})_{3}   +V_{3}(\hat{\phi}_3,x,y)\right] \,,
\end{align}
where the $\hat{F}_{ij,3}^{a}$ and $(\hat{D}_{i}\hat{\phi})_{3}$ do not include the gauge coupling anymore. 

While the 3D scalar potential at tree-level is given by the one in Eq.~\eqref{eq:S3d} and contains only quadratic and quartic terms in the field, for our purposes it is useful to consider a more general form, which also includes a cubic term 
\begin{equation}
\label{eq:general_cubic_potential}
V_{3}(\hat{\phi}_3,x,y)=y\hat{\phi}_3^\dagger\hat{\phi}_3-q (\hat{\phi}_3^\dagger\hat{\phi}_3)^{3/2}+x (\hat{\phi}_3^\dagger\hat{\phi}_3)^2 \,,
\end{equation}
where $y$ and $x$ are defined in Eq.~\eqref{eq:xy-definition}.

At this stage, for generality we keep the value of the numerical coefficient $q$ undetermined in equations that follow, yet we strongly emphasize that it is not arbitrary variable, but will be directly fixed by requirement that Eq.~\eqref{eq:general_cubic_potential} consistently describes the leading order effective potential of a given situation.
Concretely, below we will directly focus on two cases:
$q=0$ for a second-order transition (that we apply to study the sphaleron rate for the SM crossover scenario); while for a first-order phase transition 
$q=\frac{1}{4\sqrt{2}\pi}$.
This particular value arises from the consistent treatment of gauge field modes in case of first-order phase transitions, and in particular provides gauge-invariant treatment, as we shall discuss in Sec.~\ref{sec:fopt}. 
Furthermore, for now we work with the cubic potential of Eq.~\eqref{eq:general_cubic_potential} as our \textit{starting} point, and return later to discuss in detail which contributions should be accounted in the sphaleron energy to contribute to the Boltzmann factor of the sphaleron rate, see Sec.~\ref{sec:fopt}.

Within the 3D EFT, the classical sphaleron ansatz for scalar and gauge fields reads as%
\footnote{
The gauge field configuration used here is not unique, and the $U(1)$ field has been neglected. For a discussion of sphaleron ansatz under alternative conventions and the inclusion of the $U(1)$ field, see Ref.~\cite{Wu:2023mjb} and references therein.
}
\begin{align} \label{eq_app:scalar_gauge_sphaleron_confg_in_3DEFT}
\hat{\phi}_{3}&=h(\xi)\left (\begin{array}{c} 
 0  \\ 
 v_3(x,y)
 \end{array}\right) \,, \\
\label{eq_app:gauge_sphaleron_asatz_in_3DEFT}
\hat{A}_{i,3}^{a} T^{a} \mathrm{d}x^{i} &= [1-f(\xi)] \sum_{i=1}^{3} F_{i} T^{i} \,,
\end{align}
where $v_3(x, y)$ is the classical background field at the minimum of the 3D effective potential $V_3(\phi_3, x, y)$, and $h(\xi)$ is the radial profile function.
Hereafter hatted fields are used to denote the sphaleron ansatz.
For the gauge field ansatz, $f(\xi)$ denotes the radial profile function and the 1-form $F_i$  are functions of the spherical coordinates $\theta$ and $\varphi$
as well as a topological loop parameter $\mu$ \cite{Manton:1983nd,Klinkhamer:1993hb}.
We demonstrate that our 3D gauge ansatz is consistent with the original ansatz of \cite{Klinkhamer:1990fi} in Appendix \ref{app:Properties of sphaleron ansatz, action, and equation of motion}, and relegate there additional technical details about the sphaleron action, while focusing on the key features in the main body.   

Using Eq.~\eqref{eq:general_cubic_potential}, the scalar field minimum $v_3(x,y)$ then reads%
\footnote{
Note that we choose $v_3(x,y)$ to be the vacuum expectation value of the complex scalar field $\hat{\phi}_3$.
}
\begin{align}\label{eq:definition_vev}
v_3(x,y)=\frac{3q+\sqrt{-32xy+9q^2}}{8x} \,.
\end{align}
Although the action in Eq.~(\ref{eq:S3D_dimensionless_version}) has already been scaled into a dimensionless form, it is further advantageous to rescale the fields and the radial coordinate with the dimensionless $v_3(x,y)$, that is
\begin{align}
\label{eq:rescaling}
    \hat{\phi}_3 = v_3(x,y)\hat{\phi}_3^\prime \,,\ \ \hat{A}_3 = v_3(x,y)\hat{A}_3^\prime \,,\ \ \xi=\frac{1}{v_3(x,y)}\xi^\prime \,.
\end{align}
The benefit of this rescaling will be demonstrated in the context of our numerical computations below. We stress that our rescaling with $v_3(x,y)$ here is intrinsically different with the $v$-scaling in Ref.~\cite{Carson:1990jm}. First, $v_3$ is dimensionless, while $v$ in Ref.~\cite{Carson:1990jm} is dimensional, and second, $v_3$ is manifestly gauge-invariant (it is the minimum of the leading order effective potential, which is independent of gauge fixing), and while $v$ used in Ref.~\cite{Carson:1990jm} is the minimum of the tree-level potential (with the thermal mass), and hence also gauge-invariant, multiple later works (see discussion in Sec.~\ref{sec:discussion}) have used results of \cite{Carson:1990jm} and have included loop corrections to $v$ in a manner which is gauge-dependent. The motivation of our rescaling here is simply to facilitate numerical determination of the action, as we shall observe that the sphaleron action in case of a first-order phase transition is approximately equal to $v_3(x,y)\times \text{constant}$, which will greatly simplify the computation of the washout factor in Sec.~\ref{sec:BNPC}.

Consequently, the sphaleron action transforms into
\begin{align}
\label{eq:Shat}
\hat{S}_{\text{3D}} = v_3(x,y) \mathcal{C}_{\text{sph}}(x,y) \ ,
\end{align}
where
\begin{align} 
\label{eq:general_C_sph-main}
\mathcal{C}_{\text{sph}}(x,y) = \int \mathrm{d}^{3}\xi \left[\frac{1}{4}\hat{F}_{ij,3}^{a}\hat{F}^{a}_{ij,3} + (\hat{D}_{i}\hat{\phi})_{3}^{\dagger}(\hat{D}^{i}\hat{\phi})_{3} + \overline{V}_{3}(\hat{\phi}_3,x,y)\right] \ \ ,
\end{align}
and the transformed potential is
\begin{align} \label{eq:vev_transformed_cubic_potential}
    \overline{V}_{3}(\hat{\phi}_3,x,y) = \frac{y}{v_3^2(x,y)} \hat{\phi}_3^\dagger \hat{\phi}_3 - \frac{q}{v_3(x,y)} (\hat{\phi}_3^\dagger \hat{\phi}_3)^{3/2} + x (\hat{\phi}_3^\dagger \hat{\phi}_3)^2 \, ,
\end{align}
where we have omitted the prime on the fields and the coordinates to reduce notational clutter. Henceforth, $\hat A_3$ and $\hat\phi_3$ will denote fields scaled by $v_3$, and $\overline{V}_3$ will denote the scalar potential scaled by $v_3$.
It is important to note, that the Yang-Mill term and the second term in the effective action, which we refer as the covariant derivative term, do not explicitly depend on $x,y$, while $\overline{V}_{3}(\hat{\phi}_3,x,y)$ does.
After the rescaling, the sphaleron ansatze for the gauge fields remain unchanged, while the ansatz for the scalar field becomes
\begin{align}
\hat{\phi}_{3}&=h(\xi)\left (\begin{array}{c} 
 0  \\ 1
 \end{array}\right) ,
\end{align}
which only depends on the radial profile function. 
The equations of motion (EOMs) for the sphaleron can be obtained by applying the action principle to the scaled action $\mathcal{C}_{\text{sph}}$, which leads into 
\begin{align}
f^{\prime \prime} - \frac{2f(-1+f)(-1+2f)}{\xi^2}-\frac{2}{3}J(1+J)h^2 = 0 \,, \\
h^{\prime \prime}+\frac{2h^\prime}{\xi}-\frac{8}{3\xi^2}J(1+J)(-1+f)^2h-\frac{1}{2} \frac{\partial \overline{V}_3}{\partial h}=0 \,,
\end{align}
where prime denotes derivative with respect to the radial parameter.
We observe that the EOM of $f$ does not directly depend on $x,y$, while the EOM of $h$ depends on $x,y$ through $V_3$. 
We conclude that $x,y$ enter into the sphaleron action and the EOMs only through $\overline{V}_{3}(\hat{\phi}_3,x,y)$. 
In the following, we will discuss applications for the sphaleron rate at temperatures below SM crossover and first-order phase transitions, separately.

\subsubsection{Second-Order Phase Transitions}
\label{sec:2nd-order}

\begin{figure}[t]
\centering
\includegraphics[width=0.5\textwidth]{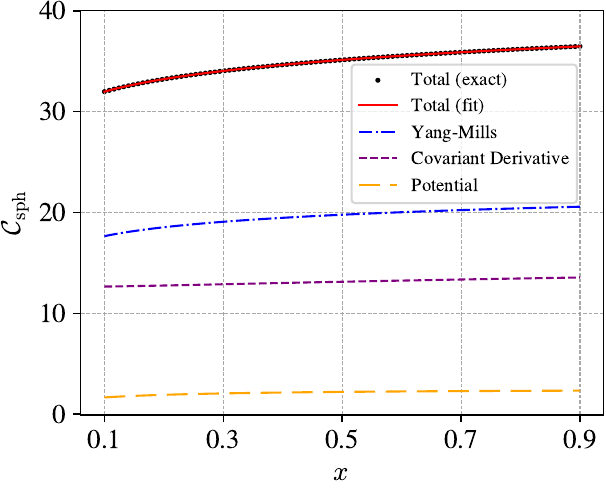}
\caption{
Scale-transformed sphaleron action $\cal{C}_{\text{sph}}$ as function of $x$. The black dots present the exact numerical computation of $\mathcal{C}_\text{sph}$, and the red line is the fitting curve, the Eq.~(\ref{eq:fit-crossover}).
These two results align on top of each other, and are almost indiscernible.
Dashed curves illustrate different contributions to $\mathcal{C}_{\text{sph}}$: blue, purple, and orange curves depict contributions from the Yang-Mills, covariant derivative and potential terms in Eq.\,\eqref{eq:general_C_sph-main}, respectively. 
}
\label{fig:sph_SM_vev_scaling_fit}
\end{figure}

The tree-level scalar potential, which corresponds to setting $q=0$, describes a second order phase transition, in which the second derivative of the pressure is discontinuous at the critical temperature, which occurs when the effective mass changes its sign, i.e. at $y=0$. For $y<0$, we can compute the sphaleron rate in the Higgs phase.
In such case,   
the $v_3(x,y)$ reads
\begin{align}
\label{eq:minimum-2nd-order}
v_3(x,y)=\sqrt{\frac{-y}{2x}},
\end{align}
and the scaled (tree-level) potential $\overline{V}_3$ depends only on $x$:
\begin{align}
\overline{V}_3(\hat{\phi}_3,x)=-2x\hat{\phi}_3^\dagger\hat{\phi}_3+ x (\hat{\phi}_3^\dagger\hat{\phi}_3)^2 \,.
\end{align}
Note that here both terms depend linearly on $x$: for the first term, this form arises due to the scale transformation.
We can then solve sphaleron EOMs numerically, and consequently compute the sphaleron action in Eq.~\eqref{eq:general_C_sph-main} numerically. The result for $\mathcal{C}_{\text{sph}}(x)$ can be expressed as a fit
\begin{align}
\label{eq:fit-crossover}
\mathcal{C}_{\text{sph}}(x)=A+B x^C \log(Dx),
\end{align}
with coefficients
\begin{align}
A=26.12,\ B=-2.145,\ C=0.4237,\ D=0.00717.
\end{align}
In Fig.~\ref{fig:sph_SM_vev_scaling_fit}, we present a comparison between the fitting formula and the exact numerical result of the rescaled sphaleron action $\mathcal{C}_{\text{sph}}$, as well as contributions from different terms in Eq.~\eqref{eq:general_C_sph-main}.

From Fig.~\ref{fig:sph_SM_vev_scaling_fit}, we can make an important observation: after the scaling transformation, the direct contribution from the scalar potential to $\mathcal{C}_{\text{sph}}$ is completely subleading compared to kinetic terms. In our formal power counting, all terms in the action contribute as
 $\int \mathrm{d}^3x (g^2 T)^{-3} \sim 1/g$, but after the scale transformation and numerical determination of the action, we see that the potential has only a minor effect, except that it dictates the form of the scalar background field at the minimum, i.e. $v_3$ in Eq.~\eqref{eq:minimum-2nd-order}, while to a great approximation $\mathcal{C}_{\text{sph}}$ is governed by the kinetic terms only. Below we will take full advantage of this observation, when we consider first-order phase transitions.  

The fitting expression in Eq.~\eqref{eq:fit-crossover} is really powerful: it precisely provides the sphaleron action of any given theory as long as the theory can be mapped onto the SU(2) Higgs 3D EFT for which leading order minimum at the Higgs phase is described by Eq.~\eqref{eq:minimum-2nd-order} at leading order, for any $y < 0$.
The temperature dependence is solely captured by the dimensional reduction matching relations through $x$ and $y$, which can be calculated e.g. by automated package {\tt DRalgo} \cite{Ekstedt:2022bff,Fonseca:2020vke}.

\subsubsection{First-Order phase transitions}
\label{sec:1st-order}

\begin{figure}[!t]
\centering
\includegraphics[width=0.4\textwidth]{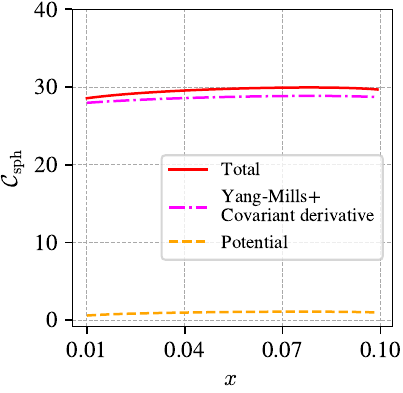}
\includegraphics[width=0.4\textwidth]{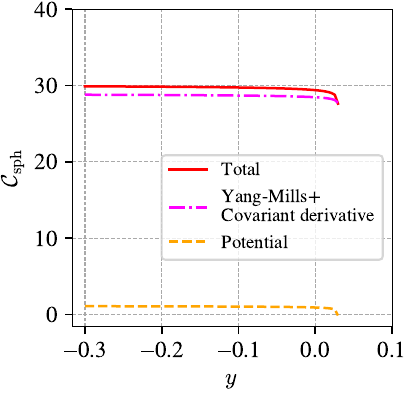} \\
\centering \includegraphics[width=0.5\textwidth]{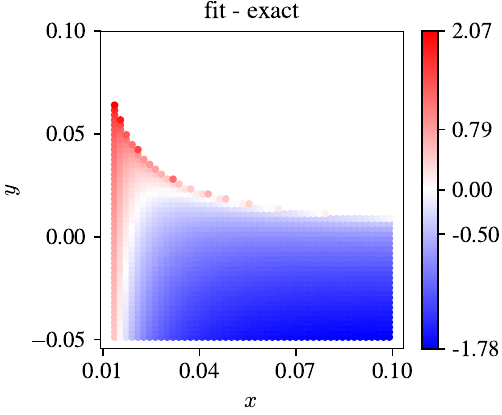}
\caption{ 
Top row: $v_3$-scaled sphaleron action in the case of cubic potential. 
Left: $\mathcal{C}_{\text{sph}}$ as function of $x$ with fixed for $y=0.008$. The red curve depicts the total action, magenta curve shows contributions from the Yang-Mills and covariant derivative terms, and orange curve contributions of the the scalar potential term. Right: similarly $\mathcal{C}_{\text{sph}}$ as function of $y$ with fixed $x=0.03$. 
Bottom row: Difference of the approximate sphaleron action (Eq.~\eqref{eq:vev_scaled_sphaleron_action_fitting}) and the full numerical action in the $(x,y)$-plane. This plot demonstrates that our approximate sphaleron action is in good agreement with the numerically obtained exact sphaleron action across a broad range of $x$ and $y$.
}
\label{fig:sph_cubic_scan_under_x_or_y}
\end{figure}

For a first-order transition, in perturbation theory the leading-order potential has to contain a barrier between the two phases, and such a potential then sets the expression for the Higgs phase minimum. 
For now, we concretely fix 
$q=\frac{1}{4\sqrt{2}\pi}$, hence both $x$ and $y$ appear in the $\overline{V}_3$. This value of $q$ corresponds to the leading order scalar potential of the SU(2) gauge-Higgs theory \cite{Ekstedt:2022zro}, and we explain in more detail in Sec.~\ref{sec:fopt} how this term arises. 
In this case, a general, closed form fitting expression for the dependence of $\mathcal{C}_{\text{sph}}(x,y)$ over two, instead of one, variables $x$ and $y$ turns out to be hard to obtain.
However, it turns out that the contribution from the potential term is very small (in analogy to previous case of a second-order transition), and $\mathcal{C}_{\text{sph}}$ is, to a good approximation, a constant.
In Fig.~\ref{fig:sph_cubic_scan_under_x_or_y} (top row), we demonstrate results for the scaled sphaleron action in fixed benchmarks with $x$ or $y$ fixed respectively. 
We see that the contribution from the potential term is negligible, and it is indeed a good approximation to regard the $v_3$-scaled sphaleron action as a constant, and we furthermore observe this trend to be general.  

To that end, we propose an approximation in which the sphaleron action is approximately equal to $v_3(x,y)$ multiplied with a constant $\mathcal{C}_{\text{sph}}(x,y) \approx 29$,  i.e.
\begin{align} 
\label{eq:vev_scaled_sphaleron_action_fitting}
\hat{S}_{\text{3D}} = v_3(x,y) \times 29,
\end{align}
where $v_3(x,y)$ is defined in Eq.~\eqref{eq:definition_vev}
with $q=\frac{1}{4\sqrt{2}\pi}$.
We further justify 
Eq.~\eqref{eq:vev_scaled_sphaleron_action_fitting}, by comparing with the result of exact numerical computation in the $(x,y)$-plane, which is shown in 
Fig.~\ref{fig:sph_cubic_scan_under_x_or_y} (bottom row).
The consistency between the approximation and the full numerical result indicates that the effect of $x$ and $y$ to the equations of motion and $\mathcal{C}_{\text{sph}}$ is almost negligible, especially in the most interesting region where $y>0$, as we will discuss in Sec.~\ref{sec:fopt}.

\subsection{Sphaleron Rate after Standard Model Crossover}
\label{sec:SM-crossover}

As a direct application of previous section, we next compute the Higgs phase sphaleron rate for temperatures below the Standard Model crossover. In the case of a crossover, there are no two separate phases, but only a one phase, and consequently all temperature derivatives of the pressure of the system are continuous.  Perturbation theory cannot describe the crossover properly,
as it predicts a second-order phase transition exactly at $y=0$.
This does not, however, affect out computation in the Higgs phase, with $y<0$. Hence, we assume a semi-classical picture in which the scalar potential is governed by the tree-level potential, i.e. our description for the second-order phase transition case, and consider region $y<0$.       

\begin{figure}[t]
  \centering
  \includegraphics[width=0.5\textwidth]{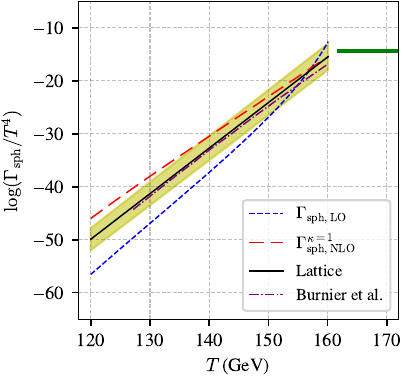}
  \caption{
  Sphaleron rate as function of temperature for the Standard Model crossover.
  Results of our perturbative computations in the Higgs phase based on two-loop dimensional reduction ($\Gamma_{\text{sph,LO}}$, $\Gamma^{\kappa=1}_{\text{sph,NLO}}$
  defined in the main text), contrasted with the previous perturbative results of Burnier et. al.~\cite{Burnier:2005hp} and non-perturbative lattice simulation of \cite{Annala:2023jvr} (black solid line with statistical error band in yellow). Green band depicts the confinement phase sphaleron rate, determined from the lattice.
}
\label{fig:sphaleron_rate_SM_compare}
\end{figure}

We depict our result in Fig.~\ref{fig:sphaleron_rate_SM_compare}. 
The SM crossover occurs at pseudo-critical temperature around $\sim 160$ GeV \cite{DOnofrio:2014rug,Laine:2015kra,Annala:2023jvr}. 
Here one should keep in mind, that as there is no phase transition, there is no order parameter to distinguish any physical phases.%
\footnote{Instead, physical properties, such as the Higgs field susceptibility or the sphaleron rate, change smoothly from their high-$T$ to low-$T$ behaviour around the pseudo-critical temperature. }
In a temperature window $120 < T/\text{GeV} < 160$ the value of $x(T) \approx 0.29$ is approximately a constant. 

At leading order, our result, which is shown as the blue dashed curve in Fig.~\ref{fig:sphaleron_rate_SM_compare}, is 
\begin{align}
\Gamma_{\text{sph,LO}}=T^4\times e^{-v_3(x,y)\mathcal{C}_{\text{sph}}(x)},
\end{align}
i.e. we estimate the prefactor solely based on dimensional grounds with $A_{\text{dyn}} \times \text{det}_{\text{sph}} \sim T^4$; the expressions for $v_3(x,y)$ and $\mathcal{C}_\text{sph}(x)$ are given by Eq.~(\ref{eq:minimum-2nd-order}) and Eq.~(\ref{eq:fit-crossover}), respectively.
This naive estimate for the prefactor can be improved upon by evaluating the fluctuation determinant more rigorously (see Appendix \ref{sec:sphaleron-determinant}), which leads to the NLO result 
\begin{align}
    \Gamma_{\text{sph,NLO}}=T\times \kappa(x) \times  \mathcal{N}_{\text{tr}}(\mathcal{N}\mathcal{V})_{\text{rot}} \, v_3^9(x,y) \, g_3^6  \, e^{-v_3(x,y) \, \mathcal{C}_{\text{sph}}(x)} \,.
\end{align}
Note that in this expression $g_3^6$ has the mass dimension $T^3$. 
In general, $\mathcal{N}_{\text{tr}}(\mathcal{N}\mathcal{V})_{\text{rot}}$ is a function of $x$, and we can approximate it as $\mathcal{N}_{\text{tr}}(\mathcal{N}\mathcal{V})_{\text{rot}} \approx \exp(8.813)$ using a result reported in Ref.\,\cite{Carson:1989rf}, for $x \approx 0.29$ in our case. 
Contributions of positive modes are encoded by the function $\kappa(x)$.
For simplicity, we do not compute it but merely set it to be unity, as a first approximation (in order to compare against the result of \cite{Burnier:2005hp}).
The $T$ factor in $\Gamma_{\text{sph,NLO}}$ again comes from our rough estimation $A_{\text{dyn}} \sim T$.%
\footnote{
As an exploration, one could also approximate the dynamic prefactor as $A_{\text{dyn}} \sim \frac{\omega_-}{2\pi}$ in terms of a negative eigenmode $\omega_-$, for which result of \cite{Carson:1989rf}
gives $\omega_{-} \approx 2\pi \exp(-2.05)$.   
However, as already discussed in Sec.~\ref{sec:dynamic-part}, this is not a consistent treatment for the non-equilibrium dynamic prefactor in terms of the effective HTL description in accord of \cite{Bodeker:1998hm,Arnold:1996dy} (see also recent \cite{Ekstedt:2022tqk}).
}
Under all these approximations, $\Gamma_{\text{sph,NLO}}$ is shown as the red dashed curve in Fig.~\ref{fig:sphaleron_rate_SM_compare}.
%

In addition, in Fig.~\ref{fig:sphaleron_rate_SM_compare}, we contrast our computations to the previous perturbative result by Burnier et al. \cite{Burnier:2005hp} (marked in purple dot-dashed curve), as well as non-perturbative lattice simulations of \cite{Annala:2023jvr,DOnofrio:2014rug} (marked in solid black curve, with statistical error estimates in yellow).
Additionally we show the results in the confinement phase by lattice simulations of \cite{Annala:2023jvr,Moore:2000mx} (in green) above the pseudo-critical temperature. 

We observe that our LO/NLO results have the same qualitative behaviour with results from more rigorous evaluations.
We emphasize, that this reasonable agreement of our purely perturbative computations (at LO and NLO) with the lattice, follows from a virtue that our treatment for the large thermal corrections from the hard scale, i.e. the dimensional reduction mapping, is the same as that has been used for the lattice analysis \cite{Annala:2023jvr} 
(as well as that in perturbative result of \cite{Burnier:2005hp}).%
\footnote{
For the Standard Model dimensional reduction, see \cite{Kajantie:1995dw}. However, this pioneering computation does not include an effect from the temporal gluon sector, that we discuss in Appendix \ref{sec:temporal-gluon-effect}.
}

\begin{figure}[t]
  \centering
  \includegraphics[width=0.4\textwidth]{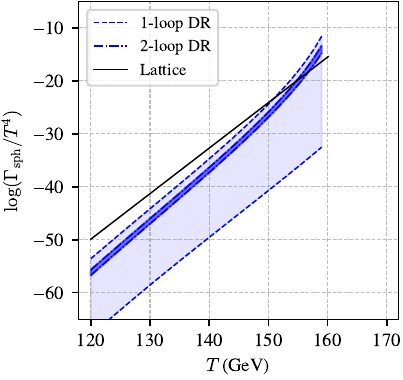}
  \includegraphics[width=0.4\textwidth]{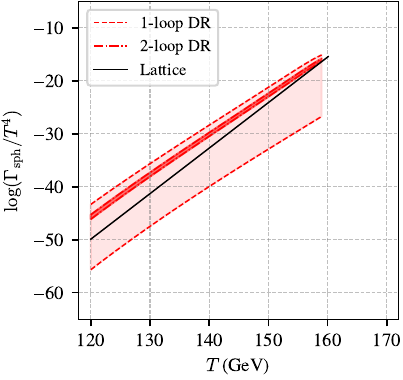}
  \caption{
Similar to fig.~\ref{fig:sphaleron_rate_SM_compare}, but varying
the effect of the renormalization scale $\mu$ at different dimensional reduction matching orders for our perturbative results $\Gamma_{\text{sph,LO}}$ (left plot) and $\Gamma^{\kappa=1}_{\text{sph,NLO}}$ (right plot).
}
\label{fig:sphaleron_rate_SM_vary_scale}
\end{figure}

To further highlight the importance of performing the dimensional reduction at two-loop order to accurately capture these thermal effects, in the right panel of Fig.~\ref{fig:sphaleron_rate_SM_vary_scale}, we replot our perturbative results using different orders for the dimensional reduction matching results.%
\footnote{
We note that while Fig.~\ref{fig:sphaleron_rate_SM_compare} uses the exactly same dimensional reduction mapping as used in \cite{Annala:2023jvr}, in Fig.~\ref{fig:sphaleron_rate_SM_vary_scale} we use the SM part of such mapping from \cite{Niemi:2021qvp}. The slight difference between the two is caused by different implementation of the zero temperature vacuum renormalization for the input parameters, c.f.~\cite{Kajantie:1995dw,Niemi:2021qvp}.  
}
For both the LO and NLO sphaleron rates, the results from mere one-loop dimensional reduction are shown as bands with dashed lines, while the two-loop results are shown as bands with solid lines.
The band is a consequence of the variation in the renormalization scale $\Lambda$ used in the dimensional reduction between $\Lambda = 0.5 \pi T$  and $\Lambda = 4\pi e^{\gamma} T$,%
\footnote{
Renormalization scale of the 3D EFT is kept fixed at $\Lambda_3 = T$, and varying it leads to negligible change. 
} 
(where $\gamma$ is the Euler–Mascheroni constant)
and it is clear that while the two-loop mapping depends only mildly on the renormalization scale (c.f. narrow bands in-between solid lines), by contrast the result at one-loop comes with a significant theoretical uncertainty (bands in-between dashed lines) \cite{Gould:2021oba}.

These comparisons of different results in the case of the SM crossover demonstrate that our perturbative computation, based on the use of 3D EFT, manages to capture large thermal effects (from the hard scale) in the Higgs phase  sphaleron rate. Hence, higher order \textit{soft} effects -- beyond the Bolzmann factor and the prefactor -- are not as large, let alone purely non-perturbative effects: as we have already discussed, the characteristic scale of the sphaleron in the Higgs phase is set by the gauge boson mass, and is hence the soft scale. 
As result, we have demonstrated that the sphaleron rate in the Higgs phase can be computed fairly accurately in perturbation theory.   
That being said, the fact that the soft scale effects captured by the statistical prefactor lead to quantitatively important corrections, motivates to extend our computation to fully compute the fluctuation determinant, i.e. the full effect of $\kappa(x)$. 
We leave such computation as well as studying the non-equilibrium effects captured by the dynamical prefactor to future works. 
%

\subsection{Sphaleron Rate after First-Order Electroweak Phase Transition}
\label{sec:fopt}

To accomodate a first-order phase transition at leading order in perturbation theory, the dimensionless scalar potential at leading order includes a cubic term from integrating out the soft gauge bosons in the Higgs phase as \cite{Hirvonen:2021zej,Ekstedt:2022zro}
\begin{align}
\label{eq:V-1st-order}
\overline{V}_{\text{LO}}(\hat{\phi}_3) &= y\hat{\phi}^{\dagger}_{3}\hat{\phi}_{3} + x(\hat{\phi}_{3}^{\dagger}\hat{\phi}_{3})^{2} - \frac{1}{2\pi} \Big( \frac{\hat{\phi}^{\dagger}_{3}\hat{\phi}_{3}}{2} \Big)^{\frac32}.
\end{align}
In terms of the scale hierarchies of Fig.~\ref{fig:energy-scales}, this potential corresponds to the leading order potential of the EFT at the supersoft scale \cite{Gould:2021ccf, Gould:2023ovu, Ekstedt:2024etx}. Furthermore, Eq.~\eqref{eq:V-1st-order} is gauge-invariant \cite{Croon:2020cgk,Gould:2021ccf}, and thereby resolves the problems discussed in \cite{Patel:2011th}.  
The cubic potential of Eq.~\eqref{eq:V-1st-order} obeys the power counting of Eq.~\eqref{eq:1st-order-scaling}: 
in the vicinity of the critical temperature, power counting for the mass of the scalar field yields $\mu^2_3 \sim \frac{g^3}{\pi} T^2$, while the mass of the gauge boson is soft 
$m_{W,3} = g_3 \sqrt{\phi_3^\dagger \phi_3} \sim g T$,
and hence the one-loop contribution $-\frac{1}{6\pi} m^3_{W,3} \sim \frac{g^3}{\pi} T^3$ from the gauge bosons -- which is gauge-invariant -- contributes at the LO, while all other one-loop diagrams contribute at higher orders (and could introduce spurious gauge-dependency, if not treated properly). 

Due to the cubic term, the potential of Eq.~\eqref{eq:V-1st-order} has a barrier between the origin and the Higgs phase minimum, and at the critical temperature (c.f. Fig.~\ref{fig:yplot})
\begin{align}
\label{eq:yc}
y(T_c) = y_c(x) = \frac{1}{128\pi^2 x} \,,
\end{align}
the two minima have degenerate free energies.
We note that Eq.~\eqref{eq:yc} describes the leading order result in the 3D EFT perturbation theory, but for Eq.~\eqref{eq:V-1st-order} to accurately describe the leading order effective potential, $x$ has to be sufficiently small, as otherwise gauge bosons are not heavy enough to be integrated out reliably in the Higgs phase.

At non-perturbative level, $y_c(x)$ has been determined in \cite{Kajantie:1995kf,Gould:2024chm}, and in stark contrast to perturbation theory 
 -- which always describes a first-order transitions (when leading order potential has the cubic term and hence a barrier) -- a first-order transition exists only for $0 < x \lesssim 0.1$, i.e. at large $x$ there is no transition but a crossover. 
In perturbation theory, $y_c(x)$ has been computed perturbatively at first five orders%
\footnote{
Any order beyond that receives contributions from non-perturbative magnetic mass of the gauge boson, and is hence beyond the reach of perturbartion theory, c.f. Linde's IR problem \cite{Linde:1980ts}.
}
as an expansion in $\sqrt{x}$, i.e. up to next-to-next-to-next-to-next-leading order (N$^4$LO) \cite{Ekstedt:2024etx}, and the result matches that from the lattice remarkably well until $x\sim 0.055$.%
\footnote{
However, nothing in perturbation theory alone indicates any kind of breakdown at $x \gtrsim 0.05$, so it remains elusive why the second-order end point occurs at $x\sim 0.1$ and not at some other, smaller or larger value. To shed light into this, lattice simulations could be used to reveal the details of the magnetic scale confinement phenomena, that presumably cause the termination of first-order transitions (see also \cite{Buchmuller:1994kk,York:2014ada}). 
We also note, that the value of $x_c$ where the end-point occurs depends on the representation of the scalar field under the gauge symmetry, e.g. $x_c \sim 0.3$ in the case of an adjoint scalar c.f.~\cite{Hart:1996ac,Kajantie:1997tt,Kajantie:1998yc} or more recent \cite{Niemi:2022bjg,Catumba:2024jau,Bonati:2024sok}.
}

To compute the sphaleron action, we can now directly follow the prescription of Sec.~\ref{sec:sphaleron_ansatz_action_equation_of_motion}. There is, however, a subtlety that we need to address: the leading-order scalar potential in Eq.~\eqref{eq:V-1st-order} with the cubic term, is formally the potential of the EFT within the supersoft scale, because this is the scale where the transition takes place \cite{Gould:2023ovu}. Sphaleron, on the otherhand, lives in the soft scale, and it is clear in technical level that in order to describe the sphaleron in the semi-classical approximation, we cannot integrate out the soft gauge boson. Hence, one might ask, is it consistent to use the cubic potential in Eq.~\eqref{eq:V-1st-order} to compute the sphaleron action? 

An attempt to resolve this by limiting to the mere (soft-scale) tree-level potential to describe the sphaleron fails immediately: the Higgs phase minimum of the tree-level potential does not exist for $y>0$, yet this is the region where the $y_c(x)$ curve lies, as well as the region where the bubble nucleation takes place (c.f.~Fig.\ref{fig:yplot}). Hence, we desire to be able to compute the sphaleron rate for $y>0$.

A resolution to this is already provided in Sec.~\ref{sec:sphaleron_ansatz_action_equation_of_motion}: after the $v_3$-scaling transformation, it is clear that the major role of the scalar potential to the sphaleron action comes only from the boundary condition to the sphaleron scalar profile function (as is apparent before the scale tranformation), while the direct effect of the potential to the sphaleron action is negligible, as it is dominated by the kinetic terms. While the sphaleron lives in the soft scale, this boundary condition for the scalar field minimum is set by the supersoft scale physics that dictates how the Higgs minimum $v_3$ -- or to be precise beyond the leading-order semi-classical picture -- the Higgs condensate, behaves in $(x,y)$-plane \cite{Ekstedt:2024etx}. The scalar potential that corresponds to $v_3$ (or the Higgs condensate) in leading order perturbation theory, is the one in Eq.~\eqref{eq:V-1st-order}. 

Furthermore, we note that the scalar potential in Eq.~\eqref{eq:V-1st-order} scales as $\sim g^3 T^3/\pi$ according to our power counting in Eq.~\eqref{eq:1st-order-scaling}. Hence, it is parametrically of same order as the fluctuation determinant for spatially varying gauge and scalar modes, and hence its direct effect to the sphaleron energy contributes at NLO, rather than at LO, in the ``soft scale expansion'' in $g/\pi$. 
A numerical consistency check of the above power counting can clearly be observed in  Fig.~\ref{fig:sph_cubic_scan_under_x_or_y}, and a comparison to Fig.~\ref{fig:sph_SM_vev_scaling_fit} shows that in the small $x$ regime the contribution of the potential to the sphaleron energy is even more negligible than at large $x$.

In summary, once the boundary condition to the scalar field radial profile function is correctly identified from the supersoft scale scalar potential for first-order transitions, that describes how the scalar field behaves at large distance scales $(g^{\frac{3}{2}} T)^{-1}$ (and hence at spatial infinity in sphaleron EOMs), the potential itself contributes only at NLO to the sphaleron rate, and its effect can be omitted in the semi-classical approximation.%
\footnote{
In order to concretely scrutinize the consistency of our perturbative expansion for the sphaleron rate, one should make sure that a treatment of all NLO contributions -- both the spatially varying fields contributing to the determinant and the role of the scalar potential, and furthermore the role of the potential in computing the determinant -- avoids any double counting. We leave such study for future work, yet speculate, that such computation might include a resummation by adding and subtracting the cubic term to the tree-level potential in analogy to \cite{Kierkla:2025qyz}. In this reference the bubble nucleation rate was computed at higher orders in terms of the bubble fluctuation determinant without resorting to the derivative expansion, and such resummation was used to eliminate the double counting related to the cubic term, which is essential for the bounce to exist at the leading order, yet is also contained in the determinant. 
}
In practical terms, we can compute the sphaleron rate using Eq.~\eqref{eq:vev_scaled_sphaleron_action_fitting}, as the difference between the exact solution (with the direct contribution of the entire potential relegated to NLO and hence omitted at LO) and the fit is numerically of the same order as difference of exact solutions with and without the potential term.

Before moving on, we point out that since first-order phase transitions take place for smaller values of $x$ (compared to the SM crossover), we can expect that in such cases our leading perturbative computation for the Higgs phase sphaleron rate is even more accurate than our results in the case of the SM crossover in Fig.~\ref{fig:sphaleron_rate_SM_compare}, since $x$ is the soft scale expansion parameter. This expectation could be put to test by fresh lattice simulations for the Higgs phase sphaleron rate, in the small $x$ regime that governs first-order phase transitions, and treating $y$ as an input parameter, rather than fixing it by any dimensional reduction mapping.       

\section{Baryon Number in First-Order Electroweak Phase Transition}
\label{sec:BNPC}

Perturbative description of first-order phase transitions in SU(2) + Higgs 3D EFT was originally presented in \cite{Arnold:1992rz,Farakos:1994kx}, and recently reformulated in \cite{Ekstedt:2022zro,Ekstedt:2024etx} in order to consistently add higher order corrections in perturbative expansion in $x \equiv \lambda_3/g^2_3$. 
In \cite{Ekstedt:2021kyx, Ekstedt:2022ceo} the bubble nucleation rate \cite{Moore:2000jw,Gould:2022ran} was computed in similar formulation. 
In this section we build upon these works, by including computation of the sphaleron rate and finding conditions in $(x,y)$-plane  for the sphaleron decoupling (or freeze-out) and washout. 
%

\subsection{Sphaleron Decoupling Temperature}
\label{sec:sphaleron-decoupling-T}

Now that we have computed the sphaleron rate in the Higgs phase (for any $x$ and $y$ using Eq.~\eqref{eq:vev_scaled_sphaleron_action_fitting}), let us use cosmology to impose a condition for the sphaleron decoupling (or freeze-out). 
We define the sphaleron decoupling by a condition that the baryon number violation rate $\Gamma_B$ equals the Hubble parameter $H$, or equivalently the mean free path for baryon preservation is equal to the horizon size $H^{-1}$. 
Hence, the decoupling temperature $T_*$ is defined as 
\begin{align} \label{eq:decoupling_condition}
\Gamma_B(T_*) = H(T_*) \,,
\end{align}
and for $T<T_*$ the baryon number is conserved. 
In the radiation dominated period, the relation of Hubble parameter with temperature reads
\begin{equation}
	H(T)=\sqrt{\frac{4 g_* \pi^3}{45}}\frac{T^2}{m_{\text{Pl}}} \,,
\end{equation}
with $m_{\text{Pl}}=1.22 \times 10^{19}\,$GeV the Planck mass,
$g_*=106.75+N$ the number of relativistic degrees of freedom with $N$ encoding the BSM physics. 
The relation between $\Gamma_B$ and the sphaleron rate $\Gamma_{\text{sph}}$ reads as \cite{Khlebnikov:1988sr,Arnold:1987mh,Burnier:2005hp}%
\footnote{
We note that the diffusion rate used in \cite{Burnier:2005hp} is twice the sphaleron rate. The sphaleron rate is a response rate to the chemical potential \cite{Arnold:1987mh}, while the diffusion rate is the Chern-Simons diffusion rate, $\Gamma_{\text{diff}(T)}\equiv \underset{V,t\rightarrow \infty}{\text{lim}} \frac{\langle Q^2(t)\rangle_T}{Vt}$, where $Q(t)\equiv N_{\text{CS}}(t)-N_{\text{CS}}(0)$ \cite{Moore:1998swa,Burnier:2005hp}. 
}
\begin{align} \label{eq:relation_B_var_and_sph}
\Gamma_B(T) = 2 N_\rho \frac{\Gamma_{\text{sph}}}{T^3} \,,
\end{align}
where $N_\rho=n^2_G \rho$; $\rho$ is a coefficient relating the baryon number to the free energy \cite{Khlebnikov:1996vj}, $n_G = 3$ is the number of generations of the fermions and we are taking the limit that $\frac{\Gamma_{\text{sph}}}{T^4}$ is rapidly and monotonicly decreasing with temperature. We will use the approximation $N_\rho\approx \frac{13}{4} \times 3$ in this work \cite{Burnier:2005hp}.

\begin{figure}[t]
\centering
\includegraphics[width=0.65\textwidth]{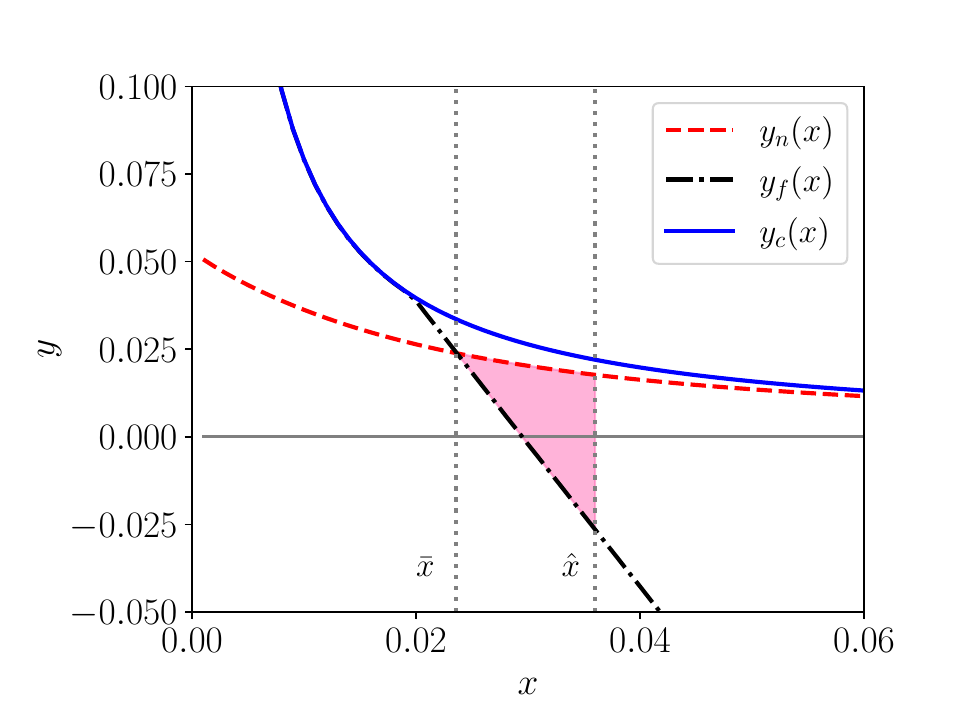}%
\caption{
Universal curves $y(T_c) = y_c(x)$ for the critical, $y(T_n) = y_n(x)$ for the nucleation and $y(T_*) = y_f(x)$ for sphaleron freeze-out temperatures  
 in the $(x,y)$-plane, at leading order in the perturbation theory. Vertical lines are drawn to $\bar{x} \approx 0.025$ at intersection of $y_n$ and $y_f$, as well as at $\hat{x} \simeq 0.036$.
For $x<\bar{x}$ sphalerons are decoupled immediately upon nucleation, while survival from baryon number wash-out is possible in the region $\bar{x} < x < \hat{x}$ (depicted in pink). For $x>\hat{x}$ washing out all possible baryon asymmetry cannot be realistically prevented. BSM theories are mapped into this plane by dimensional reduction mapping $\{ x(T), y(T) \}$, and crossing with $y_c$, $y_n$ and $y_*$ determines $T_c$, $T_n$ and $T_*$, respectively.
}
\label{fig:yplot}
\end{figure}

In order to translate our cosmological condition into a condition on 3D parameters, we first write the sphaleron rate (see the beginning of Sec.~\ref{sec:formulation}) as
\begin{align}
    \Gamma_\text{sph} = A_\text{dyn}\times [\text{det}]_\text{sph}e^{-\hat{S}_\text{3D}(x,y)} \ \ ,
\end{align}
then, we convert the decoupling condition (Eq.~(\ref{eq:decoupling_condition})) into
\begin{align}
\label{eq:Hxy}
\hat{S}_{\text{3D}}(x,y) = \ln \Big( 
\frac{A_{\text{dyn}}
\times [\text{det}]_{\text{sph}}}{T^3}
\frac{2 N_\rho}{H(T)} \Big) \approx \mathcal{S}_0 \equiv 39 \,,
\end{align}
at $T=T_*$.
That is,
by simply estimating $A_{\text{dyn}} \times [\text{det}]_{\text{sph}} \sim T^4$ using dimensional analysis, we find that the logarithm of Eq.~(\ref{eq:Hxy}) is approximately a constant, $\mathcal{S}_0$. We can thus approximate the $\hat{S}_{\text{3D}}(x,y)$ as a $T$-independent constant $\mathcal{S}_0 \approx 39$.
Subsequently, we can invert the equation $\hat{S}_{\text{3D}}(x,y(T_*)) = \mathcal{S}_{0}$ numerically to find a curve $y_f(x)$, i.e. the value of $y(T_*)$ for each $x$ that satisfies eq.~\eqref{eq:Hxy}.%
\footnote{
For $y_f$, we choose subscript to stand for sphaleron \textit{freeze-out} at $T_*$.
}
This is in complete analogy to treatment for nucleation condition $y_n(x)$ in \cite{Ekstedt:2024etx}, and we show $y_f$ in Fig.~\ref{fig:yplot}, together with $y$ of $x$ curves at the critical temperature $y_c \equiv y(T_c)$ and the nucleation temperature $y_n \equiv y(T_n)$.
%

In Fig.~\ref{fig:yplot}, regions above (below) the $y_f(x)$ curve correspond to baryon number violation rate that exceed (fall below) the Hubble constant. The area between the $y_n$ and $y_f$ curves represents the baryon washout region inside the bubble. Depending on the value of $x$, three scenarios arise: (i) no baryon washout occurs ($x<\bar{x}$); (ii) the generated baryon asymmetry is partly washed out ($\bar{x}<x<\hat{x}$); and (iii) practically all of the generated baryon asymmetry is erased ($x>\hat{x}$). The calculations supporting cases (ii) and (iii) are presented in the next subsection. 
Here we comment further on scenario (i), where the curves for the nucleation and sphaleron decoupling intersect at $\bar{x} \approx 0.025$. 
This implies, that for the very strong transitions with $x<\bar{x} \approx 0.025$ sphalerons are immediately decoupled when the transition starts, since $T_* > T_n$. 
Since $x$ depends on temperature only weakly \cite{Kajantie:1995dw}, this means that for strong transitions with $x_c \sim x_n < \bar{x} \approx 0.025$, wash-out is not possible at all, and all baryon excess generated in front of the bubble wall is immediately preserved in the Higgs phase.

\subsection{Baryon Number Preservation Criteria}
\label{sec:bnpc}

Baryon number density $n_B$ satisfies a rate equation \cite{Arnold:1987mh,Patel:2011th}%
\footnote{
On the left-hand-side of Eq.~(\ref{eq:baryon_washout_inside_bubble}), we have dropped the expansion term, $3Hn_B$, under the assumption $\Gamma_B(T)\gg H$. This approximation can fail in the narrow neighborhood of $\bar{x}$, where the $y_n$ curve is very close to $y_f$ curve ($y_f$ is defined by $\Gamma_B(T)=H$). However, this potentially problematic region is very narrow and for $x>\bar{x}$ beyond this neighborhood, the condition $\Gamma_B(T)\gg H$ holds and our approximation is therefore justified.
}
\begin{align} \label{eq:baryon_washout_inside_bubble}
\frac{\partial}{\partial t} n_B = - \Gamma_B(T) n_B \, ,   
\end{align}
where $\Gamma_B(T)$ is given by Eq.~(\ref{eq:relation_B_var_and_sph}). Upon integrating over time, we have
\begin{align} \label{eq:Rbt_washout_definition}
\mathcal{R}_B(t)\equiv\frac{n_B(t)}{n_B(0)}=\exp\left(-2 N_\rho \int_{t=0}^{ t}\frac{\Gamma_{\text{sph}}(T(t))}{T^3(t)}\mathrm{d}t \right) \equiv e^{-\mathcal{W}} \,,
\end{align}
where $t=0$ corresponds to $T=T_n$ and $t>0$ corresponds to lower temperature epoch. 
The quantity $\mathcal{R}_B(t)$, the ratio of remaining baryon number at time $t$ to the initial baryon number at $T = T_n$, measures a washout factor of the created baryon asymmetry and decreases as time goes.  
The upper limit of $t$, $t_*$ is given by the temperature where the sphaleron decouples $T(t_*)=T_*$, which has been discussed in the previous section.

Assuming radiation-dominated universe (with the Friedmann-Robertson-Walker metric), we can switch from the time $t$ to the temperature $T$ using
\begin{equation}
	\frac{\mathrm{d}}{\mathrm{d}t}=-\frac{\sqrt{24\pi}}{m_{\text{Pl}}}\frac{\sqrt{e(T)}}{\mathrm{d} \log s(T)/\mathrm{d}T}\frac{\mathrm{d}}{\mathrm{d}T} \,,
\end{equation}
where $s(T)$ is the entropy density and $e(T)$ is the energy density of the universe. For a discussion on non-standard cosmologies, see \cite{Barenboim:2012nh}.
Using the ideal gas approximation%
\footnote{
In principle, one can add higher order corrections to the pressure, see e.g.~\cite{Tenkanen:2022tly}, but such corrections are model dependent.
}
$p(T) = g_* \pi^2 T^4/90$ and thermodynamic relations $s(T)=p^\prime (T)$ and $e(T) = Ts(T)-p(T)$, we can write
\begin{align}
\mathrm{d}t = - \frac{1}{H(T)} \frac{\mathrm{d}T}{T} \,.
\end{align}
We can then convert the time integral in Eq.(\ref{eq:Rbt_washout_definition}) into an integration of $T$ as
\begin{align}
\mathcal{R}_B(T_*)=\frac{n_B(T_*)}{n_B(T_n)}=\exp\left( -2 N_\rho \int_{T^*}^{T_n} \frac{\Gamma_{\text{sph}}}{T^3}\frac{1}{H(T)} \frac{\mathrm{d}T}{T}\right) \,.
\end{align}
Thereby, we get the baryon number washout exponent factor
\begin{align}\label{eq:washout_exponent_definition}
	\mathcal{W} = 2N_\rho \int_{T^*}^{T_n} \frac{\Gamma_{\text{sph}}}{T^3}\frac{1}{H(T)} \frac{\mathrm{d}T}{T} \,,
\end{align}
and consequently
\begin{align}
\label{eq:baryon-number}
n_B(T_*) = e^{-\mathcal{W}} n_B(T_n) \,. 
\end{align}

In this prescription, should some C- and CP-violating processes have generated a baryon excess $n_B(T_n)$ at vicinity of the bubble wall at the onset of the phase transition, in the Higgs phase sphalerons will dilute the baryon density by a factor of $e^{-\mathcal{W}}$ until sphaleron processes decouple at $T=T_*$ and a final baryon density $n_B(T_*)$ remains, which can be used to compare with the observed value in the Eq.~\eqref{eq:BAU}. 
Furthermore, the washout starts at the time when the first generation of baryons moves into the bubble.
This time is identified as the formation time of the bubble, which in our approximation corresponds to $T_n$. 
In principle, as the bubble expands, new net baryons that have been created continuously outside the bubble wall, are being swept inside the bubble. 
However, in our setup the initial baryon number is the one  created at $T=T_n$, i.e. $n_B(T_n)$, and we have assumed that the bubble expansion is very rapid. 
In other words, the difference between temperature where bubbles form and the percolation temperature when the transition completes is very small compared to the difference between $T_n$ and the sphaleron decoupling temperature $T_*$.%
\footnote{
Again, this assumption may break down in the region near $\bar{x}$, as we discussed in a similar situation in the footnote of Eq.~(\ref{eq:baryon_washout_inside_bubble}). Here, we assume that the region where the approximation fails is very narrow. A more comprehensive analysis of this approximation is deferred to future work.
} 

Dividing both sides of eq.~\eqref{eq:baryon-number} by the photon number density $n_\gamma$ we can solve the initial baryon to photon density ratio
\begin{align}
\label{eq:BAU-Tn}
\frac{n_B(T_n)}{n_\gamma(T_n)} = e^{\mathcal{W}} \eta,
\end{align}
where $\eta$ is the observed value in the Eq.~\eqref{eq:BAU}.
Eq.~\eqref{eq:BAU-Tn} can be used to put a constraint on C- and CP-violating processes that generate an over-abundance of initial asymmetry, which is then washed out to produce the experimentally observed value.    

Next, our task is to compute the wash-out exponent $\mathcal{W}$. 
For this, we use the leading order sphaleron rate
\begin{align}
 \Gamma_{\text{sph}} \approx T^4 e^{-\hat{S}_{\text{3D}}(x,y)} \,.
\end{align}
Futher, we define 
\begin{align}
\eta_y \equiv T\frac{\mathrm{d}y}{\mathrm{d}T} \,, \qquad \eta_x \equiv T \frac{\mathrm{d}x}{\mathrm{d}T} \,, 
\end{align}
and assume $\eta_y \gg \eta_x$ (see \cite{Gould:2019qek}), i.e. the $T$-dependence of $x$ is subleading, and hence can be neglected. 
Upon change of variables from $T$ to $y$, the wash-out exponent $\mathcal{W}$ can be estimated as
\begin{align}
\label{eq:washout-UV-IR}
    \mathcal{W} \approx 2 N_\rho \frac{\bar{T}}{H(\bar{T})} \eta_y^{-1}(\bar{T}) \int_{y_f}^{y_n} e^{-\hat{S}_{\text{3D}}(x,y)}\mathrm{d}y
\end{align}
where $\bar{T} \equiv \frac{1}{2}(T_* + T_n)$.
Here we have taken the factor $\bar{T}/H(\bar{T})$%
\footnote{
This quantity is numerically large and slowly varying with temparture; for example, $\bar{T}/H(\bar{T})\approx e^{36.5}$ when $\bar{T}\simeq 100$ GeV.
} 
outside of the integral, given that the integration is dominated by the exponential, and hence this factor can be approximated as a constant, as well as the factor $\eta_y^{-1}$.

\begin{figure}
\centering
\includegraphics[width=0.49\textwidth]{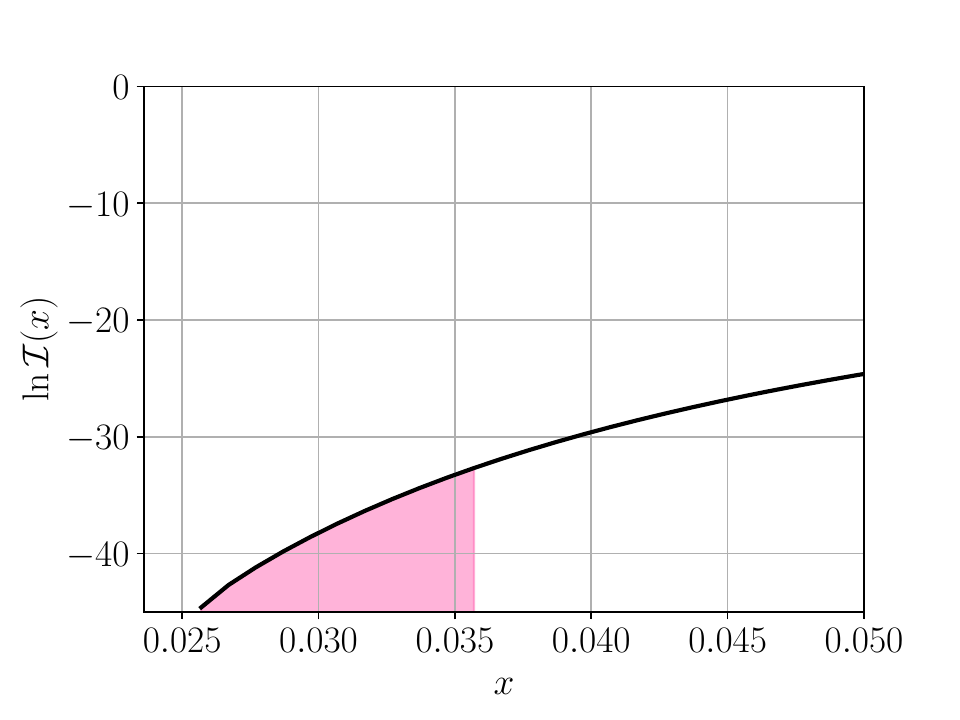}
\includegraphics[width=0.49\textwidth]{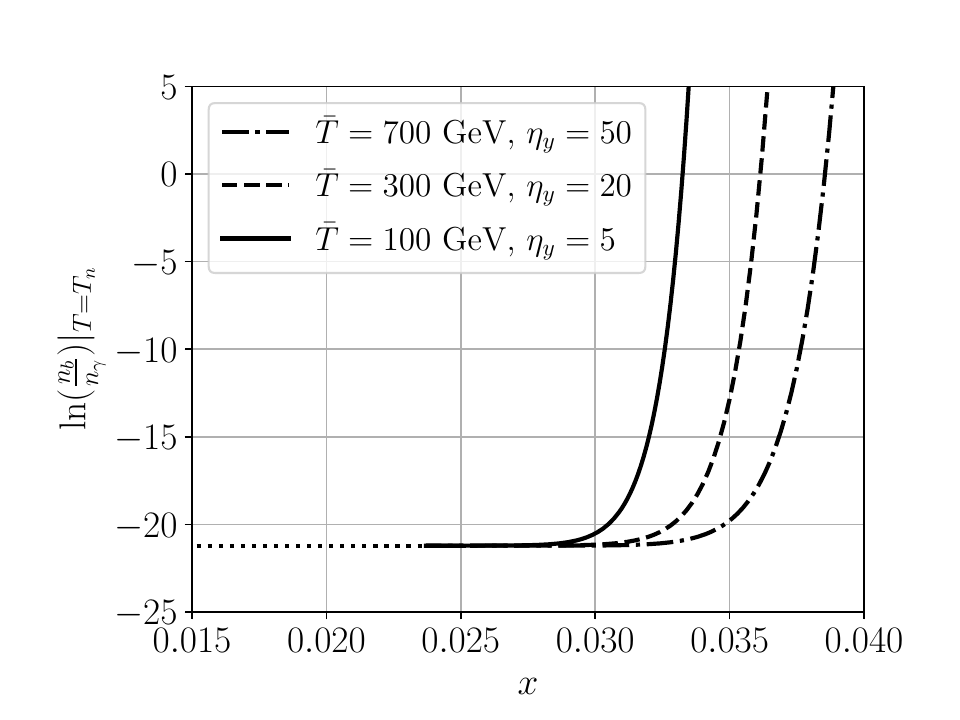}
\caption{
Left: logarithm of the 3D washout integral $\mathcal{I}(x)$. Since the function increases exponentially rapidly, survival of any baryon asymmetry is limited to a small range of $\bar{x} <x < \hat{x} \lesssim 0.036$ (pink). Right: the required initial baryon asymmetry at the nucleation temperature, that leads to the observed value of $\eta$, as function of $x$ (i.e. the $y-$axis is given by Eq.~\eqref{eq:weak-BNPC}). The horizontal dotted line  represents the logarithm of the observed baryon asymmetry, the Eq.~\eqref{eq:BAU}. For $x>\bar{x}=0.025$, the washout exponent $\mathcal{W}$ grows exponentially with $x$, resulting in an exponential rise along the $y-$axis.
}
\label{fig:logIx}
\end{figure}

The usefulness of the expression Eq.~\eqref{eq:washout-UV-IR} lies in a fact that we have factorized UV and IR contributions, i.e. thermal contributions of a parent theory and contributions of the 3D EFT, respectively.
For the pure 3D part of the wash-out integral we obtain a closed form expression
\begin{align}
\label{eq:Ix}
\mathcal{I}(x) &\equiv \int_{y_*}^{y_n} e^{-\hat{S}_{\text{3D}}(x,y)}\mathrm{d}y 
= \Biggl[ e^{-\hat{S}_{\text{3D}}(x,y)} \Biggl( \frac{4x}{841} + \frac{\sqrt{\frac{9}{2} -512 \pi^2 x y}}{232 \pi} \, \Biggr)    \Biggr]_{y_f}^{y_n},
\end{align}
This result for $\mathcal{I}(x)$ is depicted in Fig.~\ref{fig:logIx} (left). 

In summary, the washout exponent can be computed using
\begin{align}
\label{eq:S-washout}
\mathcal{W} \approx 2 N_\rho  \frac{\bar{T}}{H(\bar{T}) \eta_y(\bar{T}) } \times \mathcal{I}(x), 
\end{align}
where terms are factorized into UV and IR parts,
respectively, that are evaluated using the dimensional reduction mapping into the $(x,y)$-plane. 
For any 3D EFT mapping $( x(T), y(T) )$, crossing with $y_c$, $y_n$ and $y_*$ determines $T_c$, $T_n$ and $T_*$, respectively, c.f. Fig.~\ref{fig:yplot}.  

As we have already seen, for very strong transitions with $x<\bar{x} \approx 0.025$ sphalerons are immediately shut off when the transition starts, since $T_* > T_n$.
This allows us to define a strong version of the baryon number preservation condition (BNPC)
\begin{align}\label{eq:strong_BNPC}
\text{``Strong BNPC'':} \quad x(T_n) < \bar{x} \approx 0.025,
\end{align}
at the leading order of the perturbation theory.
We note, that typically in dimensional reduction mapping $x$ depends on temperature weakly, and hence $x(T_n) \approx x(T_c)$.%
\footnote{
Translating this bound to the Higgs expectation value gives  
$\frac{v_c}{T_c} = \sqrt{2} \frac{g_3 v_3(y_c)}{\sqrt{T_c}} = \frac{g}{8\pi x_c} > 1.03$, for $x_c < \bar{x}$ and $g^2 =0.42$, which agrees fairly well with \cite{Moore:1998swa}, yet we point out that here $\frac{v_c}{T_c}$ is the leading order result for the Higgs condensate, which is gauge-independent.
}
For parameter points which satisfy the strong BNPC, C- and CP-violating processes are constrained to produce the experimentally observed amount of baryon asymmetry, without any washout.

For slightly larger values of $x$, the baryon number density gets washed out by a factor of $e^{-\cal W}$.
Since the exponent $\cal W$ increases exponentially with $x$, the baryon number density before the bubble nucleation needs to be incredibly huge to survive the washout.
Instead of investigating which mechanism could produce such a large baryon number density, we put an upper bound on ${\cal W}$ by hand as ${\cal W} \lsim 100$ which results in an upper bound on $x < \hat{x}$  and we get
a weaker form of BNPC:
\begin{align}
\label{eq:weak-BNPC}
\text{``Weak BNPC'':} \quad 
\ln(\frac{n_b}{n_\gamma})|_{T=T_n} = \mathcal{W} +  \ln \eta
\implies  \bar{x}< x(T_n) < \hat{x} \approx 0.036 \,.
\end{align}
This ties together the required initial baryon asymmetry at $T_n$, and $x$, as well as $\eta_y$ and $\bar{T}$, both of which are determined through temperature evolution of $y(T)$.
This relationship is illustrated in the Fig.~\ref{fig:logIx} (right). 
We note, that since $\mathcal{I}(x)$ is exponentially rapidly increasing,  regardless of $T$-dependence of $y(T)$, the weak BNPC requires $x_n < \hat{x} \approx 0.036$.
According to the weaker form of BNPC, if C- and CP-violating processes generate over-abundance of baryon asymmetry, it is still possible to wash it out to produce experimentally observed asymmetry. 
For larger values $\hat{x} < x \lesssim 0.1$ that still govern first-order transitions, exponential washout can be expected to dilute any potential initial asymmetry. 

In summary, for any BSM theory that maps onto the SU(2) + Higgs 3D EFT at high temperatures and has a first-order phase transition, indicated by $x_c \lesssim 0.1$, preserving any generated baryon number necessarily requires $x_c \lesssim 0.036$ (assuming $x_n \approx x_c$).
If $x_c < \bar{x} \approx 0.025$ then all generated baryon number is automatically preserved (strong BNPC), while for $\bar{x}< x_c < \hat{x} \approx 0.036$ the Eq.~\eqref{eq:weak-BNPC} has to be satisfied (weak BNPC).
We note, that despite us working merely at leading order in the semi-classical approximation, we have obtained bounds on $x$ that agree fairly well with seminal studies in \cite{Kajantie:1995kf,Moore:1998ge,Moore:1998swa}.
%

\section{Application to Real Triplet-Extended Standard Model}
\label{sec:bsm}

Finally, we turn to a concrete BSM application where a real triplet scalar is used to catalyze a first-order phase transition.%
\footnote{
This model does not introduce new sources for CP-violation, and as such can only be used as a probe or theoretical template for studying sphaleron-related aspects of the EWBG.
}
We choose to work with this particular BSM extension, as its dimensional reduction has been well-documented \cite{Niemi:2018asa}, and its phase structure has been previously studied in terms of non-perturbative lattice simulations \cite{Niemi:2020hto}. In particular, we can integrate out the triplet in wide regions of the model parameter space (c.f.~\cite{Niemi:2018asa}), and hence use tools developed in the previous sections, to study sphaleron freeze-out and washout in the Higgs phase. We stress, that any other Higgs portal model can be studied along these lines, at least in a part of its full parameter space. 

In this model, the Standard Model Lagrangian is augmented with 
\begin{align} \label{eq:SigmaSM_Lagrangian}
\mathcal{L}_{\text{triplet}}(\phi,\Sigma) =& 
 \frac{1}{2} (D_\mu \Sigma^a)^2  + \frac12 \mu_\Sigma^2 \Sigma^a \Sigma^a + \frac{b_4}{4} (\Sigma^a\Sigma^a)^2 
 + \frac{a_2}{2} \phi^\dagger\phi \Sigma^a \Sigma^a \,,
\end{align} 
where the triplet $\Sigma^a$ transforms, as the name suggests, as an adjoint under SU(2), with isospin index $a=1,2,3$, and trivially under SU(3) and U(1) groups. 
Our definition of the covariant derivative aligns with \cite{Niemi:2018asa}, and convention for the sign of the mass term that of \cite{Gould:2023ovu}.
Electroweak phase transition in this model has been previously studied perturbatively in \cite{Patel:2012pi,Niemi:2018asa,Friedrich:2022cak,Cao:2022ocg,Gould:2023ovu} and using non-perturbative lattice simulations in \cite{Niemi:2020hto}.    
 
At high temperatures, we work in a setup where the triplet field is sufficiently heavy at the soft scale 3D EFT, so it can be further integrated along with temporal modes of the gauge fields. Details of this 3D EFT and its dimensional reduction mapping can be found in \cite{Niemi:2018asa,Niemi:2018juv} (and can be reproduced in automated fashion using {\tt DRalgo} package \cite{Ekstedt:2022bff,Fonseca:2020vke}). In short, thermal masses (couplings) are determined at two-loop (one-loop), as is customary for dimensional reduction \cite{Kajantie:1995dw}. 
This setup in which we integrate out the soft triplet allows us to study triplet-induced one-step phase transitions.  
When soft modes are integrated, at one-loop order
\begin{align}
\label{eq:lam-1loop}
\lambda_3 &= \lambda_{3,\text{hard}} - \frac{1}{8\pi} \frac{h^2_{3}}{m_{D}} - \frac{1}{8\pi} \frac{3}{4} \frac{a^2_{2,3}}{\mu_{\Sigma,3}} \,,
\end{align}
where we $\lambda_{3,\text{hard}}$ is the 3D self-coupling before integrating out the soft modes,%
\footnote{
Note, that our $\lambda_{3,\text{hard}}$ is denoted as $\lambda_3$, and our $\lambda_3$ as $\bar{\lambda}_3$ in \cite{Niemi:2018asa}, and we have chosen our notation here to be consistent with the earlier sections.  
}
and results for the SU(2) Debye mass $m_D$ and the Higgs-temporal portal coupling $h_3$ can be found in \cite{Niemi:2018asa}. 
We did not write contributions from U(1) temporal modes (see \cite{Niemi:2018asa}), for simplicity, as it is not essential to our following discussion. The main mechanism for the triplet to reduce the effective Higgs self-coupling $\lambda_3$ (and hence $x$)%
\footnote{
Triplet also contributes to the effective gauge coupling $g^2_3$, but these effects are completely subdominant in $x$, compared to the triplet effects in $\lambda_3$, see \cite{Niemi:2018asa}.
}
is the soft triplet contribution in Eq.~\eqref{eq:lam-1loop}, while triplet contributions in $\lambda_{3,\text{hard}}$ have subdominant effects: triplet affects $\lambda_{3,\text{hard}}$ by corrections from hard modes, as well as modifying the $\overline{\text{MS}}$ value of $\lambda$ at initial $M_Z$-scale through one-loop zero temperature corrections \cite{Niemi:2018asa}. 

\begin{figure}[t]
\centering
\includegraphics[width=0.75\textwidth]{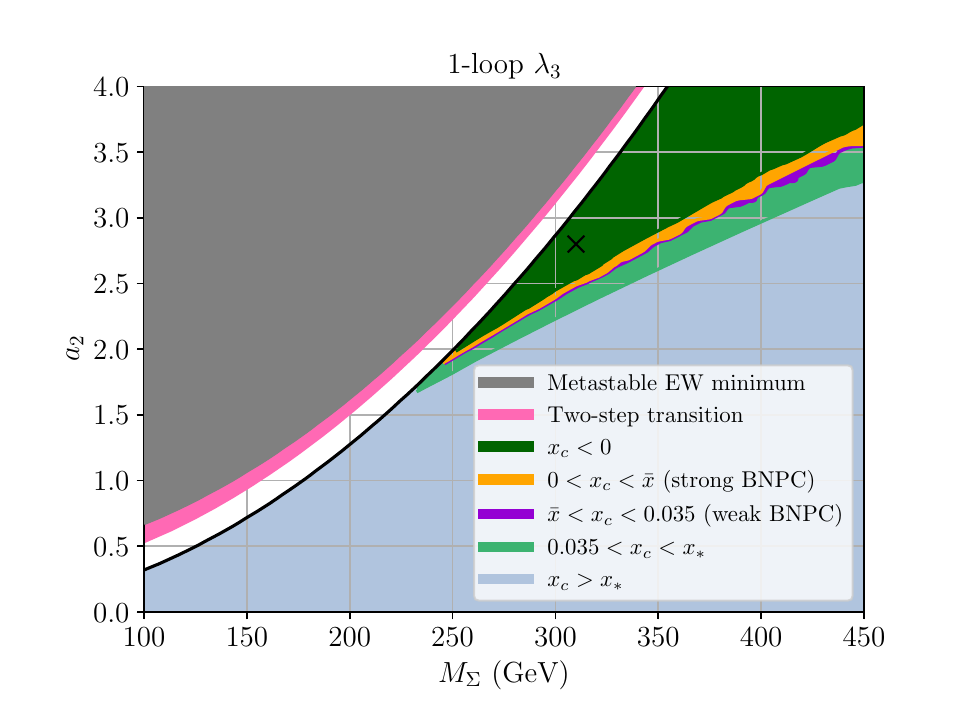}
\caption{
Phase structure of the triplet-extended SM in $(M_\Sigma,a_2)$-plane with $b_4=0.25$ based on one-loop result for $\lambda_3$, depicting different regions where BNPC can be satisfied in the one-step transition region. In white region, triplet mass parameter $\mu^2_\Sigma <0$ and the triplet cannot (necessarily) be integrated out. In region with $x_c < 0$ (dark green) 3D EFT construction fails, which signals that some higher order contributions are missing. 
}
\label{fig:triplet-plane-loop1}
\end{figure}

To proceed with numerics, we fix $b_4 = 0.25$ and compute the critical value $x_c = x(T_c)$, where the critical temperature $T_c$ is determined through leading order perturbative result $y(T_c) = y_c(x_c)$, in a  uniform scan in the $(M_\Sigma,a_2)$-plane, where $M_\Sigma$ is the physical triplet pole mass. Our result is shown in Fig.~\ref{fig:triplet-plane-loop1}.
In the gray region, the electroweak Higgs phase is not global minimum at zero temperature, and is hence excluded. 
Solid black line denotes a contour of $\mu^2_\Sigma = 0$, and above this line -- in white region -- $\mu^2_\Sigma$ is negative, indicating that the triplet can undergo a phase transition of its own at some temperature range, so in this region we do not integrate out the triplet. In principle, there can also be first-order one-step electroweak phase transitions in this region, but our approximation does not allow us to describe them accurately.%
\footnote{
In region with $\mu^2_\Sigma < 0$, one can integrate out the triplet in a different manner, following \cite{Gould:2023ovu,Ekstedt:2024etx}: in a temperature window where the triplet mass eigenvalue $m^2_\Sigma \equiv \mu^2_3 + \frac{1}{2} a_{2,3} v_3^2$ is dominated by the Higgs background field $v_3$,
one can study the phase transition in terms of a supersoft EFT expansion. Here, we do not perform such analysis. 
}
In the pink region, the electroweak phase transition proceeds in two steps, according to leading order perturbation theory, c.f. \cite{Niemi:2020hto}.
In light blue region, $x_c > x_* \approx 0.1$, and there is no phase transition but a crossover.

\begin{figure}[t]
\centering
\includegraphics[width=0.48\textwidth]{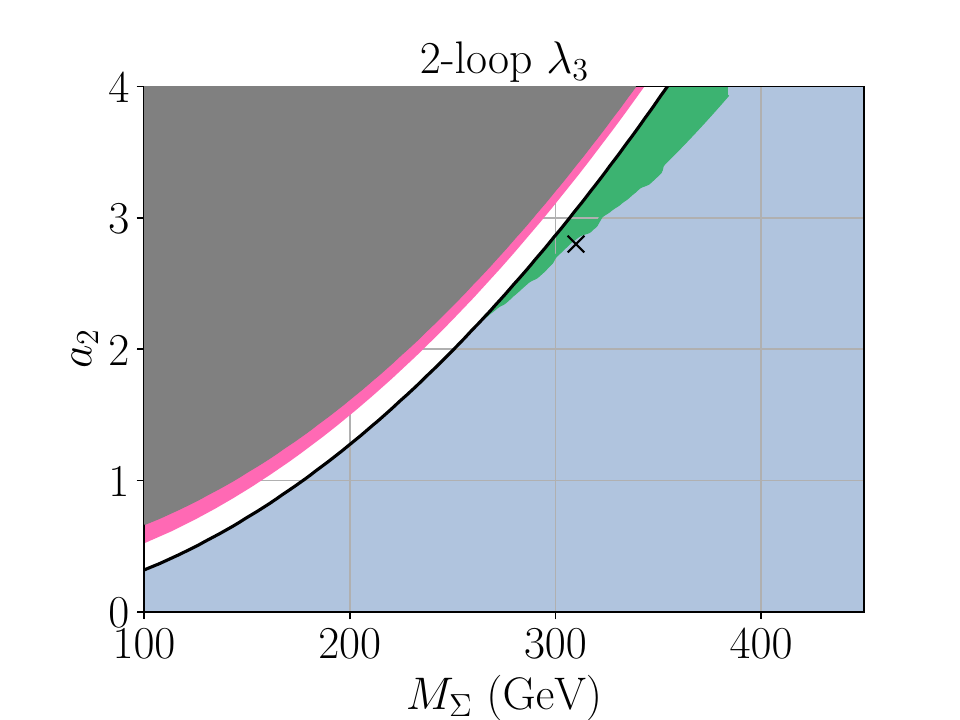}
\includegraphics[width=0.48\textwidth]{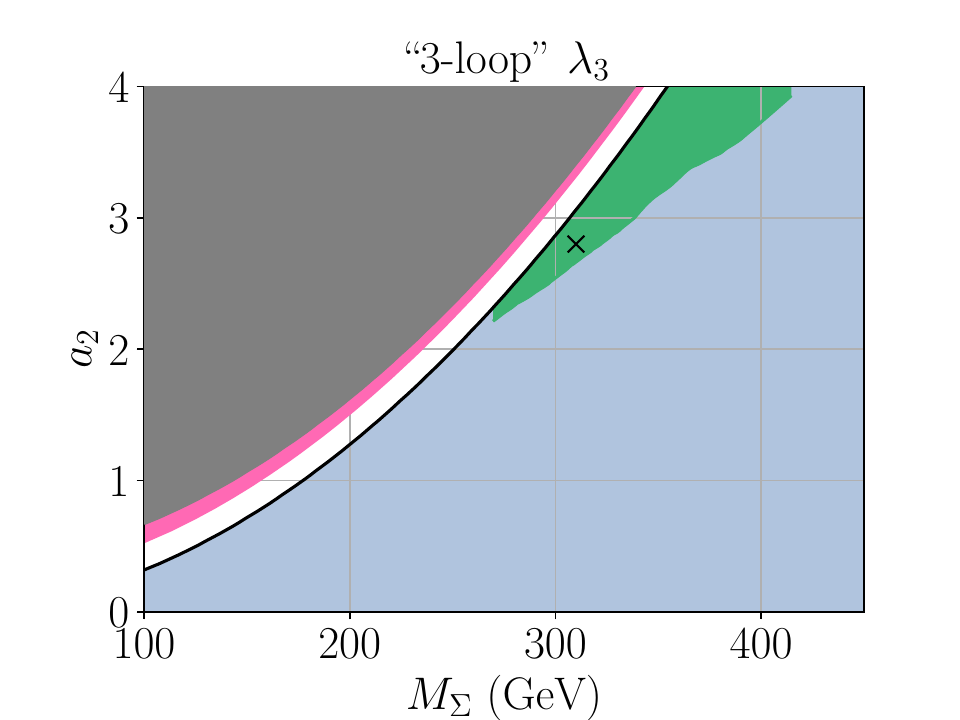} \\
\includegraphics[width=0.48\textwidth]{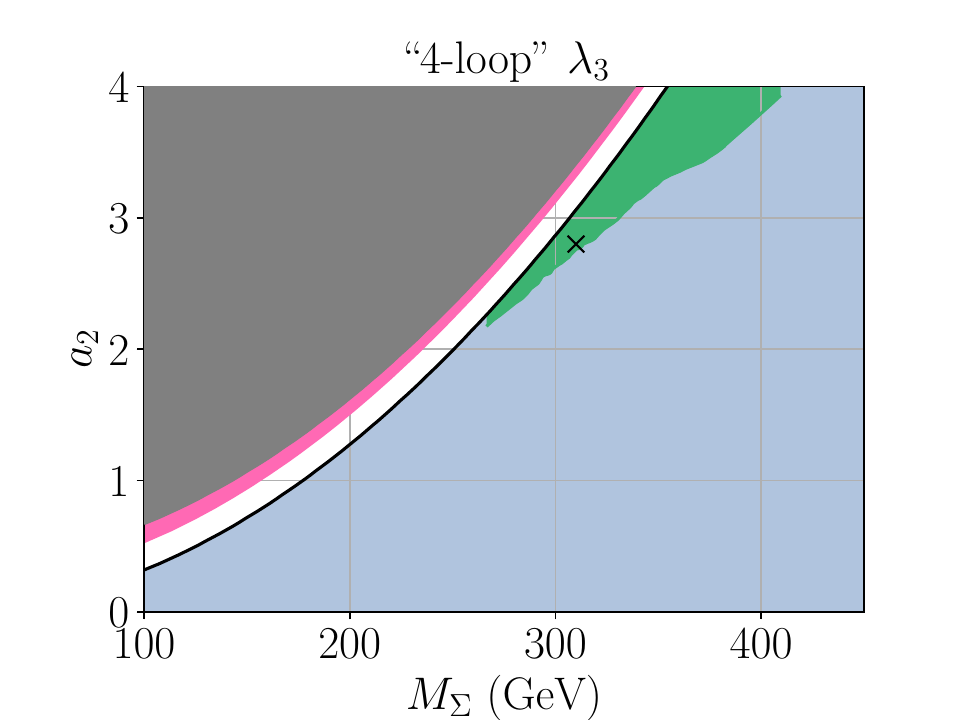}
\includegraphics[width=0.48\textwidth]{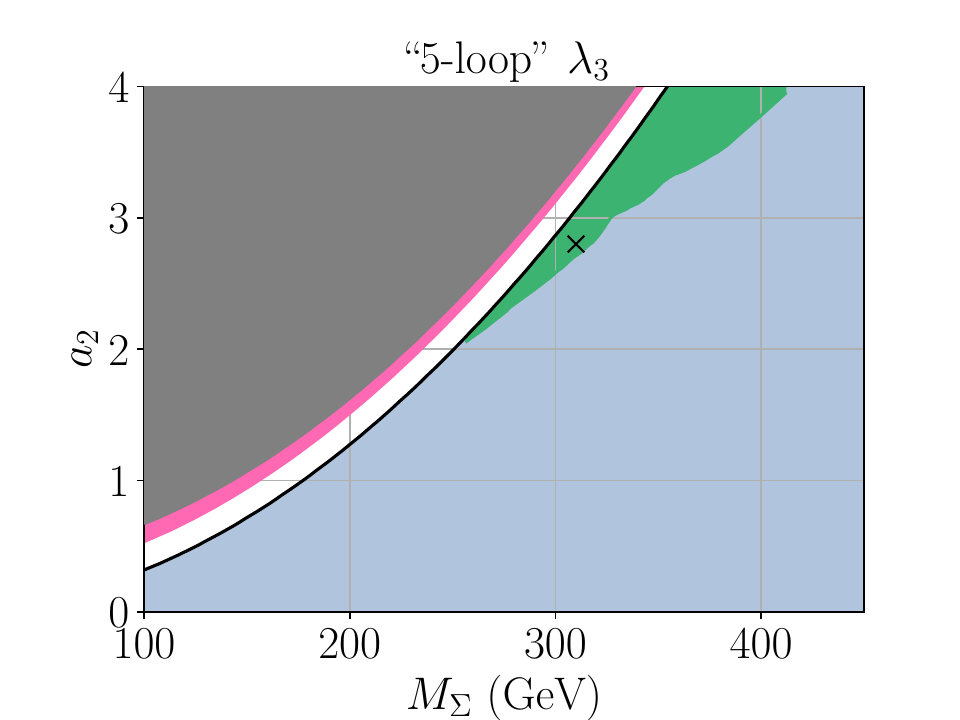}
\caption{
Similar to Fig.~\ref{fig:triplet-plane-loop1}, but with full two-loop corrections to $\lambda_3$ from Eq.~\eqref{eq:lam-2loop} (top left) and estimated higher loop effects of Eq.~\eqref{eq:lam-nloop}. Crucially, contrary to Fig.~\ref{fig:triplet-plane-loop1} there are no regions where the BNPC can be satisfied (as $x_c > 0.06$ for all the points), when full two-loop effects are accounted.
}
\label{fig:triplet-plane-corrections}
\end{figure}

For our analysis, the most interesting regions are those with $x_c < x_*$ that display a first-order one-step transition, albeit in the dark green region $x_c < 0$ and our final 3D EFT is not valid, as its scalar potential is not bounded from below due to the negative effective Higgs self-coupling. 
This is a clear signal that some higher order corrections are missing, either higher-order loop corrections to $\bar{\lambda}_3$, or higher-dimensional, marginal operators. 
We return to this issue below, but point out, that non-perturbative simulations of \cite{Niemi:2020hto} confirm that there indeed is a first-order phase transition in the point $M_\Sigma = 310$ GeV and $a_2=2.8$ ($\textsf{x}$ marks the spot in Figs.~\ref{fig:triplet-plane-loop1} and \ref{fig:triplet-plane-corrections}).
In light green region, we find a first-order phase transitions that are too weak to prevent the sphaleron washout, while regions shown in purple and orange satisfy weak and strong BNPC, respectively. We observe, that these regions are relatively narrow, especially the regime of the weak BNPC. This region (purple) is also prone to artificial numerical issues (wobbles) related to interpolation, which however could be eliminated by an extremely dense (and hence much slower) scan.  

\begin{figure}[t]
\centering
\includegraphics[width=0.6\textwidth]{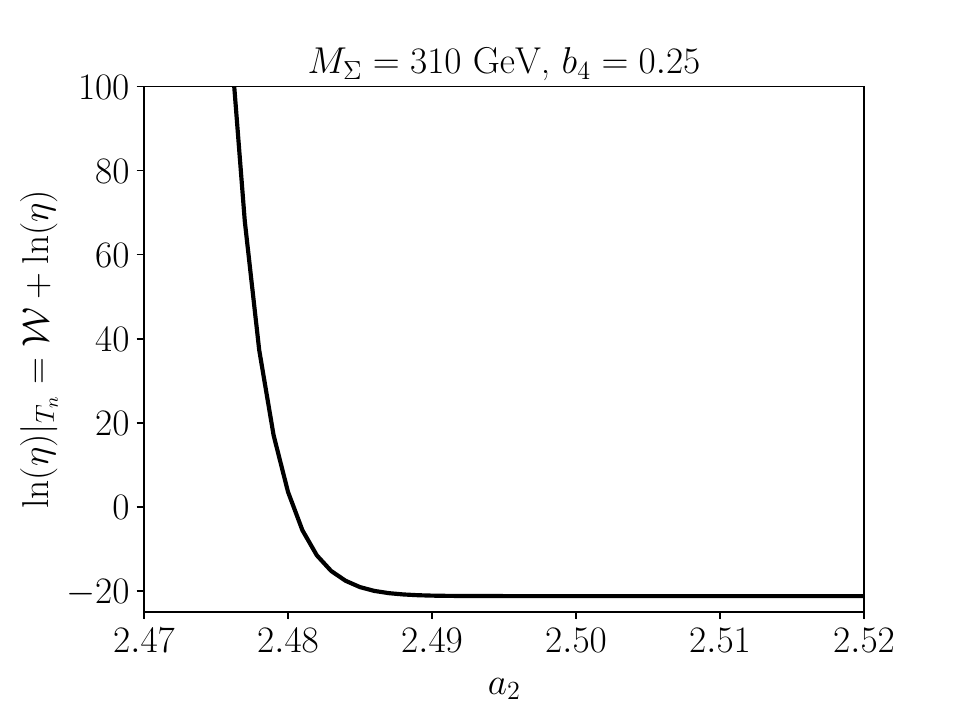}
\caption{
Initial baryon asymmetry $\ln\eta$ at $T_n$, that produces the observed value, as dictated by the washout exponent as function of $a_2$.
}
\label{fig:washout-a2}
\end{figure}

We illustrate the sphaleron washout further in Fig.~\ref{fig:washout-a2}, by fixing $M_\Sigma = 310$ GeV and plotting the initial asymmetry $\ln(\eta)$ at $T_n$ at the onset of the transition, that is required to produce the observed asymmetry, as function of $a_2$ (we compute the sphaleron washout exponent $\mathcal{W}$ in accord to Eq.~\eqref{eq:S-washout}).   
In this example case, when we vary $a_2 \in [2.47,2.52]$ we have $x_c\in [0.036,0.011]$ and $\eta_y \approx 5.75$ for all $a_2$. Interpretation of this result is clear: for $a_2 \gtrsim 2.497$ strong BNPC holds and there is no washout, which would enforce a constraint to produce the exact observed asymmetry at the bubble wall (should this be a realistic model for the EWBG with extra sources for CP-violation, as explained earlier). For slightly smaller values of $a_2$ the washout sets a tight constraint to produce much larger asymmetry at the bubble wall. This example illustrates, that if there would be an additional constraint from the CP-violation side of the computation for an upper bound for asymmetry that could be generated, it would put a tight constraint on the lower bound for the portal coupling. This is, of course, a simplified interpretation as in reality values of $M_\Sigma$ and $b_4$ also effect the situation, yet illustrates the fact that we have now a tool for such analyses.       

Our discussion so far exemplifies how the minimal SU(2) gauge-Higgs 3D EFT can be used to study the sphaleron rate and to impose the BNPC in a part of parameter space of a BSM theory. Until now, we have used the exactly same dimensional reduction mapping as in \cite{Niemi:2018asa} and in particular, determined $\lambda_3$ (Eq.~\eqref{eq:lam-1loop}) at one-loop (this is the same accuracy used in e.g.~\cite{Andersen:2017ika,Gorda:2018hvi} in the context of the Two-Higgs Doublet Model).   

However, using the mere one-loop result of Eq.~\eqref{eq:lam-1loop} has a following crucial caveat, in a context general to all Higgs portal models as pointed out in \cite{Ekstedt:2024etx,Niemi:2024vzw}. When the soft triplet is integrated out, its dominant contributions to the Higgs effective self-coupling come as a perturbative expansion in $a_{2,3}/(4\pi \mu_{\Sigma,3})$. The triplet 3D mass can indeed be large enough to justify to integrate the triplet out, e.g. $\mu_{\Sigma,3} > m_D$ compared to the Debye mass of the temporal scalar, but the 3D portal coupling $a_{2,3}$ is also required to be large enough to make it possible to reduce $x_c$ enough compared to its SM value $x_c \approx 0.29$. For this reason, the perturbative expansion for the soft triplet modes converges much slower compared to the expansion for temporal gauge field modes, which is expansion in $h_{3}/(4\pi m_D)$. For example, for the parameter space point $M_\Sigma = 310$ GeV and $a_2=2.8$ (marked by $\textsf{x}$ in Fig.~\ref{fig:triplet-plane-loop1}) around the critical temperature we find $h_{3}/(4\pi m_D) \approx 0.01$ but for the triplet $a_{2,3}/(4\pi \mu_{\Sigma,3}) \approx 0.25$. This implies, that while further corrections from the temporal gauge field mode are miniscule, triplet corrections at two-loop and beyond can have a significant effect. 

To scrutinize this, we compute such two-loop corrections from the soft triplet for the first time and find overall
\begin{align}
\label{eq:lam-2loop}
\lambda_3 & = \lambda_{3,\text{hard}} - \frac{1}{8\pi} \frac{h^2_{3}}{m_{D}} - \frac{1}{8\pi} \frac{3}{4} \frac{a^2_{2,3}}{\mu_{\Sigma,3}} \nonumber \\
& \hspace{15pt} + \frac{1}{(4\pi)^2 \mu^2_{\Sigma,3}} \biggl( \frac{3}{4}a^3_{2,3} - \frac{3}{2} a^2_{2,3} g^2_3 + \frac{3}{8} a_{2,3} g^4_3 - \frac{1}{128}\bigl(3 g^6_3 + g^4_3 {g_3'}^2 \bigr) \biggr) \,.
\end{align}
Within the two-loop corrections on the second line, the term proportional to $a^3_{2,3}$ indeed dominates over the other terms. 

In Fig.~\ref{fig:triplet-plane-corrections} (top left) we depict our results similar to Fig.~\ref{fig:triplet-plane-loop1}  but with two-loop corrections in accord with Eq.~\eqref{eq:lam-2loop}.
Strikingly, these two-loop corrections have a rather dramatic effect: the small $x_c$ region indicating first-order phase transitions in Fig.~\ref{fig:triplet-plane-loop1} shrinks significantly. 
This vividly highlights the fact that even predicting the character of a phase transition correctly, i.e. to ensure the existence of a first-order phase transition, is far more non-trivial than most often realized in the literature of phase transitions and their cosmological ramifications. In other words, predictions for all parameter space points in Fig.~\ref{fig:triplet-plane-loop1} that indicate first-order phase transitions, are not just quantitatively inaccurate, but qualitatively incorrect as there is no phase transitions but a crossover.
Furthermore all points displaying first-order transitions have $x_c \gtrsim 0.06$, i.e. the most strongest transitions (as well as the region with $x_c<0$) are completely absent. This indicates, that all one-step transitions in the region that can be studied by integrating out the soft triplet are rather weak, and hence can not prevent sphaleron washout (and hence would exclude the EWBG even if one would device additional C- and CP-violations to produce the initial asymmetry). 

In order to further inspect the convergence, we envision a following form for further higher order corrections
\begin{align}
\label{eq:lam-nloop}
-c_3 \frac{1}{(4\pi)^3} \frac{a^4_{2,3}}{\mu^3_{\Sigma,3}} + c_4 \frac{1}{(4\pi)^4} \frac{a^5_{2,3}}{\mu^4_{\Sigma,3}} - c_5 \frac{1}{(4\pi)^5} \frac{a^6_{2,3}}{\mu^5_{\Sigma,3}} \,,
\end{align}
at three-, four- and five-loops based on the expansion parameter, where $c_3$, $c_4$ and $c_5$ are numerical coefficients that would need to be determined by direct computation. We do not perform such computations here, but merely use the above form as a heuristic, semi-quantitative probe for convergence, and for this we fix each numerical coefficient to unity: a choice which is arbitrary, but it turns out that our conclusions are not terribly sensitive to this particular choice. 

We show the result of these heuristic estimates in other panels of Fig.~\ref{fig:triplet-plane-corrections}, and observe that while the boundary of first-order transition and crossover regions slightly shift, our conclusion from adding two-loop corrections remains intact: none of the transitions are strong enough to prevent the sphaleron washout. Furthermore, by inspecting heuristically the higher order corrections, we gain confidence that the full two-loop result we have used could be fairly accurate. This confidence is boosted by the non-perturbative confirmation of the first-order one-step transition in a single test point, marked by $\textsf{x}$ in Fig.~\ref{fig:triplet-plane-corrections}.

This brings us to our final conclusion for (radiatively-generated) first-order, one-step phase transitions in Higgs portal models, which we summarize as
\begin{itemize}
    \item Reliable analyzes of phase transition thermodynamics --- including ensuring the existence of the transition -- and sphaleron properties, require to consistently include \textit{all two-loop} thermal corrections related to the hard thermal scale (thermal resummations in dimensional reduction) and corrections from the soft scale perturbation theory (within the 3D EFT), if soft-scale extra scalar fields are used to trigger the phase transition.
\end{itemize}
This conclusion aligns with those previously obtained in e.g.~\cite{Kainulainen:2019kyp,Croon:2020cgk, Niemi:2024vzw} (see also \cite{Niemi:2024axp}). In this context, ubiquitous one-loop studies of the EWBG in the literature, let alone studies that do not consider thermal resummations at all but resort to mere leading order high-temperature expansions (perhaps with a motive to avoid spurious gauge-dependence, which would be an outcome of incorrect formulation of the perturbative expansion) could be not merely quantitatively inaccurate, but qualitatively flawed. This would be the case, if there is no phase transition but a crossover, or if a first-order phase transition exists but is far too weak to prevent the washout. 

We remark, that based on our study in this work at hand alone, above conclusion can not be drawn for multi-step phase transitions that are not necessarily induced radiatively by the soft-scale loops. However, even for such multi-step phase transitions similar soft scale two-loop corrections have found to be crucially important for quantitatively reliable perturbative analyses, c.f.~\cite{Gould:2023ovu}.

\section{Conclusions and Perspectives}
\label{sec:conlusions}

We conclude by summarizing our present work and by providing further discussion and outlook.

\subsection{Summary} 
\label{sec:summary}

In this article, working with high temperature effective field theory (3D EFT), we have reformulated the perturbative computation of (the statistical part of) the Higgs phase sphaleron rate. 
This reformulation was designed to ensure that 
(i) large thermal resummations are incorporated,
(ii) results are gauge-invariant,
and 
(iii) calculations are based on a consistent perturbative expansion that avoids double counting at higher orders (for the sphaleron determinant).

We concretely demonstrated features (i) and (ii) (bringing clarity to issues raised in \cite{Patel:2011th}), but relegated the aspect (iii) to future work, postponing the computation of the sphaleron determinant. We remark, that at present the dynamical, real-time part of the sphaleron rate is entirely out-of-reach of our formulation.  

Besides solely theoretical motives to seek a sound perturbative formulation for the Higgs phase sphaleron rate, there is also a clear practical motive: for more complicated BSM theories that include several light scalar fields within the 3D EFT, at present there are no lattice simulations for the sphaleron rate. Furthermore, due to the expense of such lattice simulations, scanning of wide free parameter spaces are are not feasible without a perturbative approach. In this work at hand, we have worked towards such a general perturbative formulation, yet limited our investigations for the minimal 3D EFT used previously for the SM at high temperatures.

Concretely, within the 3D EFT 
that captures large thermal corrections (thermal resummations), we resorted to a semi-classical approximation for the sphaleron rate.  We found that the sphaleron action is dominated by the Yang-Mills and scalar covariant kinetic terms, while the scalar potential itself results directly in a sub-leading effect. By rescaling the sphaleron action, we found a powerful formulation of the action in terms of the expectation value of the scalar condensate (c.f.~\cite{Qin:2024dfp}), and we could further approximate the rate by simple analytic fitting formulae.     

To put our semi-classical computation to test, we compared it directly to previous non-perturbative lattice simulations of the sphaleron rate for temperatures below the Standard Model crossover and found a reasonable agreement. In the same SU(2) gauge-Higgs 3D EFT 
a first-order phase transition requires the ratio of the effective thermal Higgs self-coupling and the effective gauge coupling to be small enough, i.e. $x\equiv \lambda_3/g^2_3 \lesssim 0.1$ at the critical temperature. While this condition is not met in the SM, in many BSM extensions with extra fields this criterion could be realized  (when extra scalars are sufficiently heavy at high temperatures, and can hence be integrated out from the 3D EFT). 

For such first-order phase transitions, we combined our computation of the sphaleron rate in the Higgs phase with perturbative treatment of the bubble nucleation rate, and formulated conditions for avoiding the sphaleron washout that would erase any baryon asymmetry generated in front of expanding bubble walls. Concretely, we found that the washout is not possible at all for $x \lesssim 0.025$ as sphalerons decouple in the Higgs phase immediately when bubbles can nucleate, while for $0.025 \lesssim x \lesssim 0.036$ sphalerons can (exponentially) washout more or less any amount of asymmetry while still providing the observed asymmetry, yet such balancing sets a tight constraint on the initial asymmetry and $x$.%
\footnote{
Note, that here one assumes that $x$ is a constant within a range of temperatures from $T_c$ to $T_n$ and $T_f$, i.e. critical, nucleation and sphaleron decoupling temperatures. This assumption could be relaxed, which leads to more specific, model-dependent conditions between $x_c$, $x_n$ and $x_*$, dictated by $x=x(T)$ dimensional reduction mapping.
}

As a direct application, we demonstrated these calculations for the real triplet-extended Standard Model in parts of the parameter space that can be mapped into the 3D EFT we have considered. Crucially, we found out that taking into account two-loop order corrections for $x$ by integrating out the heavy triplet, none of one-step the transitions (that can be found in this particular 3D EFT framework and have been reported in the past) were strong enough to prevent total washout. Such result demonstrates that this part of the parameter space would be ruled out as a possible explanation for the baryogenesis, even if it would be incorporated with a mechanism to add new sources of C- and CP-violation to generate the initial asymmetry across the bubble wall. 

Furthemore, we ephasized that to reach this conclusion, aforementioned two-loop corrections to $x$ (or $\lambda_3$) -- that were not considered in earlier studies -- studies, were crucial. This conclusion does not come as a surprise, as by now there are ample examples in the literature that have highlighted the importance of thermal two-loop effects. This further raises serious doubts on reliability of low-order computations for the sphaleron rate and phase transition thermodynamics in general, in myriad studies of the past and present literature. 

\subsection{Discussion and Outlook}
\label{sec:discussion}

Next, let us provide some perspectives by reflecting our present work to the past literature, and envisioning future directions.

\paragraph{$v_c/T_c$: gauge-dependence and convergence}
We stress that ensuring the existence of a first-order phase transition is more non-trivial than often appreciated in the vast literature. 
Indeed, myriads of EWBG studies in BSM theories (c.f. e.g.~\cite{Ahriche:2014jna,Fuyuto:2014yia,Fuyuto:2015jha,Carena:1996wj,Bodeker:1996pc,Quiros:1999jp,Huber:2000mg,Grojean:2004xa,Menon:2004wv,Delaunay:2007wb,Espinosa:2007qk,Profumo:2007wc,Barger:2008jx,Espinosa:2008kw,Funakubo:2009eg,Espinosa:2011ax,Cline:2013gha,Dorsch:2014qja,Curtin:2014jma,Basler:2016obg,Dorsch:2016nrg,Beniwal:2017eik,Bruggisser:2018mrt,Cline:2021iff}) resort to a condition $v_c/T_c \gtrsim 1$ for the existence of a strong enough first-order phase transition (strong enough to prevent sphalerons washing out the baryon asymmetry in the Higgs phase), where $v_c$ is the Higgs field expectation value at the critical temperature $T_c$. 
Such approaches are plagued by two problems: gauge dependence and (at first sight) bad convergence.

As discussed in detail in \cite{Laine:1994zq,Patel:2011th}, the quantity $v_c/T_c$ is gauge-dependent and hence unphysical: according to the Nielsen-Fuguda-Kugo identities \cite{Nielsen:1975fs,Fukuda:1975di}, only the values of the effective potential at its extrema are gauge invariant, but the field values of extrema can be shifted around in terms of a gauge fixing parameter. 
More often than not, this fact has been ignored, and otherwise serious calculations have been performed simply in Landau gauge as a leap of faith. 
Landau gauge greatly simplifies computations of the thermal effective potential, 
and $v_c/T_c$ (in Landau gauge) has been observed to positively correlate with the phase transition strength in terms of the latent heat, a physical observable obtainable by gauge-invariant means (see e.q.~\cite{Bodeker:1996pc,Laine:2012jy,Niemi:2021qvp,Schicho:2022wty}), and hence $v_c/T_c$ has been used as a heuristic measure of existence of a first-order phase transition, which is characterized by a discontinuity in its free-energy.

However, such discontinuity exists in perturbation theory whenever two minima of the thermal effective potential are separated by a barrier, and hence in such cases perturbation theory always predicts a first-order transition. This is in stark contrast with non-perturbative simulations, that can reveal that there is no phase transition at all, but a crossover (see e.g.~\cite{Kajantie:1995kf,Gould:2022ran} and recent \cite{Niemi:2024axp} for a BSM application). 
On the otherhand, such simulations have also indicated, that transitions that are strong according to the perturbation theory, are also of first order non-perturbatively. 

On the surface, this might be taken as absolution for using $v_c/T_c$ in Landau gauge as a (heuristic) measure of the phase transition strength and limiting to large enough values, but this strategy is short-lived, upon closer examination: lots of studies have resorted to using ring-resummed thermal effective potential at one-loop order, but it has been known for a long time that large two-loop thermal corrections can jeopardize such analyses \cite{Arnold:1992rz,Farakos:1994kx,Buchmuller:1995sf}. 
In a recent work \cite{Niemi:2024vzw} it has been shown that such two-loop corrections can not only strengthen transitions compared to one-loop, but also significantly weaken them so much that there is no indication for a first-order transition. 
In general, two-loop corrections were found to lead to narrower regions of strong first-order transitions in a large-scale parameter space scan of a BSM theory (see also \cite{Kainulainen:2019kyp,Croon:2020cgk}). 

On the otherhand, one might then wonder why a two-loop result would be any more reliable than a one-loop result, if the overall convergence is bad? 
It has turned out, that convergence at high temperatures is often not bad at all, but rather \textit{slow}: thermally enhanced bosonic contributions hamper the perturbative expansion (and require resummations), and reaching the same accuracy as zero temperature analogues requires significantly more work with two-loop (and even three-loop \cite{Rajantie:1996np,Gould:2021dzl,Gould:2023jbz,Ekstedt:2024etx}) calculations \cite{Arnold:1992rz,Farakos:1994kx,Gould:2021oba}. 
These higher order calculations beyond the realm of one-loop resummations have been shown to provide remarkable agreements with lattice simulations, for strong phase transitions \cite{Gould:2021dzl,Ekstedt:2022zro,Gould:2023ovu,Ekstedt:2024etx}.%
\footnote{We stress, that both the lattice and perturbative calculations are based on high-temperature expansions, which might fail in presence of large BSM couplings, and large background field values in case of strong transitions, that lead to large masses in the Higgs phase. Hence, it is crucial to scrutinize the validity of the high-$T$ expansion, see \cite{Laine:2000kv,Laine:2017hdk}.
}

\paragraph{Previous (gauge-invariant) results for BNPC}
Contrasting our results for baryon number preservation condition in terms of bounds on $x$ to previous results in \cite{Moore:1998ge,Moore:1998swa}, we found a fair agreement, yet point out that our bound $x \lesssim 0.025$ to prevent the washout entirely is slightly stronger. This indicates that the electroweak phase transition has to be even stronger than previously believed, in order to prevent the washout completely.
We note, that while the results for the sphaleron rate in \cite{Moore:1998ge,Moore:1998swa} are non-perturbative and hence superior to our leading, semi-classical approximation, the treatment of the bubble nucleation in these references was not accounted up to same standard, with the modern 3D EFT picture for the bubble nucleation \cite{Moore:2000jw,Gould:2022ran}.

It is illuminating to further reflect our perturbative formulation to a discussion in \cite{Moore:1998ge}. 
Therein, it is argued that it is in fact easier to perform non-perturbative simulation of the sphaleron rate, rather than to perform it at ``two loops'' in perturbation theory, meaning two-loop 
corrections beyond the one-loop determinant.
Arguably, such computation indeed resembles a nightmare: one has to compute such corrections around spatially varying sphaleron background, which requires using propagators based on Green's functions that have to be determined (and eventually integrated) numerically inside two-loop diagrams (this can be compared with an analogous situation with the bubble nucleation \cite{Ekstedt:2022tqk} or vacuum tunneling \cite{Bezuglov:2018qpq}).  

Ref.~\cite{Moore:1998ge} expects such two-loop corrections to be of sizable importance, as two-loop corrections were known to be large for the expectation value of the Higgs condensate $\sqrt{\langle \phi^\dagger_3 \phi_3 \rangle}$ (in the 3D EFT), that contributes to the height of the sphaleron barrier. Ref.~\cite{Moore:1998ge} correctly addresses that the perturbative expansion for the condensate is at best an expansion in $\lambda_3/g^2_3$, and concludes   
that there is no known reason why the expansion for the sphaleron rate
should be any better, i.e. why two-loop effects for it would be less important.
But now such reason is indeed known, as it has been worked out that the expansion for the Higgs condensate is an expansion in $\sqrt{\lambda_3/g^2_3}$ \cite{Ekstedt:2022zro,Ekstedt:2024etx} and, crucially, it does not align with a loop expansion: the result for the condensate at leading order includes contribution from gauge bosons at one-loop, and their two-loop contribution gives first correction suppressed by $\lambda_3/g^2_3$ (with all further corrections suppressed by $\sqrt{\lambda_3/g^2_3}$).
Intuitively, such expansion appears since the scale of the phase transition is not the soft, but rather the supersoft scale deeper in the IR.

For the sphaleron rate on the otherhand, already the one-loop determinant is suppressed by factor $\lambda_3/g^2_3$, and based on a physical picture of \cite{Gould:2023ovu}, the expansion parameter for the sphaleron rate is $\lambda_3/g^2_3$ because sphalerons live at the soft scale, i.e. aforementioned complicated two-loop corrections around the sphaleron background are further suppressed as $(\lambda_3/g^2_3)^2$, and have parametrically same suppression as three-loop corrections to the Higgs condensate \cite{Ekstedt:2024etx}. 
Therefore, we can expect that already a one-loop computation within the 3D EFT for the sphaleron rate is fairly accurate, assuming one still accounts for two-loop level hard scale corrections in dimensional reduction mapping into the 3D EFT.
Consequently, for the sphaleron rate we expect better convergence than for the bubble nucleation rate \cite{Ekstedt:2022ceo}, as characteristic scale for the sphaleron is higher in UV than the scale of the critical bubble, and hence perturbation theory convergences faster. 

This discussion is supported by the results of \cite{Burnier:2005hp} that included the computation of the sphaleron determinant (in 3D EFT), and its great agreement with non-perturbative simulations in \cite{DOnofrio:2012phz,DOnofrio:2014rug,Annala:2023jvr}.  
This certainly motivates to push our present formulation one order further, and computate the sphaleron determinant in future works.

\paragraph{Outlook}
As for other future prospects, our findings motivate to perform a set of fresh lattice simulations for the sphaleron rate akin to \cite{Annala:2023jvr,DOnofrio:2012phz,DOnofrio:2014rug} without using any concrete dimensional reduction mapping, but for all (input) $y$ and $x$ within the 3D EFT, with $x$ in range relevant for first-order phase transitions (in analogue for simulations of the bubble nucleation rate in \cite{Gould:2022ran}).
This would ultimately test reliability of our perturbative computation%
\footnote{
In principle, one -- with enough valor -- could also strive for studying the real-time, HTL-resummed effects to the dynamical part of the sphaleron rate in BSM theories, c.f.~\cite{Ekstedt:2022tqk,Ekstedt:2023anj,Ekstedt:2023oqb,Eriksson:2024ovi,Hirvonen:2024rfg,Dashko:2024anp}. 

}
(and future computations of the sphaleron determinant), and our results for the baryon number preservation condition. We note, that since such comparison for the sphaleron rate has already provided a fair agreement in the case of $x = 0.29$ relevant for temperatures below the Standard Model crossover, we can expect similar (or better) agreement for smaller $x$ as well, since the perturbative expansion of the sphaleron rate is in positive integer powers of $x$.  

Furthermore, our perturbative formulation of the sphaleron rate can be generalized to other 3D EFTs, where BSM fields that catalyze first-order phase transitions are light and are not integrated out. In such case, new fields contribute in the sphaleron equations of motion and sphaleron action, and their effect for the bubble nucleation has to be accounted as well. This is especially relevant for strong two-step phase transitions, prominent to produce strong primordial gravitational wave signatures. To study both the EWBG and gravitational wave production in such scenarios using 3D EFTs, computational tools such as {\tt DRalgo} \cite{Ekstedt:2022bff,Fonseca:2020vke} for the EFT construction can be utilized, and computations of gravitational wave thermal parameters follow those presented in \cite{Gould:2021oba,Friedrich:2022cak,Ramsey-Musolf:2024ykk,Gould:2024jjt}, and for computing previously elusive terminal velocity of the bubble wall, a new code {\tt WallGo} \cite{Ekstedt:2024fyq} could be used. 

 Finally, we envision that our perturbative formulation for the Higgs phase sphaleron rate can be used in the future in an end-to-end computation of the electroweak baryogenesis, under a concrete BSM theory that contains all necessary ingredients including new sources for C- and CP-violation and consistent predictions for generated initial asymmetry across the bubble wall. Combined with BSM collider phenomenology for such a model, this would allow to probe concrete and testable scenarios of what lies beyond the Standard Model -- reliably with refined theoretical tools for thermal bubble and sphaleron dynamics -- with aim to explain the observed baryon asymmetry of our universe.

\section*{Acknowledgments}

We thank Jaakko Annala, Andreas Ekstedt, Leon Friedrich, Oliver Gould, Joonas Hirvonen, Jaakko H{\"a}llfors, Xiyuan Jin, Benoit Laurent, Johan L{\"o}fgren, Tobias Rindlisbacher, Lauri Niemi, Kari Rummukainen, Riikka Sepp{\"a}, Ville Vaskonen, Jorinde van de Vis, Guotao Xia and Jiang Zhu for illuminating discussions. 
This work was supported in part by National Natural Science Foundation of China grant 12375094 (MJRM and YW) and partly funded by the European Union (ERC, CoCoS, 101142449). TT thanks Nils Hermansson-Truedsson and University of Edinburgh, Djuna Croon and Durham University, as well as Oliver Gould and University of Nottingham, for hospitality during completion part of this work.

\appendix
\counterwithin{equation}{section}

\section{Properties of Sphaleron Ansatz, Action, and Equations of Motion}
\label{app:Properties of sphaleron ansatz, action, and equation of motion}

In this appendix, we present more details about the sphaleron ansatz, action and equations of motion.

\paragraph{Sphaleron 3D ansatz} We propose the sphaleron ansatz in 3D that reads
\begin{align}
\hat{A}_{i,3}^{a}T^{a}\mathrm{d}x^{i}&=[1-f(\xi)]\sum_{i=1}^{3}F_{i}T^{i} \,,
\end{align}
where $F_i$ are functions of the spherical coordinates $\theta$ and $\varphi$ 
as well as a loop parameter $\mu$ \cite{Klinkhamer:1993hb}
\begin{equation} \label{eq:1-form_F_a}
	\begin{aligned}
		F_1=&-(2\sin^2\mu\cos(\mu-\varphi)-\sin2\mu\cos\theta\sin(\mu-\varphi))d\theta \\
		& -(\sin2\mu\cos(\mu-\varphi)\sin\theta+\sin^2\mu\sin2\theta\sin(\mu-\varphi))d\varphi,\\
		F_2=&-(2\sin^2\mu\sin(\mu-\varphi)+\sin2\mu\cos\theta\cos(\mu-\varphi))d\theta \\
		& +(-\sin2\mu\sin(\mu-\varphi)\sin\theta+\sin^2\mu\sin2\theta\cos(\mu-\varphi))d\varphi,\\
		F_3=&-\sin2\mu\sin\theta d\theta+2\sin^2\theta \sin^2\mu d\varphi.
	\end{aligned}
\end{equation}
Technically, $\mu$ is the non-contractible loop parameter that connects topologically different vacua \cite{Manton:1983nd}.

Now, we demonstrate that this 3D ansatz is consistent with the previous one in the full four-dimensional (4D) theory \cite{Klinkhamer:1990fi}. Relations (at leading order in dimensional reduction) between 3D and 4D fields is following: in 3D $A_{i,3}^{a}=g_3\hat{A}_{i,3}^{a}$; and the 4D gauge field $A_{i}^{a}=\sqrt{T}A_{i,3}^{a}=g_3 \sqrt{T} \hat{A}_{i,3}^{a} = gT \hat{A}_{i,3}^{a}$, where $g_3=g\sqrt{T}$ at leading order.
Then, the gauge field ansatz in Eq.(\ref{eq_app:gauge_sphaleron_asatz_in_3DEFT}) implies the 4D gauge ansatz
\begin{align} 
A_{i}^{a}T^{a}\mathrm{d}x^{i}&=gT[1-f(\xi)]\sum_{i=1}^{3}F_{i}T^{i} \,, 
\end{align}
Since 
\begin{align}
\sum_{i=1}^{3}F_{i}T^{i}=i(U^{\infty -1})\partial_{i}U^{\infty} \,,
\end{align}
where $U^\infty$ is the sphaleron matrix.%
\footnote{
The construction of $U^\infty$ for a general SU(2) scalar multiplet is given in Ref.~\cite{Wu:2023mjb}
}
The $\partial_{i}$ in dimensionless 3D theory differs a factor of $1/g_3^2$ with 4D, so we need to multiply by this factor for the 4D convention. 
Hence, the 4D gauge field sphaleron ansatz reads
\begin{align}
A_{i}^{a}T^{a}\mathrm{d}x^{i}&=\frac{1}{g}[1-f(\xi)]\sum_{i=1}^{3}F_{i}T^{i} \,.
\end{align}
which is the sphaleron ansatz in conventional 4D approach \cite{Wu:2023mjb,Klinkhamer:1990fi}. Note that an additional scaling by $v_3$ (Eq.~\ref{eq:rescaling}) does not alter the above conclusion, since the factor of $v_3$ cancels out when rescaling $\hat{A}_{i,3}$ and $\partial_i$.

\paragraph{Sphaleron action} 
In the main text, the sphaleron action transforms into
\begin{align}
\label{eq:Shat}
\hat{S}_{\text{3D}} = v_3(x,y) \mathcal{C}_{\text{sph}}(x,y),
\end{align}
where
\begin{align} 
\label{eq:general_C_sph}
\mathcal{C}_{\text{sph}}(x,y) = \int \mathrm{d}^{3}\xi \left[\frac{1}{4}\hat{F}_{ij,3}^{a}\hat{F}^{a}_{ij,3} + (\hat{D}_{i}\hat{\phi})_{3}^{\dagger}(\hat{D}_{i}\hat{\phi})_{3} + \overline{V}_{3}(\hat{\phi}_3,x,y)\right].
\end{align}
where we have omitted the prime on the transformed coordinates and fields. The transformed Yang-Mills and covariant derivative terms are given by
\begin{align}
\label{eq:app:Yang_Mills_expression_under_ansatz}
   \frac{1}{4}\hat{F}_{ij,3}^{a}\hat{F}^{a}_{ij,3}&=\frac{8(-1+f)^2f^2\sin^4\mu}{\xi^4} + \frac{4f^{\prime 2} \sin^2\mu}{\xi^2}, \\ 
   \label{eq:app:COD_expression_under_ansatz}
   \bigl(\hat{D}_{i}\hat{\phi}\bigr)_{3}^{\dagger} \bigl(\hat{D}_{i}\hat{\phi}\bigr)_{3} &= \frac{\left[3J(1+J)-J_3^2+(J+J^2-3 J_3^2)\cos 2\theta \right](-1+f)^2h^2\sin^2\mu}{\xi^2}+h^{\prime 2} \,,
\end{align}
where $\mu$ and $\theta$ are aforementioned sphaleron parameters, the prime on the radial profile functions denotes the derivative with respect to $\xi$, $J$ refers to the representation dimension for scalar field under SU(2).
The covariant derivative expression is given in a form of a general dimensional SU(2) multiplet, where for the SM complex doublet, $J=1/2$ and $J_3=1/2$. 
The spherical angles $\varphi,\theta$ in the action can be integrated over, and remaining integration is over the radial parameter, such that
\begin{align}
\mathcal{C}_\text{sph}(x,y) = \int_0^\infty \mathrm{d}\xi \left[E_{\text{YM}}(\xi)+E_{\text{COD}}(\xi)+E_{\overline{V}}(\xi)
\right] \,,
\end{align}
where different terms refer to Yang-Mills, covariant derivative (\lq\lq COD\rq\rq) and potential terms, for which
\begin{align}
\label{eq:Three_seperate_terms_of_Fx}
E_{\text{YM}}(\xi)&=16\pi \sin^2\mu \left[ \frac{2(-1+f)^2 f^2 \sin^2\mu}{\xi^2}+f^{\prime 2} \right], \\
E_{\text{COD}}(\xi)&=\frac{4\pi }{3}\left[8 J(1+J)(-1+f)^2 h^2 \sin^2\mu + 3 \xi^2 h^{\prime 2} \right] \,, \\
E_{\overline{V}}(\xi)&=4\pi \xi^2 \overline{V}_{3} (h) \,,   
\end{align}
where $\mu=\pi/2$ is taken for sphaleron energy, and the $\overline{V}_{3}(h)$ is defined through
\begin{equation} \label{eq:Vpot_definition}
\overline{V}_{3} (h) = \overline{V}_{3} \biggl(h,\mu=\frac{\pi}{2}\biggr)-\overline{V}_{3} \biggl(h,\mu=-\frac{\pi}{2}\biggr) \,.
\end{equation}
where $\mu=-\frac{\pi}{2}$ corresponds to the vacuum configuration. For more details, see Ref.~\cite{Wu:2023mjb}.

\paragraph{Equations of motion}
The equations of motion for the radial profile functions can be obtained from the variation principle
\begin{align} \label{eq:sphaleron_EOM_with_VEV}
f^{\prime \prime} - \frac{2f(-1+f)(-1+2f)}{\xi^2}-\frac{2}{3}J(1+J)h^2 = 0 \,, \\
h^{\prime \prime}+\frac{2h^\prime}{\xi}-\frac{8}{3\xi^2}J(1+J)(-1+f)^2h-\frac{1}{2} \frac{\partial \overline{V}_3}{\partial h}=0 \,.
\end{align}

\section{On Sphaleron Fluctuation Determinant}
\label{sec:sphaleron-determinant}

First, let us review the derivation of the sphaleron rate prefactor given in Ref.~\cite{Arnold:1987mh}. 
We start from the sphaleron rate expression, formulated as
\eqal{1}{
\Gamma\simeq \frac{\omega_{-}}{2\pi}\text{Im}\left(\frac{\det g_{3}^{-2} \Omega_{0}^{2}}{\det g_{3}^{-2}  \Omega_{\text{sph}}^{2}} \right)^{1/2}e^{-\beta E_{\text{sph}}}
}
where $\beta \equiv 1/T$, and $\Omega_{0}^{2}$ is the quadratic operator at the vacuum, and $\Omega_{\text{sph}}^{2}$ is the quadratic operator at the sphaleron configuration. 
Translational invariance implies that $\Gamma$ must be proportional to the volume $V(\xi)$, which is $V_{\xi}=(gv)^{3}V_{r}$ ($\xi=gvr$ in Ref.~\cite{Arnold:1987mh}). 
The $V_{\xi}$ is dimensionless, and $V_{r}$ is of dimension $[m]^{-3}$. 

The zero modes must be integrated separately using the method of collective coordinates. 
When we integrate them out, we will obtain products of form $\mathcal{N}\mathcal{V}$ where $\mathcal{N}$ is a normalization factor and $\mathcal{V}$ is the volume of the symmetry group responsible for zero modes. 
There are four symmetries in the system: translations, rotations, the SU(2)$_{L}$ of the weak gauge group and the global, custodial SU(2)$_{R}$ of the Higgs sector. 
We get
\begin{itemize}
\item Translation: $\mathcal{N}_{\text{tr}}V_{\xi}=\mathcal{N}_{\text{tr}}(gv)^{3}V_{r}$,
\item Rotation: $(\mathcal{N}\mathcal{V})_{\text{rot}}$,
\item SU(2)$_{L}$: the gauge freedoms are fixed, either local or global,
\item SU(2)$_{R}$: turns out to be linear combination of others,
\end{itemize}
so that
\eqal{1}{ 
\Gamma\simeq \frac{\omega_{-}}{2\pi}\mathcal{N}_{\text{tr}}V_{r} (\mathcal{N}\mathcal{V})_{\text{rot}} (gv)^{3} g_{3}^{-N_{0}}\text{Im}\left(\frac{\det g_{3}^{-2}\Omega_{0}^{2}}{\det^{\prime} g_{3}^{-2}  \Omega_{\text{sph}}^{2}} \right)^{1/2}e^{-\beta E_{\text{sph}}}
}
where the $g_{3}^{-N_{0}}$ is from the $N_{0}$ more eigenvalues in the numerator compared to denominator. 
We can transform this expression into a more familiar form
\eqal{1}{ \label{eq:sphaleron rate original}
\Gamma_{\text{sph}}\equiv\frac{\Gamma}{V_{r}}&\simeq \frac{\omega_{-}}{2\pi}\mathcal{N}_{\text{tr}} (\mathcal{N}\mathcal{V})_{\text{rot}} (gv)^{3} g_{3}^{-6}\text{Im}\left(\frac{\det g_{3}^{-2}\Omega_{0}^{2}}{\det^{\prime} g_{3}^{-2}  \Omega_{\text{sph}}^{2}} \right)^{1/2}e^{-\beta E_{\text{sph}}} 
}
where we take $N_{0}=6$ for three translation and three rotations. 
The $(gv)^{3} g_{3}^{-6}$ can be further written as 
\eqal{1}{
(gv)^{3} g_{3}^{-6} = (4\pi \alpha_{3})^{-3} (gv)^{3} = \alpha_{3}^{-6} \left(\frac{gv}{4\pi}\right)^{3} \left(\frac{g T}{4\pi v}\right)^{3}  =  \alpha_{3}^{-6}  \left( \frac{\alpha_{W}T}{4\pi} \right)^{3},
}
where we have used the relations
\eqal{1}{ \label{eq: scaling of alpha3 in carson et al.}
\alpha_{3}=\frac{g_{3}^{2}}{4\pi}=\alpha_{W}\frac{T}{gv}\,, \quad \alpha_{W}=\frac{g^{2}}{4\pi} \,.
}
By further denoting
\eqal{1}{ \label{eq_app:kappa_definition}
\kappa \equiv \text{Im}\left(\frac{\det g_3^{-2} \Omega_{0}^{2}}{\det^{\prime} g_3^{-2}  \Omega_{\text{sph}}^{2}} \right)^{1/2},
}
the sphaleron rate expression of Eq.~(\ref{eq:sphaleron rate original}) can be expressed as 
\eqal{1}{
\Gamma_{\text{sph}}&\simeq \frac{\omega_{-}}{2\pi}\mathcal{N}_{\text{tr}} (\mathcal{N}\mathcal{V})_{\text{rot}} \alpha_{3}^{-6}  \left( \frac{\alpha_{W}T}{4\pi} \right)^{3} \kappa e^{-\beta E_{\text{sph}}} \,.
}

Next, we want to reformulate this expression using the language of the 3D EFT. 
Such derivation follows the same lines as above.
We start from the $v_3(x,y)$-scaled action
\begin{align} \hat{S}_{\text{3D}}=v_3(x,y)\mathcal{C}_{\text{sph}}(x,y).
\end{align}
First, we need to notice that $\mathcal{N}_{\text{tr}}V_\xi=\mathcal{N}_{\text{tr}}v_3^3(x,y)g_3^6V_r$.
Second, we expect an additional $v_3^6(x,y)$ factor come from the six zero modes.  
Hence, the sphaleron rate can be written as
\begin{equation} \label{eq_app:sphaleron_rate_to_one_loop}
	\Gamma_{\text{sph}}=\frac{\omega_{-}}{2\pi}\mathcal{N}_{\text{tr}}(\mathcal{N}\mathcal{V})_{\text{rot}}v_3^9(x,y) g_3^6  \kappa e^{-v_3(x,y)\mathcal{C}_{\text{sph}}}
\end{equation}
where the definition of $\kappa$ is same with Eq.(\ref{eq_app:kappa_definition}).  
The $\omega_-$ has mass dimension one, $g_3^6$ is of dimension three, $\Gamma_{\text{sph}}$ is of dimension four and other quantities are dimensionless. 
The $\omega_{-}$ and $\mathcal{N}_{\text{tr}}(\mathcal{N}\mathcal{V})_{\text{rot}}$ are functions of $x$ and $y$.

If we only keep the leading order result, the sphaleron rate is expressed as
\begin{align}
    \Gamma_{\text{sph}}=T^4\times e^{-v_3(x,y)\mathcal{C}_{\text{sph}}} \,.
\end{align}
If we keep the leading order and the zero mode contributions to the determinant, the result for the sphaleron rate is
\begin{align}
    \Gamma_{\text{sph}}=T\times\mathcal{N}_{\text{tr}}(\mathcal{N}\mathcal{V})_{\text{rot}}v_3^9(x,y) g_3^6  e^{-v_3(x,y)\mathcal{C}_{\text{sph}}} \,.
\end{align}
These are the approximations we have used in Sec.~\ref{sec:SM-crossover}.

Finally, let us address the power counting for the determinant in Eq.~\eqref{eq:power-counting}.
Formally,
\begin{align}
\ln \frac{[\text{det}]_\text{sph}}{T^3} \sim \operatorname{Tr} \ln {\cal Q} \bigl(\hat\phi_3, \hat A_3\bigr),
\end{align}
where ${\cal Q} \bigl(\hat\phi_3, \hat A_3\bigr)$, depending on the sphaleron background $\bigl(\hat\phi_3, \hat A_3\bigr)$, is the second functional derivative of the 3D action $S_{\text{3D}}$ in Eq.\,\eqref{eq:S3d} or Eq.\,\eqref{eq:S3D_dimensionless_version} with the potential replaced by the tree-level one.
To obtain ${\cal Q} \bigl(\hat\phi_3, \hat A_3\bigr)$, one simply shifts the fields by $\hat{\phi}_3 \to \hat{\phi}_3 + \hat{\eta}_3$ and $\hat{A}_3^a \to \hat{A}_3^a + \hat{a}_{3,i}^a$ and collects the terms quadratic in $\bigl(\hat\phi_3, \hat A_3\bigr)$, where new $\hat\phi_3, \,\hat A_3$ are the sphaleron background fields and $\hat{\eta}_3, \,\hat{a}_{3,i}$ are the dimensionless fluctuating fields.
The explicit form of ${\cal Q}$ can be found in Eq. (40) in \cite{Carson:1989rf} or Eq. (12) in \cite{Carson:1990jm} and we present it here as well for completeness:
\begin{align}
    \delta^2 S_{\text{3D}} = & \, \int \mathrm{d}^3\xi \,\biggl\{ \frac{1}{2}\bigl(\hat D_i \hat a_{j}\bigr)_3^a\bigl(\hat D_i \hat a_{j}\bigr)_3^a + \epsilon^{abc}\hat F_{3,ij}^a \hat a_{3,i}^b \hat a_{3,j}^c + \frac{1}{4}\hat\phi_3^\dagger\hat\phi_3 \hat a_{3,i}^a \hat a_{3,i}^a \nonumber \\
    & + 2i \Bigl[ \hat\eta_3^\dagger \hat a_{3,i}^a \tau^a \bigl(\hat D_i \hat\phi\bigr)_3 - \bigl(\hat D_i \hat \phi\bigr)_3^\dagger \hat a_{3,i}^a \tau^a \hat \eta_3^a \Bigr] + \bigl(\hat D_i \hat \eta\bigr)_3^\dagger \bigl(\hat D_i \hat \eta\bigr)_3^\dagger \nonumber \\
    & + 
    y\hat\eta_3^\dagger\hat\eta_3 + \biggl(
    2x + \frac{1}{2}\biggr)\hat\phi_3^\dagger\hat\phi_3\hat\eta_3^\dagger\hat\eta_3
    + \biggl(
    x - \frac{1}{8}\biggr) \bigl(\hat\phi_3^\dagger\hat\eta_3 + \hat\eta_3^\dagger\hat\phi_3\bigr)^2 \biggr\} \nonumber \\
    = & \, \int \mathrm{d}^3\xi \, \Bigl( \hat{a}_{3,i}^a \quad \hat{\eta}_3^\dagger \quad \hat{\eta}_3^T \Bigr) \Bigl[ -\hat{D}^2 - M^2 \bigl(\hat\phi_3, \hat A_3\bigr) \Bigr] \left(
    \begin{array}{c}
        \hat{a}_{3,i}^a \\
        \hat{\eta} \\
        \hat{\eta}^*
    \end{array}
    \right) ,
\end{align}
where we have reformulated various terms into ${\cal Q} \bigl(\hat\phi_3, \hat A_3\bigr) = \hat{D}^2 + M^2 \bigl(\hat\phi_3, \hat A_3\bigr)$ in the last line.
The derivative independent part, $M^2 \bigl(\hat\phi_3, \hat A_3\bigr)$, hence scales like $M^2 \bigl(\hat\phi_3, \hat A_3\bigr) \sim \hat F_{3} \sim \hat\phi_3^\dagger\hat\phi_3 \sim \bigl(\hat D_i \hat\phi\bigr)_3 \sim g^{-2}$, using $\phi_3 = g_3 \hat \phi_3$, $F_3 = g_3^3 \hat{F}_3$ and $D_i = g_3^2 \hat{D}_i$ as well as the scaling of terms in Sec.\,\ref{sec:power_counting}.
In the dimensionless coordinates, the characteristic length scale of the sphaleron is $\mathrm{d}^3\xi = g_3^6 \, \mathrm{d}^3 x \sim g^{3}$.

In analogy to the calculation of the Coleman-Weinberg effective potential, the size of the fluctuation determinant can be estimated by
\begin{align}
    \ln \frac{[\text{det}]_\text{sph}}{T^3} & \sim \operatorname{Tr} \ln \Bigl[ \hat{D}^2 + M^2 \bigl(\hat\phi_3, \hat A_3\bigr) \Bigr] \nonumber \\
    & \sim \int \mathrm{d}^3\xi \int \frac{\mathrm{d}^3k}{(2\pi)^3} \ln \Bigl[ k^2 + M^2 \bigl(\hat\phi_3, \hat A_3\bigr) \Bigr]  \nonumber \\
    & = \int \mathrm{d}^3\xi \, \frac{1}{6\pi} M^3 \bigl(\hat\phi_3, \hat A_3\bigr) \,.
\end{align}
Together with our estimation of the scaling of $M^2 \bigl(\hat\phi_3, \hat A_3\bigr)$ and $\mathrm{d}^3\xi$, we then have
\begin{align}
    \ln \frac{[\text{det}]_\text{sph}}{T^3} \sim \pi^{-1} \,.
\end{align}
Hence the contribution of the determinant is suppressed by $\frac{g}{\pi}$ compared to the leading order contribution, as stated in Eq.~\eqref{eq:power-counting}.

\section{Temporal Gluon Effect to Sphaleron Rate via Dimensional Reduction}
\label{sec:temporal-gluon-effect}

For the Standard Model crossover, lattice simulations (see Fig.~\ref{fig:sphaleron_rate_SM_compare}) of \cite{DOnofrio:2014rug,Annala:2023jvr} are based on the dimensional reduction mapping of \cite{Kajantie:1995dw}, and do not include the effect from the temporal gluon sector, as pointed out in \cite{Brauner:2016fla} (and later included in \cite{Andersen:2017ika,Gorda:2018hvi,Niemi:2018asa,Croon:2020cgk,Schicho:2021gca,Niemi:2021qvp} as well as in automated {\tt DRalgo} package \cite{Ekstedt:2022bff,Fonseca:2020vke}). 
In the 3D EFT, this sector originates from the hot SU(3) sector of the Standard Model, and couples to the Higgs field via the portal operator 
\begin{align}
\omega_3 (\phi^\dagger \phi) C^{\alpha}_0 C^{\alpha}_0 \,,     
\end{align}
with the portal coupling
\begin{align}
\omega_3 = -\frac{2}{(4\pi)^2} g^2_s g^2_Y \,,
\end{align}
which is generated through a top quark loop, where $g_Y$ is the top quark Yukawa coupling and $g_s$ is SU(3) gauge coupling. 
Temporal gluon fields $C^\alpha_0$, with SU(3) adjoint index $\alpha = 1,\ldots, 8$ have the Debye screening mass
\begin{align}
m_D''^2 = T^2 g^2_s \biggl(1 + \frac{N_f}{3} \biggr) \,,    
\end{align}
with $N_f = 3$ fermion families. 
When (soft scale) temporal gluons are integrated out, they modify the Higgs thermal mass $\bar{\mu}^2_{h,3}$ via contribution 
\begin{align}
-\frac{8}{4\pi} \omega_3 m_D'' \,.
\end{align}
Parametrically this contribution scales as $\sim g^3_s g^2_y \sim \mathcal{O}(g^5)$, and was hence omitted in \cite{Kajantie:1995dw} which truncates to $\mathcal{O}(g^4)$. 
However, this contribution can nevertheless lead to a non-negligible effect due to large numerical values of $g_Y \sim g_s \sim 1$.  

\begin{figure}[t]
  \centering
  \includegraphics[width=0.5\textwidth]{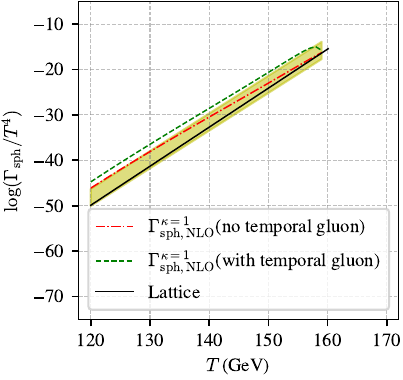}
  \caption{
 Similar to Fig.~\ref{fig:sphaleron_rate_SM_compare}, but including the effect of temporal gluons to the dimensional reduction mapping, when computing $\Gamma^{\kappa=1}_{\text{sph,NLO}}$ (in green), compared to not including it (in red). In both results renormalization scale is fixed to $\mu = 4\pi e^{-\gamma} T$.
 This effect increases the sphaleron rate, and accordingly the lattice result (in black) can be expected to increase, when this effect would be considered.
}
\label{fig:sphaleron_rate_SM_gluon-effect}
\end{figure}

In Fig.~\ref{fig:sphaleron_rate_SM_gluon-effect} we replot the Higgs phase sphaleron rate after SM crossover, depicting the effect of including the temporal gluon contribution, which modifies the $y(T)$ dimensional reduction mapping. 
We observe, that this contribution slightly increases the Higgs phase sphaleron rate, and emphasize, that since it is the dimensional reduction mapping that has changed, comparison to the lattice result is not fair. 
Indeed, we anticipate that the lattice result would increase accordingly, when the temporal gluon contribution is added. Similarly, this contribution would also result in a slight shift to the value of the SM pseudo-critical temperature.
%

\printbibliography

@article{Wu:2023mjb,
    author = "Wu, Yanda and Zhang, Wenxing and Ramsey-Musolf, Michael J.",
    title = "{Electroweak sphalerons, scalar multiplets, and symmetry breaking patterns}",
    eprint = "2307.02187",
    archivePrefix = "arXiv",
    primaryClass = "hep-ph",
    month = "7",
    year = "2023"
}

@article{Cao:2022ocg,
    author = "Cao, Qing-Hong and Hashino, Katsuya and Li, Xu-Xiang and Yue, Jiang-Hao",
    title = "{Multistep phase transition and gravitational wave from general Z2 scalar~extensions}",
    eprint = "2212.07756",
    archivePrefix = "arXiv",
    primaryClass = "hep-ph",
    doi = "10.1103/PhysRevD.111.095003",
    journal = "Phys. Rev. D",
    volume = "111",
    number = "9",
    pages = "095003",
    year = "2025"
}

@article{Funakubo:2009eg,
    author = "Funakubo, Koichi and Senaha, Eibun",
    title = "{Electroweak phase transition, critical bubbles and sphaleron decoupling condition in the MSSM}",
    eprint = "0905.2022",
    archivePrefix = "arXiv",
    primaryClass = "hep-ph",
    doi = "10.1103/PhysRevD.79.115024",
    journal = "Phys. Rev. D",
    volume = "79",
    pages = "115024",
    year = "2009"
}

@article{Arnold:1987mh,
	author = {Arnold, Peter Brockway and McLerran, Larry D.},
	doi = {10.1103/PhysRevD.36.581},
	journal = {Phys. Rev. D},
	pages = {581},
	reportnumber = {FERMILAB-PUB-87-034-T},
	title = {{Sphalerons, Small Fluctuations and Baryon Number Violation in Electroweak Theory}},
	volume = {36},
	year = {1987}}

@article{Carson:1990jm,
	author = {Carson, Larry and Li, Xu and McLerran, Larry D. and Wang, Rong-Tai},
	doi = {10.1103/PhysRevD.42.2127},
	journal = {Phys. Rev. D},
	pages = {2127--2143},
	reportnumber = {TPI-MINN-90-13-T},
	title = {{Exact Computation of the Small Fluctuation Determinant Around a Sphaleron}},
	volume = {42},
	year = {1990}}

@article{Kuzmin:1985mm,
      author         = "Kuzmin, V. A. and Rubakov, V. A. and Shaposhnikov, M. E.",
      title          = "{On the Anomalous Electroweak Baryon Number
                        Nonconservation in the Early Universe}",
      journal        = "Phys. Lett.",
      volume         = "155B",
      year           = "1985",
      pages          = "36",
      doi            = "10.1016/0370-2693(85)91028-7",
      reportNumber   = "IC/85/8",
      SLACcitation   = "%%CITATION = PHLTA,155B,36;%%"
}

@article{Morrissey:2012db,
    author = "Morrissey, David E. and Ramsey-Musolf, Michael J.",
    title = "{Electroweak baryogenesis}",
    eprint = "1206.2942",
    archivePrefix = "arXiv",
    primaryClass = "hep-ph",
    reportNumber = "NPAC-12-08",
    doi = "10.1088/1367-2630/14/12/125003",
    journal = "New J. Phys.",
    volume = "14",
    pages = "125003",
    year = "2012"
}

@article{Bodeker:2020ghk,
    author = "Bodeker, Dietrich and Buchmuller, Wilfried",
    title = "{Baryogenesis from the weak scale to the grand unification scale}",
    eprint = "2009.07294",
    archivePrefix = "arXiv",
    primaryClass = "hep-ph",
    reportNumber = "DESY 20-141, DESY-20-141",
    doi = "10.1103/RevModPhys.93.035004",
    journal = "Rev. Mod. Phys.",
    volume = "93",
    number = "3",
    pages = "035004",
    year = "2021"
}

@article{Sakharov:1967dj,
    author = "Sakharov, A. D.",
    title = "{Violation of CP Invariance, C asymmetry, and baryon asymmetry of the universe}",
    doi = "10.1070/PU1991v034n05ABEH002497",
    journal = "Pisma Zh. Eksp. Teor. Fiz.",
    volume = "5",
    pages = "32--35",
    year = "1967"
}

@article{Kajantie:1996mn,
      author         = "Kajantie, K. and Laine, M. and Rummukainen, K. and
                        Shaposhnikov, Mikhail E.",
      title          = "{Is there a hot electroweak phase transition at m(H)
                        larger or equal to m(W)?}",
      journal        = "Phys. Rev. Lett.",
      volume         = "77",
      year           = "1996",
      pages          = "2887-2890",
      doi            = "10.1103/PhysRevLett.77.2887",
      eprint         = "hep-ph/9605288",
      archivePrefix  = "arXiv",
      primaryClass   = "hep-ph",
      reportNumber   = "CERN-TH-96-126, HD-THEP-96-15, IUHET-333",
      SLACcitation   = "%%CITATION = HEP-PH/9605288;%%"
}

@article{Csikor:1998eu,
      author         = "Csikor, F. and Fodor, Z. and Heitger, J.",
      title          = "{Endpoint of the hot electroweak phase transition}",
      journal        = "Phys. Rev. Lett.",
      volume         = "82",
      year           = "1999",
      pages          = "21-24",
      doi            = "10.1103/PhysRevLett.82.21",
      eprint         = "hep-ph/9809291",
      archivePrefix  = "arXiv",
      primaryClass   = "hep-ph",
      reportNumber   = "ITP-BUDAPEST-541, KEK-TH-580, KEK-PREPRINT-98-160,
                        MS-TPI-98-16",
      SLACcitation   = "%%CITATION = HEP-PH/9809291;%%"
}

@article{Bell:1969ts,
    author = "Bell, J. S. and Jackiw, R.",
    title = "{A PCAC puzzle: $\pi^0 \to \gamma \gamma$ in the $\sigma$ model}",
    doi = "10.1007/BF02823296",
    journal = "Nuovo Cim. A",
    volume = "60",
    pages = "47--61",
    year = "1969"
}

@article{Adler:1969gk,
    author = "Adler, Stephen L.",
    title = "{Axial vector vertex in spinor electrodynamics}",
    doi = "10.1103/PhysRev.177.2426",
    journal = "Phys. Rev.",
    volume = "177",
    pages = "2426--2438",
    year = "1969"
}

@article{Klinkhamer:1984di,
    author = "Klinkhamer, Frans R. and Manton, N. S.",
    title = "{A Saddle Point Solution in the Weinberg-Salam Theory}",
    reportNumber = "NSF-ITP-84-57",
    doi = "10.1103/PhysRevD.30.2212",
    journal = "Phys. Rev. D",
    volume = "30",
    pages = "2212",
    year = "1984"
}

@article{Brauner:2012gu,
    author = "Brauner, Tomas and Taanila, Olli and Tranberg, Anders and Vuorinen, Aleksi",
    title = "{Computing the temperature dependence of effective CP violation in the standard model}",
    eprint = "1208.5609",
    archivePrefix = "arXiv",
    primaryClass = "hep-ph",
    reportNumber = "BI-TP-2012-38",
    doi = "10.1007/JHEP11(2012)076",
    journal = "JHEP",
    volume = "11",
    pages = "076",
    year = "2012"
}

@article{Brauner:2011vb,
    author = "Brauner, Tomas and Taanila, Olli and Tranberg, Anders and Vuorinen, Aleksi",
    title = "{Temperature Dependence of Standard Model CP Violation}",
    eprint = "1110.6818",
    archivePrefix = "arXiv",
    primaryClass = "hep-ph",
    doi = "10.1103/PhysRevLett.108.041601",
    journal = "Phys. Rev. Lett.",
    volume = "108",
    pages = "041601",
    year = "2012"
}

@article{Shaposhnikov:1987pf,
    author = "Shaposhnikov, M. E.",
    title = "{Structure of the High Temperature Gauge Ground State and Electroweak Production of the Baryon Asymmetry}",
    reportNumber = "NBI-HE-87-36",
    doi = "10.1016/0550-3213(88)90373-2",
    journal = "Nucl. Phys. B",
    volume = "299",
    pages = "797--817",
    year = "1988"
}

@article{Farrar:1993sp,
    author = "Farrar, Glennys R. and Shaposhnikov, M. E.",
    title = "{Baryon asymmetry of the universe in the minimal Standard Model}",
    eprint = "hep-ph/9305274",
    archivePrefix = "arXiv",
    reportNumber = "RU-93-10, CERN-TH-6729-92",
    doi = "10.1103/PhysRevLett.70.2833",
    journal = "Phys. Rev. Lett.",
    volume = "70",
    pages = "2833--2836",
    year = "1993",
    note = "[Erratum: Phys.Rev.Lett. 71, 210 (1993)]"
}

@article{Farrar:1993hn,
    author = "Farrar, Glennys R. and Shaposhnikov, M. E.",
    title = "{Baryon asymmetry of the universe in the standard electroweak theory}",
    eprint = "hep-ph/9305275",
    archivePrefix = "arXiv",
    reportNumber = "CERN-TH-6732-93, CERN-TH-6734-93, RU-93-11",
    doi = "10.1103/PhysRevD.50.774",
    journal = "Phys. Rev. D",
    volume = "50",
    pages = "774",
    year = "1994"
}

@article{Gavela:1994dt,
    author = "Gavela, M. B. and Hernandez, P. and Orloff, J. and Pene, O. and Quimbay, C.",
    title = "{Standard model CP violation and baryon asymmetry. Part 2: Finite temperature}",
    eprint = "hep-ph/9406289",
    archivePrefix = "arXiv",
    reportNumber = "CERN-TH-7263-94, LPTHE-ORSAY-94-49, HUTP-94-A015, HD-THEP-94-20, FTUAM-94-14, NSF-ITP-94-65",
    doi = "10.1016/0550-3213(94)00410-2",
    journal = "Nucl. Phys. B",
    volume = "430",
    pages = "382--426",
    year = "1994"
}

@article{Huber:2022ndk,
    author = "Huber, S. J. and Mimasu, K. and No, J. M.",
    title = "{Baryogenesis from transitional CP violation in the early Universe}",
    eprint = "2208.10512",
    archivePrefix = "arXiv",
    primaryClass = "hep-ph",
    doi = "10.1103/PhysRevD.107.075042",
    journal = "Phys. Rev. D",
    volume = "107",
    number = "7",
    pages = "075042",
    year = "2023"
}

@article{Cline:2017jvp,
    author = "Cline, James M.",
    editor = "Auge, Etienne and Dumarchez, Jacques and Tran Thanh Van, Jean",
    title = "{Is electroweak baryogenesis dead?}",
    eprint = "1704.08911",
    archivePrefix = "arXiv",
    primaryClass = "hep-ph",
    reportNumber = "CERN-TH-2017-096",
    doi = "10.1098/rsta.2017.0116",
    pages = "339--348",
    year = "2017"
}

@article{ACME:2018yjb,
    author = "Andreev, V. and others",
    collaboration = "ACME",
    title = "{Improved limit on the electric dipole moment of the electron}",
    doi = "10.1038/s41586-018-0599-8",
    journal = "Nature",
    volume = "562",
    number = "7727",
    pages = "355--360",
    year = "2018"
}

@article{Keus:2017ioh,
    author = "Keus, Venus and Koivunen, Niko and Tuominen, Kimmo",
    title = "{Singlet scalar and 2HDM extensions of the Standard Model: CP-violation and constraints from $(g-2)_\mu$ and $e$EDM}",
    eprint = "1712.09613",
    archivePrefix = "arXiv",
    primaryClass = "hep-ph",
    doi = "10.1007/JHEP09(2018)059",
    journal = "JHEP",
    volume = "09",
    pages = "059",
    year = "2018"
}

@article{Keus:2019szx,
    author = "Keus, Venus",
    title = "{Dark CP-violation through the $Z$-portal}",
    eprint = "1909.09234",
    archivePrefix = "arXiv",
    primaryClass = "hep-ph",
    reportNumber = "HIP-2019-28/TH",
    doi = "10.1103/PhysRevD.101.073007",
    journal = "Phys. Rev. D",
    volume = "101",
    number = "7",
    pages = "073007",
    year = "2020"
}

@article{Alanne:2016wtx,
    author = "Alanne, Tommi and Kainulainen, Kimmo and Tuominen, Kimmo and Vaskonen, Ville",
    title = "{Baryogenesis in the two doublet and inert singlet extension of the Standard Model}",
    eprint = "1607.03303",
    archivePrefix = "arXiv",
    primaryClass = "hep-ph",
    reportNumber = "HIP-2016-22-TH, CP3-ORIGINS-2016-031",
    doi = "10.1088/1475-7516/2016/08/057",
    journal = "JCAP",
    volume = "08",
    pages = "057",
    year = "2016"
}

@article{Ramsey-Musolf:2019lsf,
    author = "Ramsey-Musolf, Michael J.",
    title = "{The electroweak phase transition: a collider target}",
    eprint = "1912.07189",
    archivePrefix = "arXiv",
    primaryClass = "hep-ph",
    reportNumber = "ACFI-T19-14",
    doi = "10.1007/JHEP09(2020)179",
    journal = "JHEP",
    volume = "09",
    pages = "179",
    year = "2020"
}

@article{ILC:2013jhg,
    editor = "Baer, Howard and others",
    collaboration = "ILC",
    title = "{The International Linear Collider Technical Design Report - Volume 2: Physics}",
    eprint = "1306.6352",
    archivePrefix = "arXiv",
    primaryClass = "hep-ph",
    reportNumber = "ILC-REPORT-2013-040, ANL-HEP-TR-13-20, BNL-100603-2013-IR, IRFU-13-59, CERN-ATS-2013-037, COCKCROFT-13-10, CLNS-13-2085, DESY-13-062, FERMILAB-TM-2554, IHEP-AC-ILC-2013-001, INFN-13-04-LNF, JAI-2013-001, JINR-E9-2013-35, JLAB-R-2013-01, KEK-REPORT-2013-1, KNU-CHEP-ILC-2013-1, LLNL-TR-635539, SLAC-R-1004, ILC-HIGRADE-REPORT-2013-003",
    month = "6",
    year = "2013"
}

@article{CEPCStudyGroup:2018ghi,
    author = "Dong, Mingyi and others",
    editor = "Guimar\~aes da Costa, Jo\~ao Barreiro and others",
    collaboration = "CEPC Study Group",
    title = "{CEPC Conceptual Design Report: Volume 2 - Physics \& Detector}",
    eprint = "1811.10545",
    archivePrefix = "arXiv",
    primaryClass = "hep-ex",
    reportNumber = "IHEP-CEPC-DR-2018-02, IHEP-EP-2018-01, IHEP-TH-2018-01",
    month = "11",
    year = "2018"
}

@article{FCC:2018byv,
    author = "Abada, A. and others",
    collaboration = "FCC",
    title = "{FCC Physics Opportunities}: {Future Circular Collider Conceptual Design Report Volume 1}",
    reportNumber = "CERN-ACC-2018-0056",
    doi = "10.1140/epjc/s10052-019-6904-3",
    journal = "Eur. Phys. J. C",
    volume = "79",
    number = "6",
    pages = "474",
    year = "2019"
}

@article{CLIC:2018fvx,
    author = "de Blas, J. and others",
    collaboration = "CLIC",
    title = "{The CLIC Potential for New Physics}",
    eprint = "1812.02093",
    archivePrefix = "arXiv",
    primaryClass = "hep-ph",
    reportNumber = "CERN-TH-2018-267, CERN-2018-009-M, FERMILAB-TM-2795",
    doi = "10.23731/CYRM-2018-003",
    volume = "3/2018",
    month = "12",
    year = "2018"
}

@article{Apollinari:2017lan,
    editor = {Apollinari, G. and B\'ejar Alonso, I. and Br\"uning, O. and Fessia, P. and Lamont, M. and Rossi, L. and Tavian, L.},
    title = "{High-Luminosity Large Hadron Collider (HL-LHC)}: {Technical Design Report V. 0.1}",
    reportNumber = "CERN-2017-007-M",
    doi = "10.23731/CYRM-2017-004",
    volume = "4/2017",
    year = "2017"
}

@article{LISACosmologyWorkingGroup:2022jok,
    author = "Auclair, Pierre and others",
    collaboration = "LISA Cosmology Working Group",
    title = "{Cosmology with the Laser Interferometer Space Antenna}",
    eprint = "2204.05434",
    archivePrefix = "arXiv",
    primaryClass = "astro-ph.CO",
    reportNumber = "LISA CosWG-22-03, FERMILAB-PUB-22-349-SCD",
    doi = "10.1007/s41114-023-00045-2",
    journal = "Living Rev. Rel.",
    volume = "26",
    number = "1",
    pages = "5",
    year = "2023"
}

@article{Audley:2017drz,
    author = "Amaro-Seoane, Pau and others",
    collaboration = "LISA",
    title = "{Laser Interferometer Space Antenna}",
    eprint = "1702.00786",
    archivePrefix = "arXiv",
    primaryClass = "astro-ph.IM",
    month = "2",
    year = "2017"
}

@article{Fradkin:1978dv,
    author = "Fradkin, Eduardo H. and Shenker, Stephen H.",
    title = "{Phase Diagrams of Lattice Gauge Theories with Higgs Fields}",
    reportNumber = "SLAC-PUB-2238",
    doi = "10.1103/PhysRevD.19.3682",
    journal = "Phys. Rev. D",
    volume = "19",
    pages = "3682--3697",
    year = "1979"
}

@article{Elitzur:1975im,
    author = "Elitzur, S.",
    title = "{Impossibility of Spontaneously Breaking Local Symmetries}",
    doi = "10.1103/PhysRevD.12.3978",
    journal = "Phys. Rev. D",
    volume = "12",
    pages = "3978--3982",
    year = "1975"
}

@article{Manton:1983nd,
    author = "Manton, N. S.",
    title = "{Topology in the Weinberg-Salam Theory}",
    reportNumber = "NSF-ITP-83-64",
    doi = "10.1103/PhysRevD.28.2019",
    journal = "Phys. Rev. D",
    volume = "28",
    pages = "2019",
    year = "1983"
}

@article{Klinkhamer:1993hb,
    author = "Klinkhamer, Frans R.",
    title = "{Construction of a new electroweak sphaleron}",
    eprint = "hep-ph/9306295",
    archivePrefix = "arXiv",
    reportNumber = "NIKHEF-H-93-09",
    doi = "10.1016/0550-3213(93)90437-T",
    journal = "Nucl. Phys. B",
    volume = "410",
    pages = "343--354",
    year = "1993"
}

@article{tHooft:1976rip,
    author = "'t Hooft, Gerard",
    editor = "Shifman, Mikhail A.",
    title = "{Symmetry Breaking Through Bell-Jackiw Anomalies}",
    reportNumber = "PRINT-76-0254 (HARVARD)",
    doi = "10.1103/PhysRevLett.37.8",
    journal = "Phys. Rev. Lett.",
    volume = "37",
    pages = "8--11",
    year = "1976"
}

@article{tHooft:1976snw,
    author = "'t Hooft, Gerard",
    editor = "Shifman, Mikhail A.",
    title = "{Computation of the Quantum Effects Due to a Four-Dimensional Pseudoparticle}",
    reportNumber = "PRINT-76-0551 (HARVARD)",
    doi = "10.1103/PhysRevD.14.3432",
    journal = "Phys. Rev. D",
    volume = "14",
    pages = "3432--3450",
    year = "1976",
    note = "[Erratum: Phys.Rev.D 18, 2199 (1978)]"
}

@article{Arnold:1987zg,
    author = "Arnold, Peter Brockway and McLerran, Larry D.",
    title = "{The Sphaleron Strikes Back}",
    reportNumber = "FERMILAB-PUB-87-120-T",
    doi = "10.1103/PhysRevD.37.1020",
    journal = "Phys. Rev. D",
    volume = "37",
    pages = "1020",
    year = "1988"
}

@article{Carson:1989rf,
    author = "Carson, Larry and McLerran, Larry D.",
    title = "{Approximate Computation of the Small Fluctuation Determinant Around a Sphaleron}",
    reportNumber = "TPI-MINN-89-19T",
    doi = "10.1103/PhysRevD.41.647",
    journal = "Phys. Rev. D",
    volume = "41",
    pages = "647",
    year = "1990"
}

@article{Baacke:1993jr,
    author = "Baacke, J. and Junker, S.",
    title = "{Quantum corrections to the electroweak sphaleron transition}",
    eprint = "hep-ph/9306307",
    archivePrefix = "arXiv",
    reportNumber = "DO-TH-93-15",
    doi = "10.1142/S0217732393003251",
    journal = "Mod. Phys. Lett. A",
    volume = "8",
    pages = "2869--2874",
    year = "1993"
}

@article{Baacke:1993aj,
    author = "Baacke, J. and Junker, S.",
    title = "{Quantum fluctuations around the electroweak sphaleron}",
    eprint = "hep-ph/9308310",
    archivePrefix = "arXiv",
    reportNumber = "DO-TH-93-19",
    doi = "10.1103/PhysRevD.49.2055",
    journal = "Phys. Rev. D",
    volume = "49",
    pages = "2055--2073",
    year = "1994"
}

@article{Baacke:1994ix,
    author = "Baacke, J. and Junker, S.",
    title = "{Quantum fluctuations of the electroweak sphaleron: Erratum and addendum}",
    eprint = "hep-th/9402078",
    archivePrefix = "arXiv",
    reportNumber = "DO-TH-93-19-EA",
    doi = "10.1103/PhysRevD.50.4227",
    journal = "Phys. Rev. D",
    volume = "50",
    pages = "4227--4228",
    year = "1994"
}

@article{Arnold:1992rz,
    author = "Arnold, Peter Brockway and Espinosa, Olivier",
    title = "{The Effective potential and first order phase transitions: Beyond leading-order}",
    eprint = "hep-ph/9212235",
    archivePrefix = "arXiv",
    reportNumber = "UW-PT-92-18, USM-TH-60",
    doi = "10.1103/PhysRevD.47.3546",
    journal = "Phys. Rev. D",
    volume = "47",
    pages = "3546",
    year = "1993",
    note = "[Erratum: Phys.Rev.D 50, 6662 (1994)]"
}

@article{Kajantie:1995dw,
    author = "Kajantie, K. and Laine, M. and Rummukainen, K. and Shaposhnikov, Mikhail E.",
    title = "{Generic rules for high temperature dimensional reduction and their application to the standard model}",
    eprint = "hep-ph/9508379",
    archivePrefix = "arXiv",
    reportNumber = "CERN-TH-95-226, HU-TFT-95-50, IUHET-312",
    doi = "10.1016/0550-3213(95)00549-8",
    journal = "Nucl. Phys. B",
    volume = "458",
    pages = "90--136",
    year = "1996"
}

@article{Farakos:1994kx,
      author         = "Farakos, K. and Kajantie, K. and Rummukainen, K. and
                        Shaposhnikov, Mikhail E.",
      title          = "{3-D physics and the electroweak phase transition:
                        Perturbation theory}",
      journal        = "Nucl. Phys.",
      volume         = "B425",
      year           = "1994",
      pages          = "67-109",
      doi            = "10.1016/0550-3213(94)90173-2",
      eprint         = "hep-ph/9404201",
      archivePrefix  = "arXiv",
      primaryClass   = "hep-ph",
      reportNumber   = "CERN-TH-6973-94, IUHET-273",
      SLACcitation   = "%%CITATION = HEP-PH/9404201;%%"
}

@article{Braaten:1995cm,
      author         = "Braaten, Eric and Nieto, Agustin",
      title          = "{Effective field theory approach to high temperature
                        thermodynamics}",
      journal        = "Phys. Rev.",
      volume         = "D51",
      year           = "1995",
      pages          = "6990-7006",
      doi            = "10.1103/PhysRevD.51.6990",
      eprint         = "hep-ph/9501375",
      archivePrefix  = "arXiv",
      primaryClass   = "hep-ph",
      reportNumber   = "NUHEP-TH-95-2",
      SLACcitation   = "%%CITATION = HEP-PH/9501375;%%"
}

@article{Ginsparg:1980ef,
      author         = "Ginsparg, Paul H.",
      title          = "{First Order and Second Order Phase Transitions in Gauge
                        Theories at Finite Temperature}",
      journal        = "Nucl. Phys.",
      volume         = "B170",
      year           = "1980",
      pages          = "388-408",
      doi            = "10.1016/0550-3213(80)90418-6",
      reportNumber   = "SACLAY-DPh-T 80/27",
      SLACcitation   = "%%CITATION = NUPHA,B170,388;%%"
}

@article{Appelquist:1981vg,
      author         = "Appelquist, Thomas and Pisarski, Robert D.",
      title          = "{High-Temperature Yang-Mills Theories and
                        Three-Dimensional Quantum Chromodynamics}",
      journal        = "Phys. Rev.",
      volume         = "D23",
      year           = "1981",
      pages          = "2305",
      doi            = "10.1103/PhysRevD.23.2305",
      reportNumber   = "Print-81-0020 (YALE), YTP-81-01, COO-3075-203",
      SLACcitation   = "%%CITATION = PHRVA,D23,2305;%%"
}

@article{Farakos:1994xh,
      author         = "Farakos, K. and Kajantie, K. and Rummukainen, K. and
                        Shaposhnikov, Mikhail E.",
      title          = "{3-d physics and the electroweak phase transition: A
                        Framework for lattice Monte Carlo analysis}",
      journal        = "Nucl. Phys.",
      volume         = "B442",
      year           = "1995",
      pages          = "317-363",
      doi            = "10.1016/0550-3213(95)80129-4",
      eprint         = "hep-lat/9412091",
      archivePrefix  = "arXiv",
      primaryClass   = "hep-lat",
      reportNumber   = "CERN-TH-7220-94, HU-TFT-94-50, IUHET-290",
      SLACcitation   = "%%CITATION = HEP-LAT/9412091;%%"
}

@article{Kajantie:1995kf,
      author         = "Kajantie, K. and Laine, M. and Rummukainen, K. and
                        Shaposhnikov, Mikhail E.",
      title          = "{The Electroweak phase transition: A Nonperturbative
                        analysis}",
      journal        = "Nucl. Phys.",
      volume         = "B466",
      year           = "1996",
      pages          = "189-258",
      doi            = "10.1016/0550-3213(96)00052-1",
      eprint         = "hep-lat/9510020",
      archivePrefix  = "arXiv",
      primaryClass   = "hep-lat",
      reportNumber   = "CERN-TH-95-263, HD-THEP-95-44, HU-TFT-95-57, IUHET-318",
      SLACcitation   = "%%CITATION = HEP-LAT/9510020;%%"
}

@article{Matsubara:1955ws,
    author = "Matsubara, Takeo",
    title = "{A New approach to quantum statistical mechanics}",
    doi = "10.1143/PTP.14.351",
    journal = "Prog. Theor. Phys.",
    volume = "14",
    pages = "351--378",
    year = "1955"
}

@article{Hirvonen:2022jba,
    author = "Hirvonen, J.",
    title = "{Intuitive method for constructing effective field theories}",
    eprint = "2205.02687",
    archivePrefix = "arXiv",
    primaryClass = "hep-ph",
    reportNumber = "HIP-2022-6/TH",
    month = "5",
    year = "2022"
}

@article{Gould:2023ovu,
    author = "Gould, Oliver and Tenkanen, Tuomas V. I.",
    title = "{Perturbative effective field theory expansions for cosmological phase transitions}",
    eprint = "2309.01672",
    archivePrefix = "arXiv",
    primaryClass = "hep-ph",
    reportNumber = "NORDITA 2023-037",
    doi = "10.1007/JHEP01(2024)048",
    journal = "JHEP",
    volume = "01",
    pages = "048",
    year = "2024"
}

@article{Ekstedt:2024etx,
    author = "Ekstedt, Andreas and Schicho, Philipp and Tenkanen, Tuomas V. I.",
    title = "{Cosmological phase transitions at three loops: The final verdict on perturbation theory}",
    eprint = "2405.18349",
    archivePrefix = "arXiv",
    primaryClass = "hep-ph",
    reportNumber = "HIP-2024-15/TH",
    doi = "10.1103/PhysRevD.110.096006",
    journal = "Phys. Rev. D",
    volume = "110",
    number = "9",
    pages = "096006",
    year = "2024"
}

@article{Linde:1980ts,
      author         = "Linde, Andrei D.",
      title          = "{Infrared Problem in Thermodynamics of the Yang-Mills
                        Gas}",
      journal        = "Phys. Lett.",
      volume         = "96B",
      year           = "1980",
      pages          = "289-292",
      doi            = "10.1016/0370-2693(80)90769-8",
      reportNumber   = "LEBEDEV-80-106",
      SLACcitation   = "%%CITATION = PHLTA,96B,289;%%"
}

@article{Shaposhnikov:1993jh,
    author = "Shaposhnikov, Mikhail E.",
    title = "{On nonperturbative effects at the high temperature electroweak phase transition}",
    eprint = "hep-ph/9306296",
    archivePrefix = "arXiv",
    reportNumber = "CERN-TH-6918-93",
    doi = "10.1016/0370-2693(93)90666-6",
    journal = "Phys. Lett. B",
    volume = "316",
    pages = "112--120",
    year = "1993"
}

@article{Kajantie:1993ag,
    author = "Kajantie, K. and Rummukainen, K. and Shaposhnikov, Mikhail E.",
    title = "{A Lattice Monte Carlo study of the hot electroweak phase transition}",
    eprint = "hep-ph/9305345",
    archivePrefix = "arXiv",
    reportNumber = "CERN-TH-6901-93",
    doi = "10.1016/0550-3213(93)90062-T",
    journal = "Nucl. Phys. B",
    volume = "407",
    pages = "356--372",
    year = "1993"
}

@article{Gross:1980br,
    author = "Gross, David J. and Pisarski, Robert D. and Yaffe, Laurence G.",
    title = "{QCD and Instantons at Finite Temperature}",
    reportNumber = "PRINT-80-0538 (PRINCETON)",
    doi = "10.1103/RevModPhys.53.43",
    journal = "Rev. Mod. Phys.",
    volume = "53",
    pages = "43",
    year = "1981"
}

@article{Farakos:1994kj,
    author = "Farakos, K. and Kajantie, K. and Rummukainen, K. and Shaposhnikov, Mikhail E.",
    title = "{The Electroweak phase transition at m(H) approximately = m(W)}",
    eprint = "hep-ph/9405234",
    archivePrefix = "arXiv",
    reportNumber = "CERN-TH-7244-94, IUHET-279",
    doi = "10.1016/0370-2693(94)90563-0",
    journal = "Phys. Lett. B",
    volume = "336",
    pages = "494--501",
    year = "1994"
}

@article{Kajantie:1997ky,
    author = "Kajantie, K. and Laine, M. and Rummukainen, K. and Shaposhnikov, Mikhail E.",
    title = "{High temperature dimensional reduction and parity violation}",
    eprint = "hep-ph/9710538",
    archivePrefix = "arXiv",
    reportNumber = "CERN-TH-97-298, NORDITA-97-78-P",
    doi = "10.1016/S0370-2693(97)01584-0",
    journal = "Phys. Lett. B",
    volume = "423",
    pages = "137--144",
    year = "1998"
}

@article{Rummukainen:1998as,
    author = "Rummukainen, K. and Tsypin, M. and Kajantie, K. and Laine, M. and Shaposhnikov, Mikhail E.",
    title = "{The Universality class of the electroweak theory}",
    eprint = "hep-lat/9805013",
    archivePrefix = "arXiv",
    reportNumber = "CERN-TH-98-08, NORDITA-98-30-HE",
    doi = "10.1016/S0550-3213(98)00494-5",
    journal = "Nucl. Phys. B",
    volume = "532",
    pages = "283--314",
    year = "1998"
}

@article{Laine:1998wi,
    author = "Laine, M. and Rummukainen, K.",
    title = "{Higgs sector CP violation at the electroweak phase transition}",
    eprint = "hep-ph/9811369",
    archivePrefix = "arXiv",
    reportNumber = "CERN-TH-98-364, NORDITA-98-69-HE",
    doi = "10.1016/S0550-3213(99)00077-2",
    journal = "Nucl. Phys. B",
    volume = "545",
    pages = "141--182",
    year = "1999"
}

@article{Landsman:1989be,
    author = "Landsman, N. P.",
    title = "{Limitations to Dimensional Reduction at High Temperature}",
    reportNumber = "ITFA-89-101",
    doi = "10.1016/0550-3213(89)90424-0",
    journal = "Nucl. Phys. B",
    volume = "322",
    pages = "498--530",
    year = "1989"
}

@article{Nadkarni:1982kb,
    author = "Nadkarni, Sudhir",
    title = "{Dimensional Reduction in Hot QCD}",
    reportNumber = "YTP-82-21-REV, YTP-82-21",
    doi = "10.1103/PhysRevD.27.917",
    journal = "Phys. Rev. D",
    volume = "27",
    pages = "917",
    year = "1983"
}

@article{Grigoriev:1989ub,
    author = "Grigoriev, Dmitri Yu. and Rubakov, V. A. and Shaposhnikov, M. E.",
    title = "{TOPOLOGICAL TRANSITIONS AT FINITE TEMPERATURES: A REAL TIME NUMERICAL APPROACH}",
    doi = "10.1016/0550-3213(89)90553-1",
    journal = "Nucl. Phys. B",
    volume = "326",
    pages = "737--757",
    year = "1989"
}

@article{Grigoriev:1988bd,
    author = "Grigoriev, Dmitri Yu. and Rubakov, V. A.",
    title = "{Soliton Pair Creation at Finite Temperatures. Numerical Study in (1+1)-dimensions}",
    doi = "10.1016/0550-3213(88)90466-X",
    journal = "Nucl. Phys. B",
    volume = "299",
    pages = "67--78",
    year = "1988"
}

@article{Grigoriev:1989je,
    author = "Grigoriev, Dmitri Yu. and Rubakov, V. A. and Shaposhnikov, M. E.",
    title = "{Sphaleron Transitions at Finite Temperatures: Numerical Study in (1+1)-dimensions}",
    doi = "10.1016/0370-2693(89)91390-7",
    journal = "Phys. Lett. B",
    volume = "216",
    pages = "172--176",
    year = "1989"
}

@article{Ambjorn:1990wn,
    author = "Ambjorn, Jan and Askgaard, T. and Porter, H. and Shaposhnikov, M. E.",
    title = "{Lattice Simulations of Electroweak Sphaleron Transitions in Real Time}",
    reportNumber = "NBI-HE-90-18",
    doi = "10.1016/0370-2693(90)90350-F",
    journal = "Phys. Lett. B",
    volume = "244",
    pages = "479--487",
    year = "1990"
}

@article{Ambjorn:1990pu,
    author = "Ambjorn, Jan and Askgaard, T. and Porter, H. and Shaposhnikov, M. E.",
    title = "{Sphaleron transitions and baryon asymmetry: A Numerical real time analysis}",
    reportNumber = "NBI-HE-90-48",
    doi = "10.1016/0550-3213(91)90341-T",
    journal = "Nucl. Phys. B",
    volume = "353",
    pages = "346--378",
    year = "1991"
}

@article{Ambjorn:1995xm,
    author = "Ambjorn, Jan and Krasnitz, A.",
    title = "{The Classical sphaleron transition rate exists and is equal to 1.1 (alpha(w) T)**4}",
    eprint = "hep-ph/9508202",
    archivePrefix = "arXiv",
    reportNumber = "NBI-HE-95-23",
    doi = "10.1016/0370-2693(95)01157-L",
    journal = "Phys. Lett. B",
    volume = "362",
    pages = "97--104",
    year = "1995"
}

@article{Ambjorn:1997jz,
    author = "Ambjorn, Jan and Krasnitz, A.",
    title = "{Improved determination of the classical sphaleron transition rate}",
    eprint = "hep-ph/9705380",
    archivePrefix = "arXiv",
    reportNumber = "NBI-HE-97-18",
    doi = "10.1016/S0550-3213(97)00524-5",
    journal = "Nucl. Phys. B",
    volume = "506",
    pages = "387--403",
    year = "1997"
}

@article{Tang:1996qx,
    author = "Tang, Wai-Hung and Smit, Jan",
    title = "{Chern-Simons diffusion rate near the electroweak phase transition for m(H) approximates m(W)}",
    eprint = "hep-lat/9605016",
    archivePrefix = "arXiv",
    reportNumber = "ITFA-96-11",
    doi = "10.1016/S0550-3213(96)00481-6",
    journal = "Nucl. Phys. B",
    volume = "482",
    pages = "265--285",
    year = "1996"
}

@article{Arnold:1996dy,
    author = "Arnold, Peter Brockway and Son, Dam and Yaffe, Laurence G.",
    title = "{The Hot baryon violation rate is O (alpha-w**5 T**4)}",
    eprint = "hep-ph/9609481",
    archivePrefix = "arXiv",
    reportNumber = "UW-PT-96-19",
    doi = "10.1103/PhysRevD.55.6264",
    journal = "Phys. Rev. D",
    volume = "55",
    pages = "6264--6273",
    year = "1997"
}

@article{Bodeker:1995pp,
    author = "Bodeker, Dietrich and McLerran, Larry D. and Smilga, Andrei V.",
    title = "{Really computing nonperturbative real time correlation functions}",
    eprint = "hep-th/9504123",
    archivePrefix = "arXiv",
    reportNumber = "TPI-MINN-95-08-T, NUC-MINN-95-11-T, HEP-UMN-TH-1337",
    doi = "10.1103/PhysRevD.52.4675",
    journal = "Phys. Rev. D",
    volume = "52",
    pages = "4675--4690",
    year = "1995"
}

@article{Moore:1997sn,
    author = "Moore, Guy D. and Hu, Chao-ran and Muller, Berndt",
    title = "{Chern-Simons number diffusion with hard thermal loops}",
    eprint = "hep-ph/9710436",
    archivePrefix = "arXiv",
    reportNumber = "PUPT-1713, MCGILL-97-24, DUKE-TH-97-154",
    doi = "10.1103/PhysRevD.58.045001",
    journal = "Phys. Rev. D",
    volume = "58",
    pages = "045001",
    year = "1998"
}

@article{Bodeker:1999gx,
    author = "Bodeker, D. and Moore, Guy D. and Rummukainen, K.",
    title = "{Chern-Simons number diffusion and hard thermal loops on the lattice}",
    eprint = "hep-ph/9907545",
    archivePrefix = "arXiv",
    reportNumber = "NORDITA-99-45HE, MCGILL-99-22, NBI-HE-99-25",
    doi = "10.1103/PhysRevD.61.056003",
    journal = "Phys. Rev. D",
    volume = "61",
    pages = "056003",
    year = "2000"
}

@article{Bodeker:1998hm,
    author = "Bodeker, Dietrich",
    title = "{On the effective dynamics of soft nonAbelian gauge fields at finite temperature}",
    eprint = "hep-ph/9801430",
    archivePrefix = "arXiv",
    reportNumber = "HD-THEP-98-04",
    doi = "10.1016/S0370-2693(98)00279-2",
    journal = "Phys. Lett. B",
    volume = "426",
    pages = "351--360",
    year = "1998"
}

@article{Arnold:1999uy,
    author = "Arnold, Peter Brockway and Yaffe, Laurence G.",
    title = "{High temperature color conductivity at next-to-leading log order}",
    eprint = "hep-ph/9912306",
    archivePrefix = "arXiv",
    reportNumber = "UW-PT-99-24",
    doi = "10.1103/PhysRevD.62.125014",
    journal = "Phys. Rev. D",
    volume = "62",
    pages = "125014",
    year = "2000"
}

@article{Moore:1998zk,
    author = "Moore, Guy D.",
    title = "{The Sphaleron rate: Bodeker's leading log}",
    eprint = "hep-ph/9810313",
    archivePrefix = "arXiv",
    reportNumber = "MCGILL-98-28",
    doi = "10.1016/S0550-3213(99)00746-4",
    journal = "Nucl. Phys. B",
    volume = "568",
    pages = "367--404",
    year = "2000"
}

@article{Moore:2000jw,
    author = "Moore, Guy D. and Rummukainen, Kari",
    title = "{Electroweak bubble nucleation, nonperturbatively}",
    eprint = "hep-ph/0009132",
    archivePrefix = "arXiv",
    reportNumber = "UW-PT-00-01, NORDITA-2000-79-HE",
    doi = "10.1103/PhysRevD.63.045002",
    journal = "Phys. Rev. D",
    volume = "63",
    pages = "045002",
    year = "2001"
}

@article{Li:2024mts,
    author = "Li, Yuan-Zhen and Ramsey-Musolf, Michael J. and Yu, Jiang-Hao",
    title = "{Does the Electron EDM Preclude Electroweak Baryogenesis ?}",
    eprint = "2404.19197",
    archivePrefix = "arXiv",
    primaryClass = "hep-ph",
    month = "4",
    year = "2024"
}

@article{Cirigliano:2009yt,
    author = "Cirigliano, Vincenzo and Lee, Christopher and Ramsey-Musolf, Michael J. and Tulin, Sean",
    title = "{Flavored Quantum Boltzmann Equations}",
    eprint = "0912.3523",
    archivePrefix = "arXiv",
    primaryClass = "hep-ph",
    reportNumber = "NPAC-09-16, UCB-PTH-09-37",
    doi = "10.1103/PhysRevD.81.103503",
    journal = "Phys. Rev. D",
    volume = "81",
    pages = "103503",
    year = "2010"
}

@article{Lee:2004we,
    author = "Lee, Christopher and Cirigliano, Vincenzo and Ramsey-Musolf, Michael J.",
    title = "{Resonant relaxation in electroweak baryogenesis}",
    eprint = "hep-ph/0412354",
    archivePrefix = "arXiv",
    reportNumber = "CALTECH-MAP-304, CALT-68-2535",
    doi = "10.1103/PhysRevD.71.075010",
    journal = "Phys. Rev. D",
    volume = "71",
    pages = "075010",
    year = "2005"
}

@article{Cirigliano:2006wh,
    author = "Cirigliano, Vincenzo and Ramsey-Musolf, Michael J. and Tulin, Sean and Lee, Christopher",
    title = "{Yukawa and tri-scalar processes in electroweak baryogenesis}",
    eprint = "hep-ph/0603058",
    archivePrefix = "arXiv",
    reportNumber = "CALTECH-MAP-312, INT-PUB-06-03",
    doi = "10.1103/PhysRevD.73.115009",
    journal = "Phys. Rev. D",
    volume = "73",
    pages = "115009",
    year = "2006"
}

@article{Matchev:2025ivr,
    author = "Matchev, Konstantin T. and Verner, Sarunas",
    title = "{The Electroweak Sphaleron Revisited: I. Static Solutions, Energy Barrier, and Unstable Modes}",
    eprint = "2505.05607",
    archivePrefix = "arXiv",
    primaryClass = "hep-ph",
    month = "5",
    year = "2025"
}

@article{Matchev:2025irm,
    author = "Matchev, Konstantin T. and Verner, Sarunas",
    title = "{The Electroweak Sphaleron Revisited: II. Study of Decay Dynamics}",
    eprint = "2505.05608",
    archivePrefix = "arXiv",
    primaryClass = "hep-ph",
    month = "5",
    year = "2025"
}

@article{Kharzeev:2019rsy,
    author = "Kharzeev, Dmitri and Shuryak, Edward and Zahed, Ismail",
    title = "{Sphalerons, baryogenesis, and helical magnetogenesis in the electroweak transition of the minimal standard model}",
    eprint = "1906.04080",
    archivePrefix = "arXiv",
    primaryClass = "hep-ph",
    doi = "10.1103/PhysRevD.102.073003",
    journal = "Phys. Rev. D",
    volume = "102",
    number = "7",
    pages = "073003",
    year = "2020"
}

@article{Tanaka:2025cpw,
    author = "Tanaka, Masanori",
    title = "{Sphalerogenesis}",
    eprint = "2505.09984",
    archivePrefix = "arXiv",
    primaryClass = "hep-ph",
    month = "5",
    year = "2025"
}

@article{Hong:2023zrf,
    author = "Hong, Muzi and Kamada, Kohei and Yokoyama, Jun'ichi",
    title = "{Baryogenesis from sphaleron decoupling}",
    eprint = "2304.13999",
    archivePrefix = "arXiv",
    primaryClass = "hep-ph",
    reportNumber = "RESCEU-9/23",
    doi = "10.1103/PhysRevD.108.063502",
    journal = "Phys. Rev. D",
    volume = "108",
    number = "6",
    pages = "063502",
    year = "2023"
}

@article{Cirigliano:2011di,
    author = "Cirigliano, Vincenzo and Lee, Christopher and Tulin, Sean",
    title = "{Resonant Flavor Oscillations in Electroweak Baryogenesis}",
    eprint = "1106.0747",
    archivePrefix = "arXiv",
    primaryClass = "hep-ph",
    reportNumber = "MIT-CTP-4269",
    doi = "10.1103/PhysRevD.84.056006",
    journal = "Phys. Rev. D",
    volume = "84",
    pages = "056006",
    year = "2011"
}

@article{Moore:1998swa,
    author = "Moore, Guy D.",
    title = "{Measuring the broken phase sphaleron rate nonperturbatively}",
    eprint = "hep-ph/9805264",
    archivePrefix = "arXiv",
    reportNumber = "MCGILL-98-7",
    doi = "10.1103/PhysRevD.59.014503",
    journal = "Phys. Rev. D",
    volume = "59",
    pages = "014503",
    year = "1999"
}

@article{Moore:2001vf,
    author = "Moore, Guy D. and Rummukainen, Kari and Tranberg, Anders",
    title = "{Nonperturbative computation of the bubble nucleation rate in the cubic anisotropy model}",
    eprint = "hep-lat/0103036",
    archivePrefix = "arXiv",
    reportNumber = "NORDITA-2001-6-HE, UW-PT-01-05",
    doi = "10.1088/1126-6708/2001/04/017",
    journal = "JHEP",
    volume = "04",
    pages = "017",
    year = "2001"
}

@article{Bodeker:1999ey,
    author = "Bodeker, Dietrich",
    title = "{From hard thermal loops to Langevin dynamics}",
    eprint = "hep-ph/9905239",
    archivePrefix = "arXiv",
    reportNumber = "NBI-HE-99-13",
    doi = "10.1016/S0550-3213(99)00435-6",
    journal = "Nucl. Phys. B",
    volume = "559",
    pages = "502--538",
    year = "1999"
}

@article{Bodeker:2002gy,
    author = "Bodeker, Dietrich",
    title = "{Perturbative and nonperturbative aspects of the nonAbelian Boltzmann-Langevin equation}",
    eprint = "hep-ph/0205202",
    archivePrefix = "arXiv",
    reportNumber = "BI-TP-2002-08",
    doi = "10.1016/S0550-3213(02)00841-6",
    journal = "Nucl. Phys. B",
    volume = "647",
    pages = "512--538",
    year = "2002"
}

@article{Arnold:1998cy,
    author = "Arnold, Peter Brockway and Son, Dam T. and Yaffe, Laurence G.",
    title = "{Effective dynamics of hot, soft nonAbelian gauge fields. Color conductivity and log(1/alpha) effects}",
    eprint = "hep-ph/9810216",
    archivePrefix = "arXiv",
    reportNumber = "UW-PT-98-10, MIT-CTP-2779",
    doi = "10.1103/PhysRevD.59.105020",
    journal = "Phys. Rev. D",
    volume = "59",
    pages = "105020",
    year = "1999"
}

@article{Arnold:1999ux,
    author = "Arnold, Peter Brockway and Yaffe, Laurence G.",
    title = "{Nonperturbative dynamics of hot nonAbelian gauge fields: Beyond leading log approximation}",
    eprint = "hep-ph/9912305",
    archivePrefix = "arXiv",
    reportNumber = "UW-PT-99-25",
    doi = "10.1103/PhysRevD.62.125013",
    journal = "Phys. Rev. D",
    volume = "62",
    pages = "125013",
    year = "2000"
}

@article{DOnofrio:2012phz,
    author = "D'Onofrio, Michela and Rummukainen, Kari and Tranberg, Anders",
    title = "{The Sphaleron Rate through the Electroweak Cross-over}",
    eprint = "1207.0685",
    archivePrefix = "arXiv",
    primaryClass = "hep-ph",
    doi = "10.1007/JHEP08(2012)123",
    journal = "JHEP",
    volume = "08",
    pages = "123",
    year = "2012"
}

@article{DOnofrio:2014rug,
    author = "D'Onofrio, Michela and Rummukainen, Kari and Tranberg, Anders",
    title = "{Sphaleron Rate in the Minimal Standard Model}",
    eprint = "1404.3565",
    archivePrefix = "arXiv",
    primaryClass = "hep-ph",
    doi = "10.1103/PhysRevLett.113.141602",
    journal = "Phys. Rev. Lett.",
    volume = "113",
    number = "14",
    pages = "141602",
    year = "2014"
}

@article{Annala:2023jvr,
    author = "Annala, Jaakko and Rummukainen, Kari",
    title = "{Electroweak sphaleron in a magnetic field}",
    eprint = "2301.08626",
    archivePrefix = "arXiv",
    primaryClass = "hep-ph",
    doi = "10.1103/PhysRevD.107.073006",
    journal = "Phys. Rev. D",
    volume = "107",
    number = "7",
    pages = "073006",
    year = "2023"
}

@article{Moore:2010jd,
    author = "Moore, Guy D. and Tassler, Marcus",
    title = "{The Sphaleron Rate in SU(N) Gauge Theory}",
    eprint = "1011.1167",
    archivePrefix = "arXiv",
    primaryClass = "hep-ph",
    doi = "10.1007/JHEP02(2011)105",
    journal = "JHEP",
    volume = "02",
    pages = "105",
    year = "2011"
}

@article{Moore:1997cr,
    author = "Moore, Guy D. and Turok, Neil",
    title = "{Lattice Chern-Simons number without ultraviolet problems}",
    eprint = "hep-ph/9703266",
    archivePrefix = "arXiv",
    reportNumber = "PUPT-1681, DAMTP-97-17",
    doi = "10.1103/PhysRevD.56.6533",
    journal = "Phys. Rev. D",
    volume = "56",
    pages = "6533--6546",
    year = "1997"
}

@article{Moore:1998ge,
    author = "Moore, Guy D.",
    title = "{A Nonperturbative measurement of the broken phase sphaleron rate}",
    eprint = "hep-ph/9801204",
    archivePrefix = "arXiv",
    reportNumber = "MCGILL-97-36",
    doi = "10.1016/S0370-2693(98)01009-0",
    journal = "Phys. Lett. B",
    volume = "439",
    pages = "357--365",
    year = "1998"
}

@article{Moore:2000mx,
    author = "Moore, Guy D.",
    title = "{Sphaleron rate in the symmetric electroweak phase}",
    eprint = "hep-ph/0001216",
    archivePrefix = "arXiv",
    reportNumber = "UW-PT-00-01",
    doi = "10.1103/PhysRevD.62.085011",
    journal = "Phys. Rev. D",
    volume = "62",
    pages = "085011",
    year = "2000"
}

@article{Burnier:2005hp,
    author = "Burnier, Y. and Laine, M. and Shaposhnikov, M.",
    title = "{Baryon and lepton number violation rates across the electroweak crossover}",
    eprint = "hep-ph/0511246",
    archivePrefix = "arXiv",
    reportNumber = "BI-TP-2005-48",
    doi = "10.1088/1475-7516/2006/02/007",
    journal = "JCAP",
    volume = "02",
    pages = "007",
    year = "2006"
}

@article{Planck:2018vyg,
    author = "Aghanim, N. and others",
    collaboration = "Planck",
    title = "{Planck 2018 results. VI. Cosmological parameters}",
    eprint = "1807.06209",
    archivePrefix = "arXiv",
    primaryClass = "astro-ph.CO",
    doi = "10.1051/0004-6361/201833910",
    journal = "Astron. Astrophys.",
    volume = "641",
    pages = "A6",
    year = "2020",
    note = "[Erratum: Astron.Astrophys. 652, C4 (2021)]"
}

@article{Fields:2019pfx,
    author = "Fields, Brian D. and Olive, Keith A. and Yeh, Tsung-Han and Young, Charles",
    title = "{Big-Bang Nucleosynthesis after Planck}",
    eprint = "1912.01132",
    archivePrefix = "arXiv",
    primaryClass = "astro-ph.CO",
    reportNumber = "UMN--TH--3902/19, FTPI--MINN--19/25",
    doi = "10.1088/1475-7516/2020/03/010",
    journal = "JCAP",
    volume = "03",
    pages = "010",
    year = "2020",
    note = "[Erratum: JCAP 11, E02 (2020)]"
}

@article{Akiba:1988ay,
    author = "Akiba, T. and Kikuchi, H. and Yanagida, T.",
    title = "{Static Minimum Energy Path From a Vacuum to a Sphaleron in the {Weinberg-Salam} Model}",
    doi = "10.1103/PhysRevD.38.1937",
    journal = "Phys. Rev. D",
    volume = "38",
    pages = "1937--1941",
    year = "1988"
}

@article{Patel:2011th,
    author = "Patel, Hiren H. and Ramsey-Musolf, Michael J.",
    title = "{Baryon Washout, Electroweak Phase Transition, and Perturbation Theory}",
    eprint = "1101.4665",
    archivePrefix = "arXiv",
    primaryClass = "hep-ph",
    doi = "10.1007/JHEP07(2011)029",
    journal = "JHEP",
    volume = "07",
    pages = "029",
    year = "2011"
}

@article{Niemi:2024vzw,
    author = "Niemi, Lauri and Tenkanen, Tuomas V. I.",
    title = "{Investigating two-loop effects for first-order electroweak phase transitions}",
    eprint = "2408.15912",
    archivePrefix = "arXiv",
    primaryClass = "hep-ph",
    reportNumber = "HIP-2024-10/TH",
    doi = "10.1103/PhysRevD.111.075034",
    journal = "Phys. Rev. D",
    volume = "111",
    number = "7",
    pages = "075034",
    year = "2025"
}

@article{Niemi:2021qvp,
    author = "Niemi, Lauri and Schicho, Philipp and Tenkanen, Tuomas V. I.",
    title = "{Singlet-assisted electroweak phase transition at two loops}",
    eprint = "2103.07467",
    archivePrefix = "arXiv",
    primaryClass = "hep-ph",
    reportNumber = "HIP-2021-8/TH, NORDITA 2021-011",
    doi = "10.1103/PhysRevD.103.115035",
    journal = "Phys. Rev. D",
    volume = "103",
    number = "11",
    pages = "115035",
    year = "2021",
    note = "[Erratum: Phys.Rev.D 109, 039902 (2024)]"
}

@article{Gould:2021oba,
    author = "Gould, Oliver and Tenkanen, Tuomas V. I.",
    title = "{On the perturbative expansion at high temperature and implications for cosmological phase transitions}",
    eprint = "2104.04399",
    archivePrefix = "arXiv",
    primaryClass = "hep-ph",
    reportNumber = "NORDITA 2021-010",
    doi = "10.1007/JHEP06(2021)069",
    journal = "JHEP",
    volume = "06",
    pages = "069",
    year = "2021"
}

@article{Niemi:2020hto,
    author = "Niemi, Lauri and Ramsey-Musolf, Michael J. and Tenkanen, Tuomas V. I. and Weir, David J.",
    title = "{Thermodynamics of a Two-Step Electroweak Phase Transition}",
    eprint = "2005.11332",
    archivePrefix = "arXiv",
    primaryClass = "hep-ph",
    reportNumber = "HIP-2020-11/TH, ACFI-T20-05",
    doi = "10.1103/PhysRevLett.126.171802",
    journal = "Phys. Rev. Lett.",
    volume = "126",
    number = "17",
    pages = "171802",
    year = "2021"
}

@article{Kainulainen:2019kyp,
    author = "Kainulainen, Kimmo and Keus, Venus and Niemi, Lauri and Rummukainen, Kari and Tenkanen, Tuomas V. I. and Vaskonen, Ville",
    title = "{On the validity of perturbative studies of the electroweak phase transition in the Two Higgs Doublet model}",
    eprint = "1904.01329",
    archivePrefix = "arXiv",
    primaryClass = "hep-ph",
    doi = "10.1007/JHEP06(2019)075",
    journal = "JHEP",
    volume = "06",
    pages = "075",
    year = "2019"
}

@article{Niemi:2024axp,
    author = "Niemi, Lauri and Ramsey-Musolf, Michael J. and Xia, Guotao",
    title = "{Nonperturbative study of the electroweak phase transition in the real scalar singlet extended standard model}",
    eprint = "2405.01191",
    archivePrefix = "arXiv",
    primaryClass = "hep-ph",
    reportNumber = "HIP-2024-7/TH, ACFI T24-03",
    doi = "10.1103/PhysRevD.110.115016",
    journal = "Phys. Rev. D",
    volume = "110",
    number = "11",
    pages = "115016",
    year = "2024"
}

@article{Lofgren:2023sep,
    author = {L\"ofgren, Johan},
    title = "{Stop comparing resummation methods}",
    eprint = "2301.05197",
    archivePrefix = "arXiv",
    primaryClass = "hep-ph",
    doi = "10.1088/1361-6471/ad074b",
    journal = "J. Phys. G",
    volume = "50",
    number = "12",
    pages = "125008",
    year = "2023"
}

@article{Ekstedt:2022zro,
    author = {Ekstedt, Andreas and Gould, Oliver and L\"ofgren, Johan},
    title = "{Radiative first-order phase transitions to next-to-next-to-leading order}",
    eprint = "2205.07241",
    archivePrefix = "arXiv",
    primaryClass = "hep-ph",
    doi = "10.1103/PhysRevD.106.036012",
    journal = "Phys. Rev. D",
    volume = "106",
    number = "3",
    pages = "036012",
    year = "2022",
    note = "[Erratum: Phys.Rev.D 110, 019901 (2024)]"
}

@article{Gould:2024chm,
    author = "Gould, Oliver and Kormu, Anna and Weir, David J.",
    title = "{Nonperturbative test of nucleation calculations for strong phase transitions}",
    eprint = "2404.01876",
    archivePrefix = "arXiv",
    primaryClass = "hep-th",
    reportNumber = "HIP-2024-9/TH",
    doi = "10.1103/PhysRevD.111.L051901",
    journal = "Phys. Rev. D",
    volume = "111",
    number = "5",
    pages = "L051901",
    year = "2025"
}

@article{Gould:2022ran,
    author = {Gould, Oliver and G\"uyer, Sinan and Rummukainen, Kari},
    title = "{First-order electroweak phase transitions: A nonperturbative update}",
    eprint = "2205.07238",
    archivePrefix = "arXiv",
    primaryClass = "hep-lat",
    reportNumber = "HIP-2022-10/TH",
    doi = "10.1103/PhysRevD.106.114507",
    journal = "Phys. Rev. D",
    volume = "106",
    number = "11",
    pages = "114507",
    year = "2022"
}

@article{Gould:2021dzl,
    author = "Gould, Oliver",
    title = "{Real scalar phase transitions: a nonperturbative analysis}",
    eprint = "2101.05528",
    archivePrefix = "arXiv",
    primaryClass = "hep-ph",
    reportNumber = "HIP-2021-2/TH",
    doi = "10.1007/JHEP04(2021)057",
    journal = "JHEP",
    volume = "04",
    pages = "057",
    year = "2021"
}

@article{Gould:2023jbz,
    author = "Gould, Oliver and Xie, Cheng",
    title = "{Higher orders for cosmological phase transitions: a global study in a Yukawa model}",
    eprint = "2310.02308",
    archivePrefix = "arXiv",
    primaryClass = "hep-ph",
    doi = "10.1007/JHEP12(2023)049",
    journal = "JHEP",
    volume = "12",
    pages = "049",
    year = "2023"
}

@article{Schicho:2022wty,
    author = "Schicho, Philipp and Tenkanen, Tuomas V. I. and White, Graham",
    title = "{Combining thermal resummation and gauge invariance for electroweak phase transition}",
    eprint = "2203.04284",
    archivePrefix = "arXiv",
    primaryClass = "hep-ph",
    reportNumber = "HIP-2022-2/TH, NORDITA 2022-009",
    doi = "10.1007/JHEP11(2022)047",
    journal = "JHEP",
    volume = "11",
    pages = "047",
    year = "2022"
}

@article{Croon:2020cgk,
    author = "Croon, Djuna and Gould, Oliver and Schicho, Philipp and Tenkanen, Tuomas V. I. and White, Graham",
    title = "{Theoretical uncertainties for cosmological first-order phase transitions}",
    eprint = "2009.10080",
    archivePrefix = "arXiv",
    primaryClass = "hep-ph",
    reportNumber = "HIP-2020-26/TH",
    doi = "10.1007/JHEP04(2021)055",
    journal = "JHEP",
    volume = "04",
    pages = "055",
    year = "2021"
}

@article{Ekstedt:2023sqc,
    author = "Ekstedt, Andreas and Gould, Oliver and Hirvonen, Joonas",
    title = "{BubbleDet: a Python package to compute functional determinants for bubble nucleation}",
    eprint = "2308.15652",
    archivePrefix = "arXiv",
    primaryClass = "hep-ph",
    doi = "10.1007/JHEP12(2023)056",
    journal = "JHEP",
    volume = "12",
    pages = "056",
    year = "2023"
}

@article{Hirvonen:2021zej,
    author = {Hirvonen, Joonas and L\"ofgren, Johan and Ramsey-Musolf, Michael J. and Schicho, Philipp and Tenkanen, Tuomas V. I.},
    title = "{Computing the gauge-invariant bubble nucleation rate in finite temperature effective field theory}",
    eprint = "2112.08912",
    archivePrefix = "arXiv",
    primaryClass = "hep-ph",
    reportNumber = "ACFI-T21-16, HIP-2021-45/TH, NORDITA 2021-111",
    doi = "10.1007/JHEP07(2022)135",
    journal = "JHEP",
    volume = "07",
    pages = "135",
    year = "2022"
}

@article{Ekstedt:2021kyx,
    author = "Ekstedt, Andreas",
    title = "{Higher-order corrections to the bubble-nucleation rate at finite temperature}",
    eprint = "2104.11804",
    archivePrefix = "arXiv",
    primaryClass = "hep-ph",
    doi = "10.1140/epjc/s10052-022-10130-5",
    journal = "Eur. Phys. J. C",
    volume = "82",
    number = "2",
    pages = "173",
    year = "2022"
}

@article{Ekstedt:2022tqk,
    author = "Ekstedt, Andreas",
    title = "{Bubble nucleation to all orders}",
    eprint = "2201.07331",
    archivePrefix = "arXiv",
    primaryClass = "hep-ph",
    reportNumber = "DESY-22-007",
    doi = "10.1007/JHEP08(2022)115",
    journal = "JHEP",
    volume = "08",
    pages = "115",
    year = "2022"
}

@article{Ekstedt:2022ceo,
    author = "Ekstedt, Andreas",
    title = "{Convergence of the nucleation rate for first-order phase transitions}",
    eprint = "2205.05145",
    archivePrefix = "arXiv",
    primaryClass = "hep-ph",
    doi = "10.1103/PhysRevD.106.095026",
    journal = "Phys. Rev. D",
    volume = "106",
    number = "9",
    pages = "095026",
    year = "2022"
}

@article{Gould:2021ccf,
    author = "Gould, Oliver and Hirvonen, Joonas",
    title = "{Effective field theory approach to thermal bubble nucleation}",
    eprint = "2108.04377",
    archivePrefix = "arXiv",
    primaryClass = "hep-ph",
    reportNumber = "HIP-2020-19/TH",
    doi = "10.1103/PhysRevD.104.096015",
    journal = "Phys. Rev. D",
    volume = "104",
    number = "9",
    pages = "096015",
    year = "2021"
}

@article{Baacke:1999sc,
    author = "Baacke, Jurgen and Heitmann, Katrin",
    title = "{Gauge invariance of the one loop effective action of the Higgs field in the SU(2) Higgs model}",
    eprint = "hep-th/9905201",
    archivePrefix = "arXiv",
    reportNumber = "DO-TH-99-05A, DO-TH-99-08",
    doi = "10.1103/PhysRevD.60.105037",
    journal = "Phys. Rev. D",
    volume = "60",
    pages = "105037",
    year = "1999"
}

@article{Kramers:1940zz,
    author = "Kramers, H. A.",
    title = "{Brownian motion in a field of force and the diffusion model of chemical re actions}",
    doi = "10.1016/S0031-8914(40)90098-2",
    journal = "Physica",
    volume = "7",
    pages = "284--304",
    year = "1940"
}

@article{Langer:1969bc,
    author = "Langer, J. S.",
    title = "{Statistical theory of the decay of metastable states}",
    doi = "10.1016/0003-4916(69)90153-5",
    journal = "Annals Phys.",
    volume = "54",
    pages = "258--275",
    year = "1969"
}

@article{Langer:1974cpa,
    author = "Langer, J. S.",
    title = "{Metastable states}",
    doi = "10.1016/0031-8914(74)90226-2",
    journal = "Physica",
    volume = "73",
    number = "1",
    pages = "61--72",
    year = "1974"
}

@article{Linde:1980tt,
    author = "Linde, Andrei D.",
    title = "{Fate of the False Vacuum at Finite Temperature: Theory and Applications}",
    reportNumber = "LEBEDEV-80-92",
    doi = "10.1016/0370-2693(81)90281-1",
    journal = "Phys. Lett. B",
    volume = "100",
    pages = "37--40",
    year = "1981"
}

@article{Linde:1981zj,
    author = "Linde, Andrei D.",
    title = "{Decay of the False Vacuum at Finite Temperature}",
    reportNumber = "LEBEDEV-81-265",
    doi = "10.1016/0550-3213(83)90072-X",
    journal = "Nucl. Phys. B",
    volume = "216",
    pages = "421",
    year = "1983",
    note = "[Erratum: Nucl.Phys.B 223, 544 (1983)]"
}

@article{Eriksson:2024ovi,
    author = "Eriksson, M. and Laine, M.",
    title = "{Soft contributions to the thermal Higgs width across an electroweak phase transition}",
    eprint = "2404.06116",
    archivePrefix = "arXiv",
    primaryClass = "hep-ph",
    doi = "10.1088/1475-7516/2024/06/016",
    journal = "JCAP",
    volume = "06",
    pages = "016",
    year = "2024"
}

@article{Cline:2021iff,
    author = "Cline, James M. and Friedlander, Avi and He, Dong-Ming and Kainulainen, Kimmo and Laurent, Benoit and Tucker-Smith, David",
    title = "{Baryogenesis and gravity waves from a UV-completed electroweak phase transition}",
    eprint = "2102.12490",
    archivePrefix = "arXiv",
    primaryClass = "hep-ph",
    doi = "10.1103/PhysRevD.103.123529",
    journal = "Phys. Rev. D",
    volume = "103",
    number = "12",
    pages = "123529",
    year = "2021"
}

@article{Fuyuto:2014yia,
    author = "Fuyuto, Kaori and Senaha, Eibun",
    title = "{Improved sphaleron decoupling condition and the Higgs coupling constants in the real singlet-extended standard model}",
    eprint = "1406.0433",
    archivePrefix = "arXiv",
    primaryClass = "hep-ph",
    doi = "10.1103/PhysRevD.90.015015",
    journal = "Phys. Rev. D",
    volume = "90",
    number = "1",
    pages = "015015",
    year = "2014"
}

@article{Fuyuto:2015jha,
    author = "Fuyuto, Kaori and Senaha, Eibun",
    title = "{Sphaleron and critical bubble in the scale invariant two Higgs doublet model}",
    eprint = "1504.04291",
    archivePrefix = "arXiv",
    primaryClass = "hep-ph",
    doi = "10.1016/j.physletb.2015.05.061",
    journal = "Phys. Lett. B",
    volume = "747",
    pages = "152--157",
    year = "2015"
}

@article{Ekstedt:2022bff,
    author = "Ekstedt, Andreas and Schicho, Philipp and Tenkanen, Tuomas V. I.",
    title = "{DRalgo: A package for effective field theory approach for thermal phase transitions}",
    eprint = "2205.08815",
    archivePrefix = "arXiv",
    primaryClass = "hep-ph",
    reportNumber = "HIP-2022-11/TH, NORDITA 2022-030",
    doi = "10.1016/j.cpc.2023.108725",
    journal = "Comput. Phys. Commun.",
    volume = "288",
    pages = "108725",
    year = "2023"
}

@article{Kierkla:2023von,
    author = "Kierkla, Maciej and Swiezewska, Bogumila and Tenkanen, Tuomas V. I. and van de Vis, Jorinde",
    title = "{Gravitational waves from supercooled phase transitions: dimensional transmutation meets dimensional reduction}",
    eprint = "2312.12413",
    archivePrefix = "arXiv",
    primaryClass = "hep-ph",
    doi = "10.1007/JHEP02(2024)234",
    journal = "JHEP",
    volume = "02",
    pages = "234",
    year = "2024"
}

@article{Camargo-Molina:2021zgz,
    author = {Camargo-Molina, Jos\'e Eliel and Enberg, Rikard and L\"ofgren, Johan},
    title = "{A new perspective on the electroweak phase transition in the Standard Model Effective Field Theory}",
    eprint = "2103.14022",
    archivePrefix = "arXiv",
    primaryClass = "hep-ph",
    doi = "10.1007/JHEP10(2021)127",
    journal = "JHEP",
    volume = "10",
    pages = "127",
    year = "2021"
}

@article{Gould:2019qek,
    author = "Gould, Oliver and Kozaczuk, Jonathan and Niemi, Lauri and Ramsey-Musolf, Michael J. and Tenkanen, Tuomas V. I. and Weir, David J.",
    title = "{Nonperturbative analysis of the gravitational waves from a first-order electroweak phase transition}",
    eprint = "1903.11604",
    archivePrefix = "arXiv",
    primaryClass = "hep-ph",
    reportNumber = "ACFI T19-04, HIP-2019-5/TH",
    doi = "10.1103/PhysRevD.100.115024",
    journal = "Phys. Rev. D",
    volume = "100",
    number = "11",
    pages = "115024",
    year = "2019"
}

@article{Andersen:2017ika,
    author = "Andersen, Jens O. and Gorda, Tyler and Helset, Andreas and Niemi, Lauri and Tenkanen, Tuomas V. I. and Tranberg, Anders and Vuorinen, Aleksi and Weir, David J.",
    title = "{Nonperturbative Analysis of the Electroweak Phase Transition in the Two Higgs Doublet Model}",
    eprint = "1711.09849",
    archivePrefix = "arXiv",
    primaryClass = "hep-ph",
    reportNumber = "HIP-2017-26/TH, HIP-2017-26-TH",
    doi = "10.1103/PhysRevLett.121.191802",
    journal = "Phys. Rev. Lett.",
    volume = "121",
    number = "19",
    pages = "191802",
    year = "2018"
}

@article{Gorda:2018hvi,
    author = "Gorda, Tyler and Helset, Andreas and Niemi, Lauri and Tenkanen, Tuomas V. I. and Weir, David J.",
    title = "{Three-dimensional effective theories for the two Higgs doublet model at high temperature}",
    eprint = "1802.05056",
    archivePrefix = "arXiv",
    primaryClass = "hep-ph",
    reportNumber = "HIP-2018-6/TH, HIP-2018-6-TH",
    doi = "10.1007/JHEP02(2019)081",
    journal = "JHEP",
    volume = "02",
    pages = "081",
    year = "2019"
}

@article{Chala:2024xll,
    author = "Chala, Mikael and Criado, Juan Carlos and Gil, Luis and Miras, Javier L\'opez",
    title = "{Higher-order-operator corrections to phase-transition parameters in dimensional reduction}",
    eprint = "2406.02667",
    archivePrefix = "arXiv",
    primaryClass = "hep-ph",
    doi = "10.1007/JHEP10(2024)025",
    journal = "JHEP",
    volume = "10",
    pages = "025",
    year = "2024"
}

@article{Ahriche:2014jna,
	archiveprefix = {arXiv},
	author = {Ahriche, Amine and Chowdhury, Talal Ahmed and Nasri, Salah},
	doi = {10.1007/JHEP11(2014)096},
	eprint = {1409.4086},
	journal = {JHEP},
	pages = {096},
	primaryclass = {hep-ph},
	title = {{Sphalerons and the Electroweak Phase Transition in Models with Higher Scalar Representations}},
	volume = {11},
	year = {2014}}

@article{Gan:2017mcv,
    author = "Gan, Xucheng and Long, Andrew J. and Wang, Lian-Tao",
    title = "{Electroweak sphaleron with dimension-six operators}",
    eprint = "1708.03061",
    archivePrefix = "arXiv",
    primaryClass = "hep-ph",
    doi = "10.1103/PhysRevD.96.115018",
    journal = "Phys. Rev. D",
    volume = "96",
    number = "11",
    pages = "115018",
    year = "2017"
}

@article{Hu:2023gbp,
    author = "Hu, Jiahang and Yu, Bingrong and Zhou, Shun",
    title = "{Sphaleron in the Higgs Triplet Model}",
    eprint = "2307.04713",
    archivePrefix = "arXiv",
    primaryClass = "hep-ph",
    doi = "10.1007/JHEP10(2023)004",
    journal = "JHEP",
    volume = "10",
    pages = "004",
    year = "2023"
}

@article{Qin:2024idc,
    author = "Qin, Renhui and Bian, Ligong",
    title = "{First-order electroweak phase transition at finite density}",
    eprint = "2407.01981",
    archivePrefix = "arXiv",
    primaryClass = "hep-ph",
    doi = "10.1007/JHEP08(2024)157",
    journal = "JHEP",
    volume = "08",
    pages = "157",
    year = "2024"
}

@article{Laine:2015kra,
    author = "Laine, M. and Meyer, M.",
    title = "{Standard Model thermodynamics across the electroweak crossover}",
    eprint = "1503.04935",
    archivePrefix = "arXiv",
    primaryClass = "hep-ph",
    doi = "10.1088/1475-7516/2015/07/035",
    journal = "JCAP",
    volume = "07",
    pages = "035",
    year = "2015"
}

@article{Lofgren:2021ogg,
    author = {L\"ofgren, Johan and Ramsey-Musolf, Michael J. and Schicho, Philipp and Tenkanen, Tuomas V. I.},
    title = "{Nucleation at Finite Temperature: A Gauge-Invariant Perturbative Framework}",
    eprint = "2112.05472",
    archivePrefix = "arXiv",
    primaryClass = "hep-ph",
    reportNumber = "ACFI-T21-15, HIP-2021-44/TH, NORDITA 2021-110",
    doi = "10.1103/PhysRevLett.130.251801",
    journal = "Phys. Rev. Lett.",
    volume = "130",
    number = "25",
    pages = "251801",
    year = "2023"
}

@article{Brauner:2016fla,
    author = "Brauner, Tom\'a\v{s} and Tenkanen, Tuomas V. I. and Tranberg, Anders and Vuorinen, Aleksi and Weir, David J.",
    title = "{Dimensional reduction of the Standard Model coupled to a new singlet scalar field}",
    eprint = "1609.06230",
    archivePrefix = "arXiv",
    primaryClass = "hep-ph",
    reportNumber = "HIP-2016-27-TH",
    doi = "10.1007/JHEP03(2017)007",
    journal = "JHEP",
    volume = "03",
    pages = "007",
    year = "2017"
}

@article{Niemi:2018asa,
    author = "Niemi, Lauri and Patel, Hiren H. and Ramsey-Musolf, Michael J. and Tenkanen, Tuomas V. I. and Weir, David J.",
    title = "{Electroweak phase transition in the real triplet extension of the SM: Dimensional reduction}",
    eprint = "1802.10500",
    archivePrefix = "arXiv",
    primaryClass = "hep-ph",
    reportNumber = "HIP-2018-7-TH, ACFI-T18-04, HIP-2018-7/TH",
    doi = "10.1103/PhysRevD.100.035002",
    journal = "Phys. Rev. D",
    volume = "100",
    number = "3",
    pages = "035002",
    year = "2019"
}

@article{Schicho:2021gca,
    author = {Schicho, Philipp M. and Tenkanen, Tuomas V. I. and \"Osterman, Juuso},
    title = "{Robust approach to thermal resummation: Standard Model meets a singlet}",
    eprint = "2102.11145",
    archivePrefix = "arXiv",
    primaryClass = "hep-ph",
    doi = "10.1007/JHEP06(2021)130",
    journal = "JHEP",
    volume = "06",
    pages = "130",
    year = "2021"
}

@inproceedings{Quiros:1999jp,
    author = "Quiros, Mariano",
    title = "{Finite temperature field theory and phase transitions}",
    booktitle = "{ICTP Summer School in High-Energy Physics and Cosmology}",
    eprint = "hep-ph/9901312",
    archivePrefix = "arXiv",
    reportNumber = "IEM-FT-187-99",
    pages = "187--259",
    month = "1",
    year = "1999"
}

@article{Khlebnikov:1988sr,
    author = "Khlebnikov, S. Yu. and Shaposhnikov, M. E.",
    title = "{The Statistical Theory of Anomalous Fermion Number Nonconservation}",
    doi = "10.1016/0550-3213(88)90133-2",
    journal = "Nucl. Phys. B",
    volume = "308",
    pages = "885--912",
    year = "1988"
}

@article{Khlebnikov:1996vj,
    author = "Khlebnikov, S. Yu. and Shaposhnikov, M. E.",
    title = "{Melting of the Higgs vacuum: Conserved numbers at high temperature}",
    eprint = "hep-ph/9607386",
    archivePrefix = "arXiv",
    reportNumber = "PURD-TH-96-05, CERN-TH-96-167",
    doi = "10.1016/0370-2693(96)01116-1",
    journal = "Phys. Lett. B",
    volume = "387",
    pages = "817--822",
    year = "1996"
}

@article{Klinkhamer:1990fi,
    author = "Klinkhamer, Frans R. and Laterveer, R.",
    title = "{The Sphaleron at finite mixing angle}",
    reportNumber = "NIKHEF-H-90-12",
    doi = "10.1007/BF01597560",
    journal = "Z. Phys. C",
    volume = "53",
    pages = "247--252",
    year = "1992"
}

@article{Laine:2012jy,
    author = "Laine, M. and Nardini, G. and Rummukainen, K.",
    title = "{Lattice study of an electroweak phase transition at $m_h \backsimeq$ 126 GeV}",
    eprint = "1211.7344",
    archivePrefix = "arXiv",
    primaryClass = "hep-ph",
    doi = "10.1088/1475-7516/2013/01/011",
    journal = "JCAP",
    volume = "01",
    pages = "011",
    year = "2013"
}

@article{Braaten:1989mz,
    author = "Braaten, Eric and Pisarski, Robert D.",
    title = "{Soft Amplitudes in Hot Gauge Theories: A General Analysis}",
    reportNumber = "BNL-43293, FERMILAB-PUB-89-152-T, NUHEP-TH-89-7",
    doi = "10.1016/0550-3213(90)90508-B",
    journal = "Nucl. Phys. B",
    volume = "337",
    pages = "569--634",
    year = "1990"
}

@article{Braaten:1991gm,
    author = "Braaten, Eric and Pisarski, Robert D.",
    title = "{Simple effective Lagrangian for hard thermal loops}",
    reportNumber = "BNL-45096, NUHEP-TH-91-21",
    doi = "10.1103/PhysRevD.45.R1827",
    journal = "Phys. Rev. D",
    volume = "45",
    number = "6",
    pages = "R1827",
    year = "1992"
}

@article{Frenkel:1989br,
    author = "Frenkel, J. and Taylor, J. C.",
    title = "{High Temperature Limit of Thermal QCD}",
    reportNumber = "DAMTP-89-23",
    doi = "10.1016/0550-3213(90)90661-V",
    journal = "Nucl. Phys. B",
    volume = "334",
    pages = "199--216",
    year = "1990"
}

@article{Ekstedt:2023anj,
    author = "Ekstedt, Andreas",
    title = "{Two-loop hard thermal loops for vector bosons in general models}",
    eprint = "2302.04894",
    archivePrefix = "arXiv",
    primaryClass = "hep-ph",
    reportNumber = "DESY-23-016",
    doi = "10.1007/JHEP06(2023)135",
    journal = "JHEP",
    volume = "06",
    pages = "135",
    year = "2023"
}

@article{Ekstedt:2023oqb,
    author = "Ekstedt, Andreas",
    title = "{Propagation of gauge fields in hot and dense plasmas at higher orders}",
    eprint = "2304.09255",
    archivePrefix = "arXiv",
    primaryClass = "hep-ph",
    month = "4",
    year = "2023"
}

@article{Basler:2016obg,
    author = "Basler, P. and Krause, M. and Muhlleitner, M. and Wittbrodt, J. and Wlotzka, A.",
    title = "{Strong First Order Electroweak Phase Transition in the CP-Conserving 2HDM Revisited}",
    eprint = "1612.04086",
    archivePrefix = "arXiv",
    primaryClass = "hep-ph",
    doi = "10.1007/JHEP02(2017)121",
    journal = "JHEP",
    volume = "02",
    pages = "121",
    year = "2017"
}

@article{Cline:2013gha,
    author = "Cline, James M. and Kainulainen, Kimmo and Scott, Pat and Weniger, Christoph",
    title = "{Update on scalar singlet dark matter}",
    eprint = "1306.4710",
    archivePrefix = "arXiv",
    primaryClass = "hep-ph",
    doi = "10.1103/PhysRevD.88.055025",
    journal = "Phys. Rev. D",
    volume = "88",
    pages = "055025",
    year = "2013",
    note = "[Erratum: Phys.Rev.D 92, 039906 (2015)]"
}

@article{Profumo:2007wc,
    author = "Profumo, Stefano and Ramsey-Musolf, Michael J. and Shaughnessy, Gabe",
    title = "{Singlet Higgs phenomenology and the electroweak phase transition}",
    eprint = "0705.2425",
    archivePrefix = "arXiv",
    primaryClass = "hep-ph",
    reportNumber = "CALTECH-MAP-333, MADPH-07-1489",
    doi = "10.1088/1126-6708/2007/08/010",
    journal = "JHEP",
    volume = "08",
    pages = "010",
    year = "2007"
}

@article{Carena:1996wj,
    author = "Carena, Marcela and Quiros, M. and Wagner, C. E. M.",
    title = "{Opening the window for electroweak baryogenesis}",
    eprint = "hep-ph/9603420",
    archivePrefix = "arXiv",
    reportNumber = "CERN-TH-96-30, IEM-FT-126-96",
    doi = "10.1016/0370-2693(96)00475-3",
    journal = "Phys. Lett. B",
    volume = "380",
    pages = "81--91",
    year = "1996"
}

@article{Espinosa:2011ax,
    author = "Espinosa, Jose R. and Konstandin, Thomas and Riva, Francesco",
    title = "{Strong Electroweak Phase Transitions in the Standard Model with a Singlet}",
    eprint = "1107.5441",
    archivePrefix = "arXiv",
    primaryClass = "hep-ph",
    reportNumber = "CERN-PH-TH-2011-171",
    doi = "10.1016/j.nuclphysb.2011.09.010",
    journal = "Nucl. Phys. B",
    volume = "854",
    pages = "592--630",
    year = "2012"
}

@article{Grojean:2004xa,
    author = "Grojean, Christophe and Servant, Geraldine and Wells, James D.",
    title = "{First-order electroweak phase transition in the standard model with a low cutoff}",
    eprint = "hep-ph/0407019",
    archivePrefix = "arXiv",
    reportNumber = "SACLAY-T04-084, MCTP-04-37, ANL-HEP-PR-04-63, EFI-04-23",
    doi = "10.1103/PhysRevD.71.036001",
    journal = "Phys. Rev. D",
    volume = "71",
    pages = "036001",
    year = "2005"
}

@article{Barger:2008jx,
    author = "Barger, Vernon and Langacker, Paul and McCaskey, Mathew and Ramsey-Musolf, Michael and Shaughnessy, Gabe",
    title = "{Complex Singlet Extension of the Standard Model}",
    eprint = "0811.0393",
    archivePrefix = "arXiv",
    primaryClass = "hep-ph",
    reportNumber = "MADPH-08-1516, NUHEP-TH-08-06, ANL-HEP-PR-08-58, NPAC-08-21",
    doi = "10.1103/PhysRevD.79.015018",
    journal = "Phys. Rev. D",
    volume = "79",
    pages = "015018",
    year = "2009"
}

@article{Bodeker:1996pc,
    author = "Bodeker, D. and John, P. and Laine, M. and Schmidt, M. G.",
    title = "{The Two loop MSSM finite temperature effective potential with stop condensation}",
    eprint = "hep-ph/9612364",
    archivePrefix = "arXiv",
    reportNumber = "HD-THEP-96-56",
    doi = "10.1016/S0550-3213(97)00252-6",
    journal = "Nucl. Phys. B",
    volume = "497",
    pages = "387--414",
    year = "1997"
}

@article{Biondini:2017rpb,
    author = "Biondini, Simone and others",
    title = "{Status of rates and rate equations for thermal leptogenesis}",
    eprint = "1711.02864",
    archivePrefix = "arXiv",
    primaryClass = "hep-ph",
    doi = "10.1142/S0217751X18420046",
    journal = "Int. J. Mod. Phys. A",
    volume = "33",
    number = "05n06",
    pages = "1842004",
    year = "2018"
}

@article{Curtin:2014jma,
    author = "Curtin, David and Meade, Patrick and Yu, Chiu-Tien",
    title = "{Testing Electroweak Baryogenesis with Future Colliders}",
    eprint = "1409.0005",
    archivePrefix = "arXiv",
    primaryClass = "hep-ph",
    reportNumber = "YITP-SB-14-33",
    doi = "10.1007/JHEP11(2014)127",
    journal = "JHEP",
    volume = "11",
    pages = "127",
    year = "2014"
}

@article{Turok:1990in,
    author = "Turok, Neil and Zadrozny, John",
    title = "{Dynamical generation of baryons at the electroweak transition}",
    reportNumber = "PUPT-90-1183",
    doi = "10.1103/PhysRevLett.65.2331",
    journal = "Phys. Rev. Lett.",
    volume = "65",
    pages = "2331--2334",
    year = "1990"
}

@article{Menon:2004wv,
    author = "Menon, A. and Morrissey, D. E. and Wagner, C. E. M.",
    title = "{Electroweak baryogenesis and dark matter in the nMSSM}",
    eprint = "hep-ph/0404184",
    archivePrefix = "arXiv",
    reportNumber = "ANL-HEP-PR-04-038, EFI-04-09",
    doi = "10.1103/PhysRevD.70.035005",
    journal = "Phys. Rev. D",
    volume = "70",
    pages = "035005",
    year = "2004"
}

@article{Delaunay:2007wb,
    author = "Delaunay, Cedric and Grojean, Christophe and Wells, James D.",
    title = "{Dynamics of Non-renormalizable Electroweak Symmetry Breaking}",
    eprint = "0711.2511",
    archivePrefix = "arXiv",
    primaryClass = "hep-ph",
    reportNumber = "CERN-PH-TH-2007-219, MCTP-07-31, SACLAY-T07-141",
    doi = "10.1088/1126-6708/2008/04/029",
    journal = "JHEP",
    volume = "04",
    pages = "029",
    year = "2008"
}

@article{Espinosa:2007qk,
    author = "Espinosa, Jose Ramon and Quiros, Mariano",
    title = "{Novel Effects in Electroweak Breaking from a Hidden Sector}",
    eprint = "hep-ph/0701145",
    archivePrefix = "arXiv",
    reportNumber = "IFT-UAM-CSIC-07-01, UAB-FT-623",
    doi = "10.1103/PhysRevD.76.076004",
    journal = "Phys. Rev. D",
    volume = "76",
    pages = "076004",
    year = "2007"
}

@article{Huber:2000mg,
    author = "Huber, S. J. and Schmidt, M. G.",
    title = "{Electroweak baryogenesis: Concrete in a SUSY model with a gauge singlet}",
    eprint = "hep-ph/0003122",
    archivePrefix = "arXiv",
    reportNumber = "HD-THEP-00-15",
    doi = "10.1016/S0550-3213(01)00250-4",
    journal = "Nucl. Phys. B",
    volume = "606",
    pages = "183--230",
    year = "2001"
}

@article{Beniwal:2017eik,
    author = "Beniwal, Ankit and Lewicki, Marek and Wells, James D. and White, Martin and Williams, Anthony G.",
    title = "{Gravitational wave, collider and dark matter signals from a scalar singlet electroweak baryogenesis}",
    eprint = "1702.06124",
    archivePrefix = "arXiv",
    primaryClass = "hep-ph",
    reportNumber = "ADP-17-08-T1014, ADP--17--08-T1014",
    doi = "10.1007/JHEP08(2017)108",
    journal = "JHEP",
    volume = "08",
    pages = "108",
    year = "2017"
}

@article{Espinosa:2008kw,
    author = "Espinosa, J. R. and Konstandin, T. and No, J. M. and Quiros, M.",
    title = "{Some Cosmological Implications of Hidden Sectors}",
    eprint = "0809.3215",
    archivePrefix = "arXiv",
    primaryClass = "hep-ph",
    reportNumber = "CERN-PH-TH-2008-196, IFT-UAM-CSIC-08-54, UAB-FT-655",
    doi = "10.1103/PhysRevD.78.123528",
    journal = "Phys. Rev. D",
    volume = "78",
    pages = "123528",
    year = "2008"
}

@article{Dorsch:2016nrg,
    author = "Dorsch, G. C. and Huber, S. J. and Konstandin, T. and No, J. M.",
    title = "{A Second Higgs Doublet in the Early Universe: Baryogenesis and Gravitational Waves}",
    eprint = "1611.05874",
    archivePrefix = "arXiv",
    primaryClass = "hep-ph",
    reportNumber = "DESY-16-213",
    doi = "10.1088/1475-7516/2017/05/052",
    journal = "JCAP",
    volume = "05",
    pages = "052",
    year = "2017"
}

@article{Bruggisser:2018mrt,
    author = "Bruggisser, Sebastian and Von Harling, Benedict and Matsedonskyi, Oleksii and Servant, G\'eraldine",
    title = "{Electroweak Phase Transition and Baryogenesis in Composite Higgs Models}",
    eprint = "1804.07314",
    archivePrefix = "arXiv",
    primaryClass = "hep-ph",
    reportNumber = "DESY-17-229",
    doi = "10.1007/JHEP12(2018)099",
    journal = "JHEP",
    volume = "12",
    pages = "099",
    year = "2018"
}

@article{Dorsch:2014qja,
    author = "Dorsch, G. C. and Huber, S. J. and Mimasu, K. and No, J. M.",
    title = "{Echoes of the Electroweak Phase Transition: Discovering a second Higgs doublet through $A_0 \rightarrow ZH_0$}",
    eprint = "1405.5537",
    archivePrefix = "arXiv",
    primaryClass = "hep-ph",
    doi = "10.1103/PhysRevLett.113.211802",
    journal = "Phys. Rev. Lett.",
    volume = "113",
    number = "21",
    pages = "211802",
    year = "2014"
}

@article{Laine:1994zq,
    author = "Laine, M.",
    title = "{Gauge dependence of the high temperature two loop effective potential for the Higgs field}",
    eprint = "hep-ph/9411252",
    archivePrefix = "arXiv",
    reportNumber = "HU-TFT-94-46",
    doi = "10.1103/PhysRevD.51.4525",
    journal = "Phys. Rev. D",
    volume = "51",
    pages = "4525--4532",
    year = "1995"
}

@article{Nielsen:1975fs,
    author = "Nielsen, N. K.",
    title = "{On the Gauge Dependence of Spontaneous Symmetry Breaking in Gauge Theories}",
    reportNumber = "Print-75-0792 (AARHUS)",
    doi = "10.1016/0550-3213(75)90301-6",
    journal = "Nucl. Phys. B",
    volume = "101",
    pages = "173--188",
    year = "1975"
}

@article{Fukuda:1975di,
    author = "Fukuda, Reijiro and Kugo, Taichiro",
    title = "{Gauge Invariance in the Effective Action and Potential}",
    reportNumber = "RIFP-237",
    doi = "10.1103/PhysRevD.13.3469",
    journal = "Phys. Rev. D",
    volume = "13",
    pages = "3469",
    year = "1976"
}

@article{Rajantie:1996np,
    author = "Rajantie, Arttu K.",
    title = "{Feynman diagrams to three loops in three-dimensional field theory}",
    eprint = "hep-ph/9606216",
    archivePrefix = "arXiv",
    reportNumber = "HU-TFT-96-22",
    doi = "10.1016/S0550-3213(96)00474-9",
    journal = "Nucl. Phys. B",
    volume = "480",
    pages = "729--752",
    year = "1996",
    note = "[Erratum: Nucl.Phys.B 513, 761--762 (1998)]"
}

@article{Laine:2022ytc,
    author = "Laine, M. and Niemi, L. and Procacci, S. and Rummukainen, K.",
    title = "{Shape of the hot topological charge density spectral function}",
    eprint = "2209.13804",
    archivePrefix = "arXiv",
    primaryClass = "hep-ph",
    doi = "10.1007/JHEP11(2022)126",
    journal = "JHEP",
    volume = "11",
    pages = "126",
    year = "2022"
}

@article{Moore:1999fs,
    author = "Moore, Guy D. and Rummukainen, Kari",
    title = "{Classical sphaleron rate on fine lattices}",
    eprint = "hep-ph/9906259",
    archivePrefix = "arXiv",
    reportNumber = "MCGILL-99-21, NORDITA-99-33HE",
    doi = "10.1103/PhysRevD.61.105008",
    journal = "Phys. Rev. D",
    volume = "61",
    pages = "105008",
    year = "2000"
}

@article{Altenkort:2020axj,
    author = "Altenkort, Luis and Eller, Alexander M. and Kaczmarek, Olaf and Mazur, Lukas and Moore, Guy D. and Shu, Hai-Tao",
    title = "{Sphaleron rate from Euclidean lattice correlators: An exploration}",
    eprint = "2012.08279",
    archivePrefix = "arXiv",
    primaryClass = "hep-lat",
    doi = "10.1103/PhysRevD.103.114513",
    journal = "Phys. Rev. D",
    volume = "103",
    number = "11",
    pages = "114513",
    year = "2021"
}

@article{Postma:2022dbr,
    author = "Postma, Marieke and van de Vis, Jorinde and White, Graham",
    title = "{Resummation and cancellation of the VIA source in electroweak baryogenesis}",
    eprint = "2206.01120",
    archivePrefix = "arXiv",
    primaryClass = "hep-ph",
    reportNumber = "Nikhef 2022-007, DESY-22-092, IPMU22-0034",
    doi = "10.1007/JHEP12(2022)121",
    journal = "JHEP",
    volume = "12",
    pages = "121",
    year = "2022"
}

@article{Laine:2017hdk,
    author = "Laine, M. and Meyer, M. and Nardini, G.",
    title = "{Thermal phase transition with full 2-loop effective potential}",
    eprint = "1702.07479",
    archivePrefix = "arXiv",
    primaryClass = "hep-ph",
    doi = "10.1016/j.nuclphysb.2017.04.023",
    journal = "Nucl. Phys. B",
    volume = "920",
    pages = "565--600",
    year = "2017"
}

@article{Laine:2000kv,
    author = "Laine, M. and Losada, M.",
    title = "{Two loop dimensional reduction and effective potential without temperature expansions}",
    eprint = "hep-ph/0003111",
    archivePrefix = "arXiv",
    reportNumber = "CERN-TH-2000-072",
    doi = "10.1016/S0550-3213(00)00298-4",
    journal = "Nucl. Phys. B",
    volume = "582",
    pages = "277--295",
    year = "2000"
}

@article{Bezuglov:2018qpq,
    author = "Bezuglov, M. A. and Onishchenko, A. I.",
    title = "{Two-loop corrections to false vacuum decay in scalar field theory}",
    eprint = "1805.06482",
    archivePrefix = "arXiv",
    primaryClass = "hep-ph",
    doi = "10.1016/j.physletb.2018.11.005",
    journal = "Phys. Lett. B",
    volume = "788",
    pages = "122--130",
    year = "2019"
}

@article{Tenkanen:2022tly,
    author = "Tenkanen, Tuomas V. I. and van de Vis, Jorinde",
    title = "{Speed of sound in cosmological phase transitions and effect on gravitational waves}",
    eprint = "2206.01130",
    archivePrefix = "arXiv",
    primaryClass = "hep-ph",
    reportNumber = "NORDITA 2022-031, DESY-22-091",
    doi = "10.1007/JHEP08(2022)302",
    journal = "JHEP",
    volume = "08",
    pages = "302",
    year = "2022"
}

@article{Liu:2011jh,
    author = "Liu, Tao and Ramsey-Musolf, Michael J. and Shu, Jing",
    title = "{Electroweak Beautygenesis: From b {\textbackslash{}to} s CP-violation to the Cosmic Baryon Asymmetry}",
    eprint = "1109.4145",
    archivePrefix = "arXiv",
    primaryClass = "hep-ph",
    reportNumber = "IPMU11-0076, NPAC-11-10, SISSA-48-2011-EP",
    doi = "10.1103/PhysRevLett.108.221301",
    journal = "Phys. Rev. Lett.",
    volume = "108",
    pages = "221301",
    year = "2012"
}

@article{Enomoto:2022rrl,
    author = "Enomoto, Kazuki and Kanemura, Shinya and Mura, Yushi",
    title = "{New benchmark scenarios of electroweak baryogenesis in aligned two Higgs double models}",
    eprint = "2207.00060",
    archivePrefix = "arXiv",
    primaryClass = "hep-ph",
    reportNumber = "OU-HET-1147",
    doi = "10.1007/JHEP09(2022)121",
    journal = "JHEP",
    volume = "09",
    pages = "121",
    year = "2022"
}

@article{Liu:2023sey,
    author = "Liu, Songtao and Wang, Lei",
    title = "{Spontaneous CP violation electroweak baryogenesis and gravitational wave through multistep phase transitions}",
    eprint = "2302.04639",
    archivePrefix = "arXiv",
    primaryClass = "hep-ph",
    doi = "10.1103/PhysRevD.107.115008",
    journal = "Phys. Rev. D",
    volume = "107",
    number = "11",
    pages = "115008",
    year = "2023"
}

@article{Aoki:2023xnn,
    author = "Aoki, Mayumi and Shibuya, Hiroto",
    title = "{Electroweak baryogenesis between broken phases in multi-step phase transition}",
    eprint = "2302.11551",
    archivePrefix = "arXiv",
    primaryClass = "hep-ph",
    reportNumber = "KANAZAWA-23-03",
    doi = "10.1016/j.physletb.2023.138041",
    journal = "Phys. Lett. B",
    volume = "843",
    pages = "138041",
    year = "2023"
}

@article{Enomoto:2021dkl,
    author = "Enomoto, Kazuki and Kanemura, Shinya and Mura, Yushi",
    title = "{Electroweak baryogenesis in aligned two Higgs doublet models}",
    eprint = "2111.13079",
    archivePrefix = "arXiv",
    primaryClass = "hep-ph",
    reportNumber = "OU-HET-1114",
    doi = "10.1007/JHEP01(2022)104",
    journal = "JHEP",
    volume = "01",
    pages = "104",
    year = "2022"
}

@article{Kainulainen:2021oqs,
    author = "Kainulainen, Kimmo",
    title = "{CP-violating transport theory for electroweak baryogenesis with thermal corrections}",
    eprint = "2108.08336",
    archivePrefix = "arXiv",
    primaryClass = "hep-ph",
    doi = "10.1088/1475-7516/2021/11/042",
    journal = "JCAP",
    volume = "11",
    number = "11",
    pages = "042",
    year = "2021"
}

@article{Cline:2021dkf,
    author = "Cline, James M. and Laurent, Benoit",
    title = "{Electroweak baryogenesis from light fermion sources: A critical study}",
    eprint = "2108.04249",
    archivePrefix = "arXiv",
    primaryClass = "hep-ph",
    doi = "10.1103/PhysRevD.104.083507",
    journal = "Phys. Rev. D",
    volume = "104",
    number = "8",
    pages = "083507",
    year = "2021"
}

@article{Modak:2021vre,
    author = "Modak, Tanmoy and Senaha, Eibun",
    title = "{Electroweak baryogenesis via bottom transport: Complementarity between LHC and future lepton collider probes}",
    eprint = "2107.12789",
    archivePrefix = "arXiv",
    primaryClass = "hep-ph",
    doi = "10.1016/j.physletb.2021.136695",
    journal = "Phys. Lett. B",
    volume = "822",
    pages = "136695",
    year = "2021"
}

@article{Zhou:2020irf,
    author = "Zhou, Ruiyu and Bian, Ligong",
    title = "{Gravitational wave and electroweak baryogenesis with two Higgs doublet models}",
    eprint = "2001.01237",
    archivePrefix = "arXiv",
    primaryClass = "hep-ph",
    doi = "10.1016/j.physletb.2022.137105",
    journal = "Phys. Lett. B",
    volume = "829",
    pages = "137105",
    year = "2022"
}

@article{Cline:2020jre,
    author = "Cline, James M. and Kainulainen, Kimmo",
    title = "{Electroweak baryogenesis at high bubble wall velocities}",
    eprint = "2001.00568",
    archivePrefix = "arXiv",
    primaryClass = "hep-ph",
    reportNumber = "CERN-TH-2019-227",
    doi = "10.1103/PhysRevD.101.063525",
    journal = "Phys. Rev. D",
    volume = "101",
    number = "6",
    pages = "063525",
    year = "2020"
}

@article{Bell:2019mbn,
    author = "Bell, Nicole F. and Dolan, Matthew J. and Friedrich, Leon S. and Ramsey-Musolf, Michael J. and Volkas, Raymond R.",
    title = "{Electroweak Baryogenesis with Vector-like Leptons and Scalar Singlets}",
    eprint = "1903.11255",
    archivePrefix = "arXiv",
    primaryClass = "hep-ph",
    reportNumber = "ACFI-T19-02",
    doi = "10.1007/JHEP09(2019)012",
    journal = "JHEP",
    volume = "09",
    pages = "012",
    year = "2019"
}

@article{deVries:2017ncy,
    author = "de Vries, Jordy and Postma, Marieke and van de Vis, Jorinde and White, Graham",
    title = "{Electroweak Baryogenesis and the Standard Model Effective Field Theory}",
    eprint = "1710.04061",
    archivePrefix = "arXiv",
    primaryClass = "hep-ph",
    reportNumber = "Nikhef-2017-044",
    doi = "10.1007/JHEP01(2018)089",
    journal = "JHEP",
    volume = "01",
    pages = "089",
    year = "2018"
}

@article{Ramsey-Musolf:2017tgh,
    author = "Ramsey-Musolf, Michael J. and Winslow, Peter and White, Graham",
    title = "{Color Breaking Baryogenesis}",
    eprint = "1708.07511",
    archivePrefix = "arXiv",
    primaryClass = "hep-ph",
    reportNumber = "ACFI-T17-20",
    doi = "10.1103/PhysRevD.97.123509",
    journal = "Phys. Rev. D",
    volume = "97",
    number = "12",
    pages = "123509",
    year = "2018"
}

@article{Chiang:2016vgf,
    author = "Chiang, Cheng-Wei and Fuyuto, Kaori and Senaha, Eibun",
    title = "{Electroweak Baryogenesis with Lepton Flavor Violation}",
    eprint = "1607.07316",
    archivePrefix = "arXiv",
    primaryClass = "hep-ph",
    doi = "10.1016/j.physletb.2016.09.052",
    journal = "Phys. Lett. B",
    volume = "762",
    pages = "315--320",
    year = "2016"
}

@article{Inoue:2015pza,
    author = "Inoue, Satoru and Ovanesyan, Grigory and Ramsey-Musolf, Michael J.",
    title = "{Two-Step Electroweak Baryogenesis}",
    eprint = "1508.05404",
    archivePrefix = "arXiv",
    primaryClass = "hep-ph",
    reportNumber = "ACFI-T15-12",
    doi = "10.1103/PhysRevD.93.015013",
    journal = "Phys. Rev. D",
    volume = "93",
    pages = "015013",
    year = "2016"
}

@article{Shu:2013uua,
    author = "Shu, Jing and Zhang, Yue",
    title = "{Impact of a CP Violating Higgs Sector: From LHC to Baryogenesis}",
    eprint = "1304.0773",
    archivePrefix = "arXiv",
    primaryClass = "hep-ph",
    reportNumber = "CAS-KITPC-ITP-365, CALT-68-2926",
    doi = "10.1103/PhysRevLett.111.091801",
    journal = "Phys. Rev. Lett.",
    volume = "111",
    number = "9",
    pages = "091801",
    year = "2013"
}

@phdthesis{Jukkala:2022jzy,
    author = "Jukkala, Henri",
    title = "{Quantum coherence in relativistic transport theory : applications to baryogenesis}",
    eprint = "2211.11785",
    archivePrefix = "arXiv",
    primaryClass = "hep-ph",
    school = "Jyvaskyla U, Jyvaskyla U.",
    year = "2022"
}

@article{Kainulainen:2024qpm,
    author = "Kainulainen, Kimmo and Venkatesan, Niyati",
    title = "{Systematic moment expansion for electroweak baryogenesis}",
    eprint = "2407.13639",
    archivePrefix = "arXiv",
    primaryClass = "hep-ph",
    doi = "10.1088/1475-7516/2024/08/058",
    journal = "JCAP",
    volume = "08",
    pages = "058",
    year = "2024"
}

@article{Kudoh:2005as,
    author = "Kudoh, Hideaki and Taruya, Atsushi and Hiramatsu, Takashi and Himemoto, Yoshiaki",
    title = "{Detecting a gravitational-wave background with next-generation space interferometers}",
    eprint = "gr-qc/0511145",
    archivePrefix = "arXiv",
    reportNumber = "UTAP-544, RESCEU-37-05",
    doi = "10.1103/PhysRevD.73.064006",
    journal = "Phys. Rev. D",
    volume = "73",
    pages = "064006",
    year = "2006"
}

@article{Kawamura:2011zz,
    author = "Kawamura, Seiji and others",
    editor = "Buchman, Sasha and Sun, Ke-Xun",
    title = "{The Japanese space gravitational wave antenna: DECIGO}",
    doi = "10.1088/0264-9381/28/9/094011",
    journal = "Class. Quant. Grav.",
    volume = "28",
    pages = "094011",
    year = "2011"
}

@article{Yagi:2011wg,
    author = "Yagi, Kent and Seto, Naoki",
    title = "{Detector configuration of DECIGO/BBO and identification of cosmological neutron-star binaries}",
    eprint = "1101.3940",
    archivePrefix = "arXiv",
    primaryClass = "astro-ph.CO",
    doi = "10.1103/PhysRevD.83.044011",
    journal = "Phys. Rev. D",
    volume = "83",
    pages = "044011",
    year = "2011",
    note = "[Erratum: Phys.Rev.D 95, 109901 (2017)]"
}

@article{Gong:2014mca,
    author = "Gong, Xuefei and others",
    editor = "Ciani, Giacomo and Conklin, John W. and Mueller, Guido",
    title = "{Descope of the ALIA mission}",
    eprint = "1410.7296",
    archivePrefix = "arXiv",
    primaryClass = "gr-qc",
    doi = "10.1088/1742-6596/610/1/012011",
    journal = "J. Phys. Conf. Ser.",
    volume = "610",
    number = "1",
    pages = "012011",
    year = "2015"
}

@article{Hu:2017mde,
    author = "Hu, Wen-Rui and Wu, Yue-Liang",
    title = "{The Taiji Program in Space for gravitational wave physics and the nature of gravity}",
    doi = "10.1093/nsr/nwx116",
    journal = "Natl. Sci. Rev.",
    volume = "4",
    number = "5",
    pages = "685--686",
    year = "2017"
}

@article{Guo:2018npi,
    author = "Ruan, Wen-Hong and Guo, Zong-Kuan and Cai, Rong-Gen and Zhang, Yuan-Zhong",
    title = "{Taiji program: Gravitational-wave sources}",
    eprint = "1807.09495",
    archivePrefix = "arXiv",
    primaryClass = "gr-qc",
    doi = "10.1142/S0217751X2050075X",
    journal = "Int. J. Mod. Phys. A",
    volume = "35",
    number = "17",
    pages = "2050075",
    year = "2020"
}

@article{TianQin:2015yph,
    author = "Luo, Jun and others",
    collaboration = "TianQin",
    title = "{TianQin: a space-borne gravitational wave detector}",
    eprint = "1512.02076",
    archivePrefix = "arXiv",
    primaryClass = "astro-ph.IM",
    doi = "10.1088/0264-9381/33/3/035010",
    journal = "Class. Quant. Grav.",
    volume = "33",
    number = "3",
    pages = "035010",
    year = "2016"
}

@article{Hu:2017yoc,
    author = "Hu, Yi-Ming and Mei, Jianwei and Luo, Jun",
    title = "{Science prospects for space-borne gravitational-wave missions}",
    doi = "10.1093/nsr/nwx115",
    journal = "Natl. Sci. Rev.",
    volume = "4",
    number = "5",
    pages = "683--684",
    year = "2017"
}

@article{Crowder:2005nr,
    author = "Crowder, Jeff and Cornish, Neil J.",
    title = "{Beyond LISA: Exploring future gravitational wave missions}",
    eprint = "gr-qc/0506015",
    archivePrefix = "arXiv",
    doi = "10.1103/PhysRevD.72.083005",
    journal = "Phys. Rev. D",
    volume = "72",
    pages = "083005",
    year = "2005"
}

@article{Liang:2021bde,
   author = "Liang, Zheng-Cheng and Hu, Yi-Ming and Jiang, Yun and Cheng, Jun and Zhang, Jian-dong and Mei, Jianwei",
   title = "{Science with the TianQin Observatory: Preliminary results on stochastic gravitational-wave background}",
   eprint = "2107.08643",
   archivePrefix = "arXiv",
   primaryClass = "astro-ph.CO",
   doi = "10.1103/PhysRevD.105.022001",
   journal = "Phys. Rev. D",
   volume = "105",
   number = "2",
   pages = "022001",
   year = "2022"
}

@article{Musha:2017usi,
    author = "Musha, Mitsuru",
    editor = "Cugny, Bruno and Karafolas, Nikos and Sodnik, Zoran",
    collaboration = "DECIGO Working group",
    title = "{Space gravitational wave detector DECIGO/pre-DECIGO}",
    doi = "10.1117/12.2296050",
    journal = "Proc. SPIE Int. Soc. Opt. Eng.",
    volume = "10562",
    pages = "105623T",
    year = "2017"
}

@article{Ekstedt:2024fyq,
    author = "Ekstedt, Andreas and Gould, Oliver and Hirvonen, Joonas and Laurent, Benoit and Niemi, Lauri and Schicho, Philipp and van de Vis, Jorinde",
    title = "{How fast does the WallGo? A package for computing wall velocities in first-order phase transitions}",
    eprint = "2411.04970",
    archivePrefix = "arXiv",
    primaryClass = "hep-ph",
    reportNumber = "CERN-TH-2024-174, DESY-24-162, HIP-2024-21/TH",
    month = "11",
    year = "2024"
}

@article{Ramsey-Musolf:2024ykk,
    author = "Ramsey-Musolf, Michael J. and Tenkanen, Tuomas V. I. and Tran, Van Que",
    title = "{Refining Gravitational Wave and Collider Physics Dialogue via Singlet Scalar Extension}",
    eprint = "2409.17554",
    archivePrefix = "arXiv",
    primaryClass = "hep-ph",
    month = "9",
    year = "2024"
}

@article{Friedrich:2022cak,
    author = "Friedrich, Leon S. and Ramsey-Musolf, Michael J. and Tenkanen, Tuomas V. I. and Tran, Van Que",
    title = "{Addressing the Gravitational Wave - Collider Inverse Problem}",
    eprint = "2203.05889",
    archivePrefix = "arXiv",
    primaryClass = "hep-ph",
    reportNumber = "NORDITA 2022-010",
    month = "3",
    year = "2022"
}

@article{Gould:2024jjt,
    author = "Gould, Oliver and Saffin, Paul M.",
    title = "{Perturbative gravitational wave predictions for the real-scalar extended Standard Model}",
    eprint = "2411.08951",
    archivePrefix = "arXiv",
    primaryClass = "hep-ph",
    doi = "10.1007/JHEP03(2025)105",
    journal = "JHEP",
    volume = "03",
    pages = "105",
    year = "2025"
}

@article{Patel:2012pi,
    author = "Patel, Hiren H. and Ramsey-Musolf, Michael J.",
    title = "{Stepping Into Electroweak Symmetry Breaking: Phase Transitions and Higgs Phenomenology}",
    eprint = "1212.5652",
    archivePrefix = "arXiv",
    primaryClass = "hep-ph",
    doi = "10.1103/PhysRevD.88.035013",
    journal = "Phys. Rev. D",
    volume = "88",
    pages = "035013",
    year = "2013"
}

@mastersthesis{Niemi:2018juv,
    author = "Niemi, Lauri",
    title = "{Dimensional reduction in the study of the electroweak phase transition}",
    school = "U. Helsinki (main)",
    month = "4",
    year = "2018"
}

@article{Buchmuller:1995sf,
    author = "Buchmuller, W. and Fodor, Z. and Hebecker, Arthur",
    title = "{Thermodynamics of the electroweak phase transition}",
    eprint = "hep-ph/9502321",
    archivePrefix = "arXiv",
    reportNumber = "DESY-95-028",
    doi = "10.1016/0550-3213(95)00254-P",
    journal = "Nucl. Phys. B",
    volume = "447",
    pages = "317--339",
    year = "1995"
}

@article{Kajantie:1996qd,
    author = "Kajantie, K. and Laine, M. and Rummukainen, K. and Shaposhnikov, Mikhail E.",
    title = "{A Nonperturbative analysis of the finite T phase transition in SU(2) x U(1) electroweak theory}",
    eprint = "hep-lat/9612006",
    archivePrefix = "arXiv",
    reportNumber = "BI-TP-96-54, CERN-TH-96-334A, HD-THEP-96-48",
    doi = "10.1016/S0550-3213(97)00164-8",
    journal = "Nucl. Phys. B",
    volume = "493",
    pages = "413--438",
    year = "1997"
}

@article{Hart:1996ac,
    author = "Hart, A. and Philipsen, O. and Stack, J. D. and Teper, M.",
    title = "{On the phase diagram of the SU(2) adjoint Higgs model in (2+1)-dimensions}",
    eprint = "hep-lat/9612021",
    archivePrefix = "arXiv",
    reportNumber = "LSUHE-249-1996, HD-THEP-96-58, ILL-TH-96-17, OUTP-96-75-P",
    doi = "10.1016/S0370-2693(97)00104-4",
    journal = "Phys. Lett. B",
    volume = "396",
    pages = "217--224",
    year = "1997"
}

@article{Kajantie:1997tt,
    author = "Kajantie, K. and Laine, M. and Rummukainen, K. and Shaposhnikov, Mikhail E.",
    title = "{3-D SU(N) + adjoint Higgs theory and finite temperature QCD}",
    eprint = "hep-ph/9704416",
    archivePrefix = "arXiv",
    reportNumber = "CERN-TH-97-081, CERN-TH-97-81, BI-TP-97-10, HD-THEP-97-17",
    doi = "10.1016/S0550-3213(97)00425-2",
    journal = "Nucl. Phys. B",
    volume = "503",
    pages = "357--384",
    year = "1997"
}

@article{Kajantie:1998yc,
    author = "Kajantie, K. and Laine, M. and Rajantie, A. and Rummukainen, K. and Tsypin, M.",
    title = "{The Phase diagram of three-dimensional SU(3) + adjoint Higgs theory}",
    eprint = "hep-lat/9811004",
    archivePrefix = "arXiv",
    reportNumber = "CERN-TH-98-350, NORDITA-98-66-HE",
    doi = "10.1088/1126-6708/1998/11/011",
    journal = "JHEP",
    volume = "11",
    pages = "011",
    year = "1998"
}

@article{Catumba:2024jau,
    author = "Catumba, Guilherme and Hiraguchi, Atsuki and Hou, Wei-Shu and Jansen, Karl and Kao, Ying-Jer and Lin, C. -J. David and Ramos, Alberto and Sarkar, Mugdha",
    title = "{Lattice study of SU(2) gauge theory coupled to four adjoint Higgs fields}",
    eprint = "2407.15422",
    archivePrefix = "arXiv",
    primaryClass = "hep-lat",
    doi = "10.1103/PhysRevResearch.6.043172",
    journal = "Phys. Rev. Res.",
    volume = "6",
    number = "4",
    pages = "043172",
    year = "2024"
}

@article{Niemi:2022bjg,
    author = {Niemi, Lauri and Rummukainen, Kari and Sepp\"a, Riikka and Weir, David J.},
    title = "{Infrared physics of the 3D SU(2) adjoint Higgs model at the crossover transition}",
    eprint = "2206.14487",
    archivePrefix = "arXiv",
    primaryClass = "hep-lat",
    reportNumber = "HIP-2022-18/TH",
    doi = "10.1007/JHEP02(2023)212",
    journal = "JHEP",
    volume = "02",
    pages = "212",
    year = "2023"
}

@article{Hirvonen:2024rfg,
    author = "Hirvonen, Joonas",
    title = "{Nucleation Rate in a High-Temperature Quantum Field Theory with Hard Particles}",
    eprint = "2403.07987",
    archivePrefix = "arXiv",
    primaryClass = "hep-ph",
    reportNumber = "HIP-2024-6/TH",
    month = "3",
    year = "2024"
}

@article{Moore:1995jv,
    author = "Moore, Guy D.",
    title = "{Fermion determinant and the sphaleron bound}",
    eprint = "hep-ph/9508405",
    archivePrefix = "arXiv",
    reportNumber = "PUPT-1557, PUP-TH-1557",
    doi = "10.1103/PhysRevD.53.5906",
    journal = "Phys. Rev. D",
    volume = "53",
    pages = "5906--5917",
    year = "1996"
}

@article{Bernardo:2025vkz,
    author = "Bernardo, Fabio and Klose, Philipp and Schicho, Philipp and Tenkanen, Tuomas V. I.",
    title = "{Higher-dimensional operators at finite-temperature affect gravitational-wave predictions}",
    eprint = "2503.18904",
    archivePrefix = "arXiv",
    primaryClass = "hep-ph",
    reportNumber = "HIP-2025-6/TH",
    month = "3",
    year = "2025"
}

@article{Chala:2025oul,
    author = "Chala, Mikael and Gil, Luis and Ren, Zhe",
    title = "{Phase Transitions in Dimensional Reduction up to Three Loops}",
    eprint = "2505.14335",
    archivePrefix = "arXiv",
    primaryClass = "hep-ph",
    month = "5",
    year = "2025"
}

@article{Chala:2025aiz,
    author = "Chala, Mikael and Guedes, Guilherme",
    title = "{The High-Temperature Limit of the SM(EFT)}",
    eprint = "2503.20016",
    archivePrefix = "arXiv",
    primaryClass = "hep-ph",
    month = "3",
    year = "2025"
}

@article{Bonati:2024sok,
    author = "Bonati, Claudio and Pelissetto, Andrea and Vicari, Ettore",
    title = "{Three-dimensional Abelian and non-Abelian gauge Higgs theories}",
    eprint = "2410.05823",
    archivePrefix = "arXiv",
    primaryClass = "cond-mat.stat-mech",
    month = "10",
    year = "2024"
}

@article{York:2014ada,
    author = "York, Mark C. Abraao and Moore, Guy D.",
    title = "{2PI resummation in 3D SU(N ) Higgs theory}",
    eprint = "1407.3816",
    archivePrefix = "arXiv",
    primaryClass = "hep-ph",
    doi = "10.1007/JHEP10(2014)105",
    journal = "JHEP",
    volume = "10",
    pages = "105",
    year = "2014"
}

@article{Barenboim:2012nh,
    author = "Barenboim, Gabriela and Rasero, Javier",
    title = "{Electroweak baryogenesis window in non standard cosmologies}",
    eprint = "1202.6070",
    archivePrefix = "arXiv",
    primaryClass = "hep-ph",
    doi = "10.1007/JHEP07(2012)028",
    journal = "JHEP",
    volume = "07",
    pages = "028",
    year = "2012"
}

@article{Dosch:1995uz,
    author = "Dosch, Hans Gunter and Kripfganz, Jochen and Laser, Andreas and Schmidt, Michael G.",
    title = "{Bound states in the hot electroweak phase}",
    eprint = "hep-ph/9509352",
    archivePrefix = "arXiv",
    reportNumber = "HD-THEP-95-42",
    doi = "10.1016/0370-2693(95)01269-9",
    journal = "Phys. Lett. B",
    volume = "365",
    pages = "213--218",
    year = "1996"
}

@article{Buchmuller:1994kk,
    author = "Buchmuller, W. and Fodor, Z.",
    title = "{Confinement in three-dimensions and the electroweak phase transition}",
    eprint = "hep-ph/9403388",
    archivePrefix = "arXiv",
    reportNumber = "DESY-94-045",
    doi = "10.1016/0370-2693(94)90952-0",
    journal = "Phys. Lett. B",
    volume = "331",
    pages = "124--130",
    year = "1994"
}

@article{Damgaard:1985nb,
    author = "Damgaard, P. H. and Heller, Urs M.",
    title = "{Higgs and Confinement Phases in the Fundamental SU(2) Higgs Model: Mean Field Analysis}",
    reportNumber = "CERN-TH-4246/85",
    doi = "10.1016/0370-2693(85)90044-9",
    journal = "Phys. Lett. B",
    volume = "164",
    pages = "121--126",
    year = "1985"
}

@article{Hallfors:2025key,
    author = {H\"allfors, Jaakko and Rummukainen, Kari},
    title = "{Expanded ensemble method for bubble nucleation}",
    eprint = "2502.14610",
    archivePrefix = "arXiv",
    primaryClass = "hep-lat",
    month = "2",
    year = "2025"
}

@article{Rubakov:1996vz,
    author = "Rubakov, V. A. and Shaposhnikov, M. E.",
    title = "{Electroweak baryon number nonconservation in the early universe and in high-energy collisions}",
    eprint = "hep-ph/9603208",
    archivePrefix = "arXiv",
    reportNumber = "CERN-TH-96-13, INR-0913-96",
    doi = "10.1070/PU1996v039n05ABEH000145",
    journal = "Usp. Fiz. Nauk",
    volume = "166",
    pages = "493--537",
    year = "1996"
}

@article{LHCb:2024exp,
    author = "Aaij, R. and others",
    collaboration = "LHCb",
    title = "{First Evidence for Direct CP Violation in Beauty to Charmonium Decays}",
    eprint = "2411.12178",
    archivePrefix = "arXiv",
    primaryClass = "hep-ex",
    reportNumber = "LHCb-PAPER-2024-031, CERN-EP-2024-286",
    doi = "10.1103/PhysRevLett.134.101801",
    journal = "Phys. Rev. Lett.",
    volume = "134",
    number = "10",
    pages = "101801",
    year = "2025"
}

@article{Bernreuther:2002uj,
    author = "Bernreuther, Werner",
    title = "{CP violation and baryogenesis}",
    eprint = "hep-ph/0205279",
    archivePrefix = "arXiv",
    reportNumber = "PITHA-02-08",
    journal = "Lect. Notes Phys.",
    volume = "591",
    pages = "237--293",
    year = "2002"
}

@article{Bodeker:2004ws,
    author = "Bodeker, Dietrich and Fromme, Lars and Huber, Stephan J. and Seniuch, Michael",
    title = "{The Baryon asymmetry in the standard model with a low cut-off}",
    eprint = "hep-ph/0412366",
    archivePrefix = "arXiv",
    reportNumber = "BI-TP-2004-41, CERN-PH-TH-2004-258",
    doi = "10.1088/1126-6708/2005/02/026",
    journal = "JHEP",
    volume = "02",
    pages = "026",
    year = "2005"
}

@article{Fromme:2006wx,
    author = "Fromme, Lars and Huber, Stephan J.",
    title = "{Top transport in electroweak baryogenesis}",
    eprint = "hep-ph/0604159",
    archivePrefix = "arXiv",
    reportNumber = "CERN-PH-TH-2006-064, BI-TP-2006-10",
    doi = "10.1088/1126-6708/2007/03/049",
    journal = "JHEP",
    volume = "03",
    pages = "049",
    year = "2007"
}

@article{Fromme:2006cm,
    author = "Fromme, Lars and Huber, Stephan J. and Seniuch, Michael",
    title = "{Baryogenesis in the two-Higgs doublet model}",
    eprint = "hep-ph/0605242",
    archivePrefix = "arXiv",
    reportNumber = "CERN-PH-TH-2006-094, BI-TP-2006-18",
    doi = "10.1088/1126-6708/2006/11/038",
    journal = "JHEP",
    volume = "11",
    pages = "038",
    year = "2006"
}

@article{Konstandin:2013caa,
    author = "Konstandin, Thomas",
    title = "{Quantum Transport and Electroweak Baryogenesis}",
    eprint = "1302.6713",
    archivePrefix = "arXiv",
    primaryClass = "hep-ph",
    reportNumber = "DESY-13-036",
    doi = "10.3367/UFNe.0183.201308a.0785",
    journal = "Phys. Usp.",
    volume = "56",
    pages = "747--771",
    year = "2013"
}

@article{Strumia:1998nf,
    author = "Strumia, Alessandro and Tetradis, Nikolaos",
    title = "{A Consistent calculation of bubble nucleation rates}",
    eprint = "hep-ph/9806453",
    archivePrefix = "arXiv",
    reportNumber = "SNS-PH-1998-12, IFUP-TH-98-24",
    doi = "10.1016/S0550-3213(98)00804-9",
    journal = "Nucl. Phys. B",
    volume = "542",
    pages = "719--741",
    year = "1999"
}

@article{Garny:2012cg,
    author = "Garny, Mathias and Konstandin, Thomas",
    title = "{On the gauge dependence of vacuum transitions at finite temperature}",
    eprint = "1205.3392",
    archivePrefix = "arXiv",
    primaryClass = "hep-ph",
    reportNumber = "DESY-12-073, CERN-PH-TH-2012-127",
    doi = "10.1007/JHEP07(2012)189",
    journal = "JHEP",
    volume = "07",
    pages = "189",
    year = "2012"
}

@article{Wainwright:2011qy,
    author = "Wainwright, Carroll and Profumo, Stefano and Ramsey-Musolf, Michael J.",
    title = "{Gravity Waves from a Cosmological Phase Transition: Gauge Artifacts and Daisy Resummations}",
    eprint = "1104.5487",
    archivePrefix = "arXiv",
    primaryClass = "hep-ph",
    reportNumber = "NPAC-11-04",
    doi = "10.1103/PhysRevD.84.023521",
    journal = "Phys. Rev. D",
    volume = "84",
    pages = "023521",
    year = "2011"
}

@article{Profumo:2014opa,
    author = "Profumo, Stefano and Ramsey-Musolf, Michael J. and Wainwright, Carroll L. and Winslow, Peter",
    title = "{Singlet-catalyzed electroweak phase transitions and precision Higgs boson studies}",
    eprint = "1407.5342",
    archivePrefix = "arXiv",
    primaryClass = "hep-ph",
    doi = "10.1103/PhysRevD.91.035018",
    journal = "Phys. Rev. D",
    volume = "91",
    number = "3",
    pages = "035018",
    year = "2015"
}

@article{Chao:2014ina,
    author = "Chao, Wei",
    title = "{First order electroweak phase transition triggered by the Higgs portal vector dark matter}",
    eprint = "1412.3823",
    archivePrefix = "arXiv",
    primaryClass = "hep-ph",
    reportNumber = "ACFI-T14-26",
    doi = "10.1103/PhysRevD.92.015025",
    journal = "Phys. Rev. D",
    volume = "92",
    number = "1",
    pages = "015025",
    year = "2015"
}

@article{Blinov:2015sna,
    author = "Blinov, Nikita and Kozaczuk, Jonathan and Morrissey, David E. and Tamarit, Carlos",
    title = "{Electroweak Baryogenesis from Exotic Electroweak Symmetry Breaking}",
    eprint = "1504.05195",
    archivePrefix = "arXiv",
    primaryClass = "hep-ph",
    reportNumber = "IPPP-15-23, DCPT-15-46",
    doi = "10.1103/PhysRevD.92.035012",
    journal = "Phys. Rev. D",
    volume = "92",
    number = "3",
    pages = "035012",
    year = "2015"
}

@article{Zhang:2023jvh,
    author = "Zhang, Wenxing and Li, Hao-Lin and Liu, Kun and Ramsey-Musolf, Michael J. and Zeng, Yonghao and Arunasalam, Suntharan",
    title = "{Probing electroweak phase transition in the singlet Standard Model via bb\ensuremath{\gamma}\ensuremath{\gamma} and 4l channels}",
    eprint = "2303.03612",
    archivePrefix = "arXiv",
    primaryClass = "hep-ph",
    doi = "10.1007/JHEP12(2023)018",
    journal = "JHEP",
    volume = "12",
    pages = "018",
    year = "2023"
}

@article{Zhang:2023mnu,
    author = "Zhang, Wenxing and Cai, Yizhou and Ramsey-Musolf, Michael J. and Zhang, Lei",
    title = "{Testing complex singlet scalar cosmology at the Large Hadron Collider}",
    eprint = "2307.01615",
    archivePrefix = "arXiv",
    primaryClass = "hep-ph",
    doi = "10.1007/JHEP01(2024)051",
    journal = "JHEP",
    volume = "01",
    pages = "051",
    year = "2024"
}

@article{Qin:2024dfp,
    author = "Qin, Renhui and Bian, Ligong",
    title = "{First-order electroweak phase transition with a gauge-invariant approach}",
    eprint = "2408.09677",
    archivePrefix = "arXiv",
    primaryClass = "hep-ph",
    doi = "10.1103/PhysRevD.111.L051702",
    journal = "Phys. Rev. D",
    volume = "111",
    number = "5",
    pages = "L051702",
    year = "2025"
}

@article{Kozaczuk:2015owa,
    author = "Kozaczuk, Jonathan",
    title = "{Bubble Expansion and the Viability of Singlet-Driven Electroweak Baryogenesis}",
    eprint = "1506.04741",
    archivePrefix = "arXiv",
    primaryClass = "hep-ph",
    doi = "10.1007/JHEP10(2015)135",
    journal = "JHEP",
    volume = "10",
    pages = "135",
    year = "2015"
}

@article{Kotwal:2016tex,
    author = "Kotwal, Ashutosh V. and Ramsey-Musolf, Michael J. and No, Jose Miguel and Winslow, Peter",
    title = "{Singlet-catalyzed electroweak phase transitions in the 100 TeV frontier}",
    eprint = "1605.06123",
    archivePrefix = "arXiv",
    primaryClass = "hep-ph",
    reportNumber = "ACFI-T16-12, FERMILAB-PUB-16-670",
    doi = "10.1103/PhysRevD.94.035022",
    journal = "Phys. Rev. D",
    volume = "94",
    number = "3",
    pages = "035022",
    year = "2016"
}

@article{Chao:2017vrq,
    author = "Chao, Wei and Guo, Huai-Ke and Shu, Jing",
    title = "{Gravitational Wave Signals of Electroweak Phase Transition Triggered by Dark Matter}",
    eprint = "1702.02698",
    archivePrefix = "arXiv",
    primaryClass = "hep-ph",
    doi = "10.1088/1475-7516/2017/09/009",
    journal = "JCAP",
    volume = "09",
    pages = "009",
    year = "2017"
}

@article{Chen:2017qcz,
    author = "Chen, Chien-Yi and Kozaczuk, Jonathan and Lewis, Ian M.",
    title = "{Non-resonant Collider Signatures of a Singlet-Driven Electroweak Phase Transition}",
    eprint = "1704.05844",
    archivePrefix = "arXiv",
    primaryClass = "hep-ph",
    reportNumber = "ACFI-T17-07, SLAC-PUB-16951",
    doi = "10.1007/JHEP08(2017)096",
    journal = "JHEP",
    volume = "08",
    pages = "096",
    year = "2017"
}

@article{Kang:2017mkl,
    author = "Kang, Zhaofeng and Ko, P. and Matsui, Toshinori",
    title = "{Strong first order EWPT $\&$ strong gravitational waves in Z$_{3}$-symmetric singlet scalar extension}",
    eprint = "1706.09721",
    archivePrefix = "arXiv",
    primaryClass = "hep-ph",
    reportNumber = "KIAS-P17047",
    doi = "10.1007/JHEP02(2018)115",
    journal = "JHEP",
    volume = "02",
    pages = "115",
    year = "2018"
}

@article{Chiang:2017zbz,
    author = "Chiang, Cheng-Wei and Senaha, Eibun",
    title = "{On gauge dependence of gravitational waves from a first-order phase transition in classical scale-invariant $U(1)'$ models}",
    eprint = "1707.06765",
    archivePrefix = "arXiv",
    primaryClass = "hep-ph",
    reportNumber = "NCTS-PH-1723",
    doi = "10.1016/j.physletb.2017.09.064",
    journal = "Phys. Lett. B",
    volume = "774",
    pages = "489--493",
    year = "2017"
}

@article{Chiang:2017nmu,
    author = "Chiang, Cheng-Wei and Ramsey-Musolf, Michael J. and Senaha, Eibun",
    title = "{Standard Model with a Complex Scalar Singlet: Cosmological Implications and Theoretical Considerations}",
    eprint = "1707.09960",
    archivePrefix = "arXiv",
    primaryClass = "hep-ph",
    reportNumber = "NCTS-PH-1724, ACFI-T17-16",
    doi = "10.1103/PhysRevD.97.015005",
    journal = "Phys. Rev. D",
    volume = "97",
    number = "1",
    pages = "015005",
    year = "2018"
}

@article{Chiang:2018gsn,
    author = "Chiang, Cheng-Wei and Li, Yen-Ting and Senaha, Eibun",
    title = "{Revisiting electroweak phase transition in the standard model with a real singlet scalar}",
    eprint = "1808.01098",
    archivePrefix = "arXiv",
    primaryClass = "hep-ph",
    reportNumber = "CTPU-PTC-18-24",
    doi = "10.1016/j.physletb.2018.12.017",
    journal = "Phys. Lett. B",
    volume = "789",
    pages = "154--159",
    year = "2019"
}

@article{Alves:2018jsw,
    author = "Alves, Alexandre and Ghosh, Tathagata and Guo, Huai-Ke and Sinha, Kuver and Vagie, Daniel",
    title = "{Collider and Gravitational Wave Complementarity in Exploring the Singlet Extension of the Standard Model}",
    eprint = "1812.09333",
    archivePrefix = "arXiv",
    primaryClass = "hep-ph",
    doi = "10.1007/JHEP04(2019)052",
    journal = "JHEP",
    volume = "04",
    pages = "052",
    year = "2019"
}

@article{Zhou:2019uzq,
    author = "Zhou, Ruiyu and Bian, Ligong and Guo, Huai-Ke",
    title = "{Connecting the electroweak sphaleron with gravitational waves}",
    eprint = "1910.00234",
    archivePrefix = "arXiv",
    primaryClass = "hep-ph",
    doi = "10.1103/PhysRevD.101.091903",
    journal = "Phys. Rev. D",
    volume = "101",
    number = "9",
    pages = "091903",
    year = "2020"
}

@article{Kozaczuk:2019pet,
    author = "Kozaczuk, Jonathan and Ramsey-Musolf, Michael J. and Shelton, Jessie",
    title = "{Exotic Higgs boson decays and the electroweak phase transition}",
    eprint = "1911.10210",
    archivePrefix = "arXiv",
    primaryClass = "hep-ph",
    reportNumber = "ACFI T19-10",
    doi = "10.1103/PhysRevD.101.115035",
    journal = "Phys. Rev. D",
    volume = "101",
    number = "11",
    pages = "115035",
    year = "2020"
}

@article{Ekstedt:2020abj,
    author = {Ekstedt, Andreas and L\"ofgren, Johan},
    title = "{A Critical Look at the Electroweak Phase Transition}",
    eprint = "2006.12614",
    archivePrefix = "arXiv",
    primaryClass = "hep-ph",
    doi = "10.1007/JHEP12(2020)136",
    journal = "JHEP",
    volume = "12",
    pages = "136",
    year = "2020"
}

@article{Kajantie:1997hn,
    author = "Kajantie, K. and Karjalainen, M. and Laine, M. and Peisa, J.",
    title = "{Three-dimensional U(1) gauge + Higgs theory as an effective theory for finite temperature phase transitions}",
    eprint = "hep-lat/9711048",
    archivePrefix = "arXiv",
    reportNumber = "CERN-TH-97-332",
    doi = "10.1016/S0550-3213(98)00064-9",
    journal = "Nucl. Phys. B",
    volume = "520",
    pages = "345--381",
    year = "1998"
}

@unpublished{Annala:2025xxx,
  author    = {Annala, Jaakko and Rummukainen, Kari and Tenkanen, Tuomas V.I. },
  title     = {Non-perturbative determination of the sphaleron rate for first-order phase transitions},
  note      = {In preparation},
  year      = {2025}
}

@article{Kierkla:2025qyz,
    author = "Kierkla, Maciej and Schicho, Philipp and Swiezewska, Bogumila and Tenkanen, Tuomas V. I. and van de Vis, Jorinde",
    title = "{Finite-temperature bubble nucleation with shifting scale hierarchies}",
    eprint = "2503.13597",
    archivePrefix = "arXiv",
    primaryClass = "hep-ph",
    reportNumber = "CERN-TH-2025-046, HIP-2024-27/TH",
    month = "3",
    year = "2025"
}

@book{Laine:2016hma,
    author = "Laine, Mikko and Vuorinen, Aleksi",
    title = "{Basics of Thermal Field Theory}",
    eprint = "1701.01554",
    archivePrefix = "arXiv",
    primaryClass = "hep-ph",
    doi = "10.1007/978-3-319-31933-9",
    publisher = "Springer",
    volume = "925",
    year = "2016"
}

@article{Gynther:2005av,
    author = "Gynther, A. and Vepsalainen, M.",
    title = "{Pressure of the standard model near the electroweak phase transition}",
    eprint = "hep-ph/0512177",
    archivePrefix = "arXiv",
    reportNumber = "HIP-2005-55-TH",
    doi = "10.1088/1126-6708/2006/03/011",
    journal = "JHEP",
    volume = "03",
    pages = "011",
    year = "2006"
}

@article{Fonseca:2020vke,
    author = "Fonseca, Renato M.",
    title = "{GroupMath: A Mathematica package for group theory calculations}",
    eprint = "2011.01764",
    archivePrefix = "arXiv",
    primaryClass = "hep-th",
    doi = "10.1016/j.cpc.2021.108085",
    journal = "Comput. Phys. Commun.",
    volume = "267",
    pages = "108085",
    year = "2021"
}

@article{Dosch:1996ex,
    author = "Dosch, Hans Gunter and Kripfganz, Jochen and Laser, Andreas and Schmidt, Michael G.",
    title = "{Nonperturbative correlation masses in the hot electroweak phase}",
    eprint = "hep-ph/9612450",
    archivePrefix = "arXiv",
    reportNumber = "HD-THEP-96-53, DO-TH-96-26",
    doi = "10.1016/S0550-3213(97)00439-2",
    journal = "Nucl. Phys. B",
    volume = "507",
    pages = "519--546",
    year = "1997"
}

@article{Braaten:1994na,
    author = "Braaten, Eric",
    title = "{Solution to the perturbative infrared catastrophe of hot gauge theories}",
    eprint = "hep-ph/9409434",
    archivePrefix = "arXiv",
    reportNumber = "NUHEP-TH-94-24",
    doi = "10.1103/PhysRevLett.74.2164",
    journal = "Phys. Rev. Lett.",
    volume = "74",
    pages = "2164--2167",
    year = "1995"
}

@article{Dashko:2024anp,
    author = "Dashko, Andrii and Ekstedt, Andreas",
    title = "{Bubble-wall speed with loop corrections}",
    eprint = "2411.05075",
    archivePrefix = "arXiv",
    primaryClass = "hep-ph",
    reportNumber = "DESY-24-169",
    doi = "10.1007/JHEP03(2025)024",
    journal = "JHEP",
    volume = "03",
    pages = "024",
    year = "2025"
}

\end{document}